\documentclass[apj]{emulateapj}

\usepackage{times}
\usepackage{natbib}
\usepackage[backref,breaklinks,colorlinks,citecolor=cyan]{hyperref} 
\usepackage{booktabs}
\usepackage{graphicx}
\usepackage{bm}
\usepackage{multirow}
\usepackage{enumitem}
\usepackage{amsmath}
\usepackage{float}
\usepackage[caption=false]{subfig}
\usepackage{booktabs}

\shorttitle{Cosmic evolution of the \mbh-host relation}
\shortauthors{X. Ding et al.}

\begin{document}
\def\lcdm{$\Lambda$CDM}
\def\hst{{\it HST}}
\def\efr{$R_{\mathrm{eff}}$}
\def\galfit{{\sc Galfit}}
\def\mbh{$\mathcal M_{\rm BH}$}
\def\lhost{$L_{\rm host}$}
\def\jcap{Journal of Cosmology and Astroparticle Physics}
\def\halpha{${\rm H}\alpha$}
\def\hbeta{${\rm H}\beta$}
\def\sersic{S\'ersic}	
\def\lenstronomy{{\sc Lenstronomy}}
\def\Reff{{$R_{\mathrm{eff}}$}}
\def\kms{km~s$^{\rm -1}$}
\def\sigstar{{$\sigma_*$}}
\def\smass{{$M_*$}}
\def\bmass{{$M_{*, \rm {bulge}}$}}
\newcommand{\Mgii}{Mg$_{\rm II}$}
\newcommand{\Civ}{C$_{\rm IV}$}
\newcommand{\sint}{$\sigma_{\rm int}$}
\newcommand{\angstrom}{\text{\normalfont\AA}}
\newcommand{\ding}[1]{\textcolor{red}{[{\bf Xuheng}: #1]}}

\newcommand{\red}[1]{{\bf \textcolor{red}{[#1]}}}
\newcommand{\blue}[1]{{\bf \textcolor{blue}{[#1]}}}

\makeatletter
\newcommand\footnoteref[1]{\protected@xdef\@thefnmark{\ref{#1}}\@footnotemark}  
\makeatother

\title{The mass relations between supermassive black holes and their host galaxies at $1< z<2$ with \hst-WFC3}

\author{Xuheng Ding\altaffilmark{1, 2}, John Silverman\altaffilmark{3,4}, Tommaso Treu\altaffilmark{1}, Andreas Schulze\altaffilmark{5}, Malte Schramm\altaffilmark{5}, Simon Birrer\altaffilmark{1}, Daeseong Park\altaffilmark{6}, Knud Jahnke\altaffilmark{7}, Vardha N. Bennert\altaffilmark{8}, Jeyhan S. Kartaltepe\altaffilmark{9}, Anton M. Koekemoer\altaffilmark{10},  Matthew A. Malkan\altaffilmark{1}, David Sanders\altaffilmark{11}
 }
 \email{dxh@astro.ucla.edu}
\altaffiltext{1}{Department of Physics and Astronomy, University of California, Los Angeles, CA, 90095-1547, USA}
\altaffiltext{2}{School of Physics and Technology, Wuhan University, Wuhan 430072, China}
\altaffiltext{3}{Kavli Institute for the Physics and Mathematics of the Universe (WPI), The University of Tokyo, Kashiwa, Chiba 277-8583, Japan}
\altaffiltext{4}{Department of Astronomy, School of Science, The University of Tokyo, 7-3-1 Hongo, Bunkyo, Tokyo 113-0033, Japan}
\altaffiltext{5}{National Astronomical Observatory of Japan, Mitaka, Tokyo 181-8588, Japan}
\altaffiltext{6}{Korea Astronomy and Space Science Institute, Deajeon, 34055, Republic of Korea}
\altaffiltext{7}{Max-Planck-Institut f\"ur Astronomie, K\"onigstuhl 17, D-69117, Heidelberg, Germany}
\altaffiltext{8}{Physics Department, California Polytechnic State University, San Luis Obispo CA 93407, USA}
\altaffiltext{9}{School of Physics and Astronomy, Rochester Institute of Technology, 84 Lomb Memorial Drive, Rochester, NY 14623, USA}
\altaffiltext{10}{Space Telescope Science Institute, 3700 San Martin Drive, Baltimore, MD 21218, USA}
\altaffiltext{11}{Institute for Astronomy, University of Hawaii, 2680 Woodlawn Drive, Honolulu, HI 96822, USA}
\begin{abstract}

Correlations between the mass of a supermassive black hole (SMBH) and the properties of its host galaxy (e.g., total stellar mass \smass, luminosity \lhost) suggest an evolutionary connection. A powerful test of a co-evolution scenario is to measure the relations \mbh-\lhost\ and \mbh-\smass~at high redshift and compare with local estimates. For this purpose, we acquired {\it Hubble Space Telescope} imaging with WFC3 of 32 \hbox{X-ray-selected} broad-line (type-1) AGN at $1.2<z<1.7$ in deep survey fields. By applying state-of-the-art tools to decompose the \hst\ images including available ACS data, we measured the host galaxy luminosity and stellar mass along with other properties through the 2D model fitting. The black hole mass (\mbh) was determined using the broad \halpha~line, detected in the near-infrared with Subaru/FMOS, which potentially minimizes systematic effects using other indicators. We find that the {\it observed} ratio of \mbh\ to total \smass\ is $2.7\times$ larger at $z\sim1.5$ than in the local universe, while the scatter is equivalent between the two epochs. A non-evolving mass ratio is consistent with the data at the 2-3$\sigma$ confidence level when accounting for selection effects (estimated using two independent and complementary methods), and their uncertainties. The relationship between \mbh\ and host galaxy total luminosity paints a similar picture. Therefore, our results cannot distinguish whether SMBHs and their total host stellar mass and luminosity proceed in lockstep or whether the growth of the former somewhat overshoots the latter, given the uncertainties. Based on a statistical estimate of the bulge-to-total mass fraction, the ratio \mbh/\bmass~is offset from the local value by a factor of $\sim7$ which is significant even accounting for selection effects. Taken together, these observations are consistent with a scenario in which stellar mass is subsequently transferred from an angular momentum supported component of the galaxy to the pressure supported one through secular processes or minor mergers at a faster rate than mass accretion onto the SMBH.

\end{abstract}


\section{Introduction}
\label{sec:introduction}

Most galactic nuclei are thought to harbor a supermassive black hole (SMBH), whose mass (\mbh) is known to correlate with the host properties, such as luminosity (\lhost), stellar mass (\smass), and stellar velocity dispersion (\sigstar). The tightness of these correlations (also known as scaling relations) may indicate a connection between nuclear activity, and galaxy formation and evolution~\citep[e.g.,][]{Mag++98, F+M00, M+H03, Gul++09,Beifi2012, H+R04, Geb++01b, Gra++2011}. Currently, the physical mechanism that can produce such a tight relationship is unknown due to the daunting range of scales between the dynamical sphere ($\sim$pc) of the SMBH and their host galaxy ($\sim10$ kpc). On one hand, cosmological simulations of structure formation are able to reproduce the mean local correlations, possibly by invoking active galactic nucleus (AGN) feedback as the physical connection~\citep{Springel2005, Hopkins2008, Matteo2008, DeG++15} or having them share a common gas supply~\citep{Cen2015, Menci2016}, not necessarily in a direct manner.
On the other hand, there may not be a need for a physical coupling \citep{Peng2007, Jahnke2011, Hirschmann2010}, the statistical convergence from galaxy assembly alone (i.e., mergers) may reproduce the observed correlations.

To understand the nature of these correlations, it is important to study them as a function of redshift, determining how and when they emerge and evolve over cosmic time~\citep[e.g.,][]{TMB04,Sal++06,Woo++06, Jah++09,SS13,Sun2015}. During the past decade, there has been much progress on this front using type-1 AGNs. For example, it has been demonstrated that AGN host galaxies, at $z<1$ and for fixed \mbh, are under luminous compared to today's hosts \citep{Park15, Tre++07, Pen++06qsob}. Similarly, \citet{Bennert11} and \citet{Woo++08} found a positive evolution of \mbh, especially when the data are sufficiently robust to isolate the luminosity or stellar mass of the bulge or spheroidal component. At $z>1$, \citet{Merloni2010} decomposed the entire spectral energy distributions (SED) into a nuclear AGN and host-galaxy components and found a positive evolution of the mass ratios of black holes to their host galaxies. Although, the accuracy of such an approach for luminous AGNs has not yet been well established as discussed herein. If realized, such offsets can be interpreted as a scenario in which SMBHs were built up first with galaxies then growing around their deep potential wells.  A possible mechanism to account for the latter part of the growth of the galaxy without increasing \mbh\ is the transfer of stellar mass from the disk to the bulge \citep{Jah++09,Bennert++2011,SS13} through bar instabilities or minor mergers. 

However, there are studies \citep{Cisternas2011,SS13,Mechtley2016} based on \hst\  imaging of deep survey fields such as COSMOS and CDFS that report no evolution in the \mbh - \smass ~ mass ratio as compared to the local relation. In support, \citet{Sun2015} has re-analyzed the mass ratios for broad-line AGNs in the COSMOS field in a similar manner (i.e., stellar mass measurements from SED fitting) to \citet{Merloni2010} and finds no evolution when accounting for selection effects described in \citet{Schulze2014} which consider the black hole mass function and Eddington rate distribution of the sample at their respective epoch. 

To make substantial progress, it is important to construct statistical high-$z$ samples that reduce the uncertainties, and carefully consider selection effects and inherent systematic errors. First, one needs to deal with the inherent uncertainties in BH mass estimates using the so-called ``virial" method. In particular, many studies rely on \mbh\ estimates using the \Civ\ (or \Mgii) line that may have unknown systematics, such as a non-gravitational component of the gas dynamics of the broad-line region (BLR), when compared to local samples with masses based on broad Balmer lines~\citep[i.e., \halpha\ and \hbeta,][]{Schulze2018, Baskin2005, Trakhtenbrot2012}. Second, measurements of the host galaxy properties are challenging due to the overwhelming glare of the bright nuclear light. This ultimately requires careful modeling of the point spread function and, whenever possible, the use of lensed AGNs since the magnification increases the spatial resolution \citep{Pen++06qsob, Ding2017a, Ding2017b}. Especially not to be overlooked as in past studies, the selection function needs to be taken into account when interpreting the observations~\citep{Treu2007, Lauer2007}. For instance, it has been demonstrated by~\citet{Schulze2011, Schulze2014} that selecting bright AGNs at high redshift results in a  \mbh - \smass\ relation with a steeper slope than if chosen randomly, suggesting the selection effects are the culprit rather than an intrinsically faster or even an existing evolution. It is also important to consider the selection function when comparing observed scaling relations with those from simulations \citep{DeG++15}.

In this study, we aim to make progress by utilizing a large sample with high-quality data both for the measurement of the \mbh\ and their host properties, extending to higher redshifts where evolutionary effects should be strongest. In particular, samples based on 2D image analysis using \hst\ at $z>1$ are limited. To date, the infrared capabilities of \hst/WFC3 have not been fully exploited on this topic. Here, we measure the properties of 32 host galaxies with a redshift range of $1.2<z<1.7$ using \hst/WFC3 imaging data, and estimate their \mbh\ based on the robust \halpha\ ~detections, using the multi-object spectrograph Subaru/FMOS. Given the high-quality and sample size of our data, we are capable of testing whether the growth of BH predates that of the host by a factor of at least $1.7$ \citep[i.e., $\sim0.23$ dex,][]{Schulze2014}. This value is the minimum offset expected for a non-evolving mass ratio as described in Section~\ref{sec:sf_framework}. Furthermore, we collect from the literature comparison samples of intermediate-$z$ and local AGN, selected to have been analyzed with very similar methods to those applied in the distant sample. We recalibrate the relevant quantities from the literature based on a set of self-consistent recipes to ensure that our differential measurement of evolution is robust to calibration and methodological issues (Section~\ref{sec:compare_sample}).

The paper is organized as follows. We describe the sample selection and their BH masses in Section~\ref{sec:data}. We describe the new \hst/WFC3 observations, available \hst/ACS imaging, and construction of a PSF library in Section~\ref{observation}. In Section~\ref{sec:analysis}, we describe our method to decompose the rest-frame optical emission and measure the host galaxy surface photometry. In Section~\ref{sec:result}, we use the multi-band host magnitudes to infer their stellar population from which we apply to derive the rest-frame R-band \lhost\,~and \smass\ to compare with local relations. In addition, we use the information on the radial light distribution (i.e., \sersic\ index) to infer the likely fraction of stars in the bulge (\bmass) and its relation to the \mbh. The discussion and conclusions are presented in Section~\ref{sec:dis} and Section~\ref{sec:sum}. Throughout this paper, we adopt a standard concordance cosmology with $H_0= 70$ km s$^{-1}$ Mpc$^{-1}$, $\Omega{_m} = 0.30$, and $\Omega{_\Lambda} = 0.70$. Magnitudes are given in the AB system. A Chabrier initial mass function (IMF) is employed consistently.

\section{Experimental design}
\label{sec:data}
\begin{figure}
\centering
{
\includegraphics[height=0.5\textwidth]{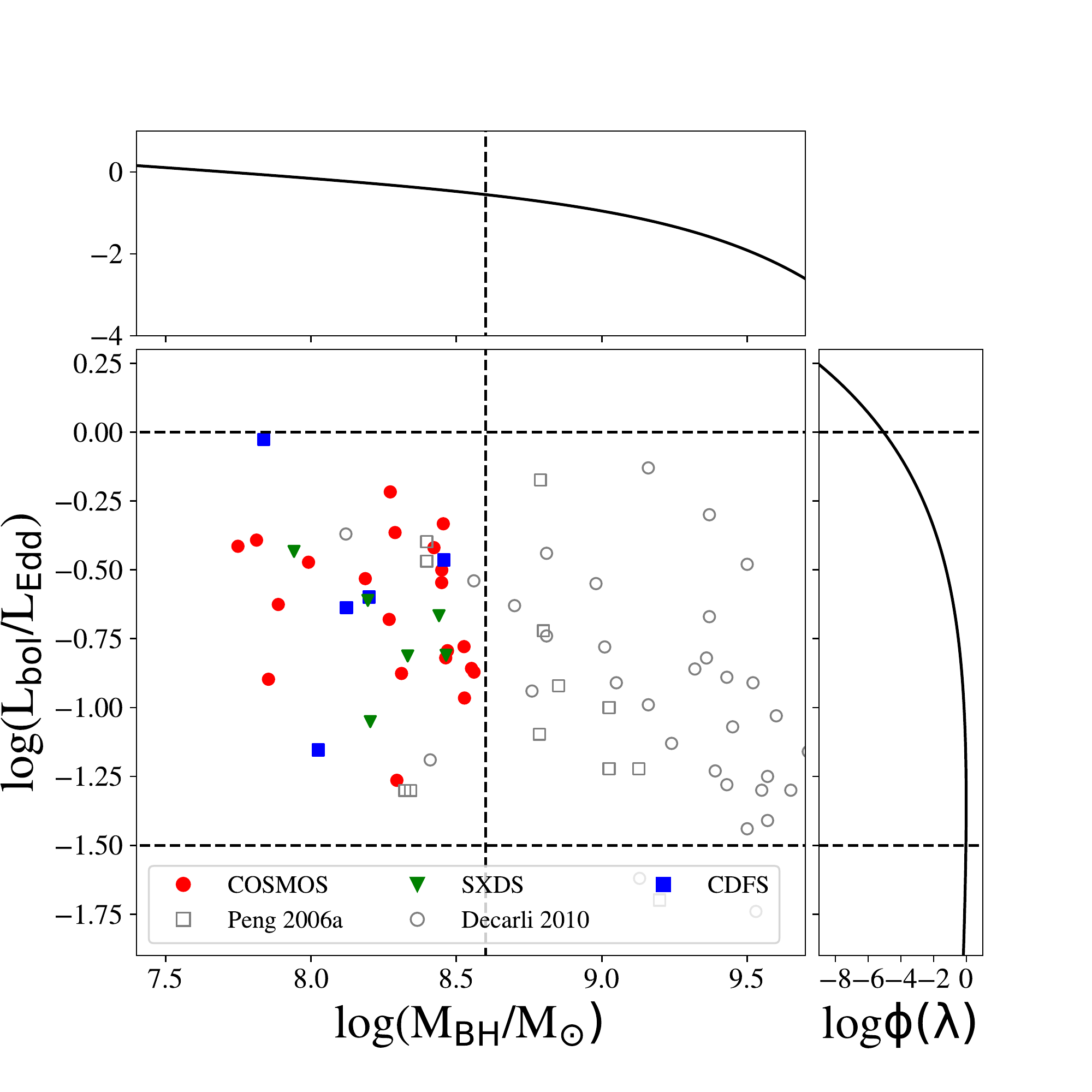}
}
\caption{\label{fig:selection} Selection window used to choose our AGN sample based on  \mbh~and Eddington ratios ($\lambda = L_{\rm Bol}/L_{\rm Edd}$). As indicated by the vertical red line, our sample (color coded) falls below the knee of BH mass function at $z=1.5$ \citep[top panel,][]{Schulze2015}. On the right, we illustrate the shape of the Eddington rate distribution at these redshifts from the same reference. For comparison, we plot the high-$z$ luminous SDSS AGN samples that have been studied with \hst\ ~\citep[grey squares and circles from][respectively]{Peng2006a, Decarli2010}, mainly at the high mass end.}
\end{figure}

We utilize a sample size of 32 broad-line (FWHM$>2000$ km s$^{-1}$; type-1) AGNs that have black hole mass measurements and fall within deep extragalactic survey fields that offer rich ancillary data. Specifically, we focused on meeting the following criteria to overcome the limitations of previous studies:

\begin{itemize}

\item Black hole mass estimates (\mbh) are based on Balmer lines (i.e., \halpha) which avoid potential systematic uncertainties in UV-based estimators \citep{Greene2005}.

\item Black holes masses \mbh/$M_{\odot}\lesssim8.7$, are below the knee of the black hole mass function to minimize selection biases (see Figure~\ref{fig:selection}, top panel). 

\item Eddington ratios above $5\%$ to further ensure homogeneity.

\item X-ray selected sample has host-to-total flux ratios typically above $30\%$ which facilitates the galaxy mass measurements. 

\item \hst/WFC3 imaging of the host galaxy at rest-frame wavelength $\sim5500$~\angstrom, which is above the $4000$~\angstrom\ break and does not include the broad \halpha\ line ($6563$~\angstrom), which ensures sensitivity to the total stellar mass content. This is somewhat coupled to the previous item in this list.

\item A large fraction of the sample has additional \hst\ imaging (i.e., ACS), providing color information to achieve reliable K-corrections and stellar mass determinations. 
\end{itemize}

\subsection{Sample selection}\label{sec:target_selection}

The 32 AGNs are initially selected by the X-ray observations of COSMOS~\citep{Civano2016}, (E)-CDFS-S~\citep{Lehmer2005, Xue2011}, and SXDS~\citep{Ueda2008} fields. In most cases, the X-ray sources are first identified as broad-line AGNs through optical spectroscopic campaigns with the VLT, Keck, and Magellan. Follow-up near-infrared spectroscopic observations of the AGNs in these fields are carried out with Subaru's Fiber Multi-Object Spectrograph~\citep[FMOS, ][]{Kimura2010, Nobuta2012,Matsuoka2013}, covering the wavelength range $0.9-1.8$~$\mu m$ which provides the favorable \halpha\ and \hbeta\ lines out to $z\sim1.7$ to estimates the \mbh. Recently, \citet{Schulze2018} present near-IR spectroscopy of a large compilation of 243 X-ray AGN in these fields. It is from this catalog that we primarily select our targets. 
The continuum fitting and emission-line modeling provided by this work are performed through a common procedure for spectral model fits. This approach first corrects the spectra for galactic extinction and shifts them to the rest-frame. The spectral regions in the near-IR that are strongly affected by OH emission are masked. The spectrum is modeled by both a narrow-line and a broad-line AGN template, and the emission line information is inferred based on $\chi^2$ minimization. We refer the reader to the aforementioned paper for further details.
Note that three CDFS objects (i.e., CDFS-1, CDFS-229 and CDFS-724) are not included in \citet{Schulze2018}; their infrared spectra are provided by \citet{Suh2015} using a similar approach.

Based on the \mbh\ estimates described below (Section~\ref{mbh}), we select targets with masses between $7.5 \lesssim {\rm log}$~(\mbh/$M_{\odot})\lesssim8.5$. The bolometric luminosities are measured from the broad AGN lines by \citet[][Section 3.3]{Schulze2018}, which are used to calculate their Eddington ratio $\lambda = L_{\rm bol}/L_{\rm Edd}$. The flux-limited nature of the sample results in a slightly higher Eddington ratio distribution at lower \mbh\, as shown in Figure~\ref{fig:selection}. Moreover, we have a higher preference to select targets which have rest-frame UV images (i.e., \hst/ACS) as provided by \citet{Scoville2007} and \citet{Koekemoer2007} in the COSMOS field. We list the 32 AGNs observed with \hst/WFC3 and analyzed in this work in Table~\ref{tab:objlist} sorted by field and redshift.

\subsection{Details on BH mass estimates}
\label{mbh}

The \mbh\ of type-1 AGNs can be determined using the so-called virial method~\citep{Peterson2004, Shen2013}. The kinematics of the BLR trace the gravitational field of the central supermassive black hole, assuming the gravity dominates the motion of the BLR gas. Under these assumptions, the width of the emission-line provides the scale of the velocity dispersion ($\Delta V$), while the AGN continuum luminosity establishes an empirical scale of the BLR size ($R_{\rm BLR}$). The estimation of \mbh\ is then achieved using these measurements, i.e., \mbh$\simeq G^{-1} R_{\rm BLR} \Delta V^2$ \citep{McLure2004}.

To avoid any systematic bias between samples in the literature, we adopt a class of self-consistent estimators for our analysis. We first compare the estimators implemented in \citet{Schulze2018} and \citet{Ding2017b}, and find very consistent \hbeta(FWHM(5100))-based masses (\mbh\ r.m.s. $<0.03$~dex). However, there is a $\sim0.2$ dex inconsistency in their \halpha\ mass estimates. Therefore, we utilize the AGNs in \citet{Schulze2018} that have both \halpha\ and \hbeta\ lines (35 AGNs in total) and carry out a cross-calibration to determine which \halpha\ estimator has the best agreement between the two lines. As a  result, the \halpha\ estimator in \citet{Schulze2018} has better agreement with \hbeta. Thus, we adopt the scheme given in \citet{Schulze2018} for all AGN samples used in this study, including the comparison samples described below.

\begin{eqnarray}
\label{eq:Ha}
\log \left(\frac{\mathcal M_{\rm BH} ({\rm H}\alpha)}{M_{\odot}}\right)&~=~& 6.71+0.48 \log \left(\frac{ \rm L _{H\alpha}}{10^{42}{\rm erg~s^{-1}}}\right) \nonumber\\
&~+~& 2.12 \log \left(\frac{\rm FWHM(H\alpha)}{1000 ~{\rm km~s^{-1}}}\right) ,
\end {eqnarray}

and

\begin{eqnarray}
\label{eq:Hb}
\log \left(\frac{\mathcal M_{\rm BH}({\rm H}\beta) }{M_{\odot}}\right)&~=~& 6.91+0.50\log \left(\frac{ \rm L _{\lambda_{5100}}}{10^{44}{\rm erg~s^{-1}}}\right) \nonumber\\
&~+~& 2.0 \log \left(\frac{\rm FWHM(H\beta)}{1000 ~{\rm km~s^{-1}}}\right) .
\end {eqnarray}

Note that these recipes are first provided by \citet{Vestergaard2006}. Using these recipes, we estimate \mbh\ by adopting the emission-line properties as measured by \citet{Schulze2018} for all 32 AGNs. While fourteen AGNs have emission line properties for both \halpha\ and \hbeta, we adopt the value of \mbh\ based on the \halpha\ emission line to have consistency across the sample. While most recipes are calibrated against \hbeta\, the line is typically weaker than H$\alpha$, hence lower signal-to-noise. We provide the \mbh\ measurements together with the properties of the emission lines in Table~\ref{tab:result_mbh}.    

It is worth noting that the absolute flux calibration of the FMOS spectra is set to match the available ground-based IR imaging, UltraVISTA in the case of COSMOS. While an initial flux calibration is performed during the reduction of the FMOS data using calibration stars, there can be differential flux loss due to aperture effects, variable seeing conditions, and minor alignment issues with the instrument. The flux normalization is effectively an aperture correction in the J or H band depending on the source redshift. We note that this procedure does not correct for intrinsic variability that may induce an additional error of 0.2 mag. We refer the reader to Section 2.2 of \citet{Schulze2018} for full details on the flux calibration.

\subsection{Expected bias from the selection function}

\label{sec:sf_framework}

Our AGN sample is primarily selected based on the value of \mbh~and Eddington ratio. It is well known that sample selection effects must be taken into account in order to  interpret the observed black hole-host relations and measure its evolution with redshift thus avoiding biases~\citep{Tre++07,Schulze2011,Bennert++2011, Schulze2014,Park15}. The main source of bias is due to the fact that active samples are necessarily selected based on properties that correlate with black hole mass estimators (such as AGN luminosity, line strength, and width) and thus one tends to favor overly massive black holes in presence of intrinsic scatter and observational errors. Correcting for observational biases requires a well-characterized selection function, such as the one we have for our sample, and a model of the black hole mass function, and the evolution of the correlations between \mbh, and other properties.

Given our selection function, we use the Bayesian framework introduced by \citet{Schulze2011,Schulze2014} to estimate the expected bias. In this framework, under the assumption of no evolution of the correlations between \mbh\ and host-galaxy properties, one can compute the expected bias for a given sample, prior to the observations. The key ingredient of this model are the local \mbh\ host-galaxy property correlations, the black hole mass function, and Eddington ratio distribution at the redshift of observation. The latter two quantities are estimated from the type-1 AGN distributions \citep{Schulze2015} and corrected to represent the parent population of all active SMBH hosts. 

Adopting our specific selection limits into the framework (i.e., $\log($\mbh$/M_{\odot})\in[7.5, 8.56]$, $\log(L_{\rm bol}/L_{\odot})\in[45.0, 46.2] $ and  $\log(\lambda) \in [-2.0, 0.5]$), we infer an expected bias of $+0.21$~dex in the observed $\Delta\log($\mbh$)$ for samples at $z\sim1.5$, assuming the baseline choices for the inputs to the model that include the local value of the mass \mbh/\smass ~ratio. To reiterate, an offset in the observed mass relations of 0.2 dex at $z\sim1.5$ can be considered consistent with no evolution in the mass ratio. This bias correction should be considered an approximate estimate, with an uncertainty that depends on the uncertainty of the inputs. Furthermore, if the scaling relations actually evolve, one needs to introduce a model for the evolution in order to infer the underlying bias-corrected trends. We will revisit these issues in Section~\ref{select_eff}.

\subsection{Comparison samples}\label{sec:compare_sample}

We make use of the \mbh-\lhost\ and \mbh-\smass\ relations in the literature for comparison with our high-$z$ data. For the local relation, we use the measurements of \citet{Ben++10} and \citet{Bennert++2011} (hereafter, B10 and B11) to define our zero-point for local AGN samples. The sample by B10 consists of 19 AGN with the \mbh-\lhost\ relation determined with reliable \mbh\ masses using reverberation mapping with an uncertainty level $\sim0.15$ dex. Note that B10 only provides the single galaxy V-band luminosity. \citet{Ding2017b} derive the galaxy R-band luminosity based on the same early-type galaxy template spectrum adopted by B10 using K-correction (see Section~3.2 therein). The work of B11 contains 25 local active AGNs, where the \mbh\ are measured using the single-epoch method (\mbh\ uncertainty level $\sim0.4$ dex). To track the local \smass\ and \mbh\  relations to higher values, we include 30 inactive galaxies (mainly ellipticals or S0) from \citet{H+R04} (hereafter HR04). It is worth noting that the local inactive sample is mainly bulge dominated, and we adopt the bulge mass for the entire local sample. In other words, our local comparison is the \mbh-\smass$_{\rm ,bulge}$ and not those involving total quantities.

We include in our analysis published samples at intermediate redshifts to understand the evolution of these correlations. We select samples that were analyzed by members of our team to ensure uniform measurements. The intermediate redshift AGNs that we include are 52 objects published by \citet{Park15} using a single band which are applicable for the \mbh-\lhost\ relation at $0.36<z<0.57$ and 27 objects published by \citet{Bennert11} and \citet{SS13} at $0.5<z<1.9$. Similar to the B10 sample, the R-band luminosities of these intermediate redshift systems (79 in total) are obtained by \citet{Ding2017b} using K-corrections from the V-band based on the same stellar template. Moreover, \citet{Bennert11} and \citet{SS13} estimate the stellar masses of their 27 objects using multi-band imaging data, and we adopt them as the \mbh-\smass\ comparison sample.
 In addition, we adopt a sample of 32 \mbh-\smass\ measurements at $0.3<z<0.9$ by \citet{Cisternas2011} to compare with our sample. Across the intermediate redshift sample, we recalibrate the \mbh\ using the self-consistent recipes introduced in Section~\ref{mbh} including those estimated from \Mgii\ using the recipe in \citet{Ding2017b}. 

The values for the all the \mbh-\lhost\ comparison samples are listed in the paper by \citet{Ding2017b} (Table~1 and Table~2 therein). The \mbh-\smass\ comparison samples are collected and summarized in the Appendix~\ref{sec:comp_sample_value}, Table~\ref{tab:comp_sample}.

\section{\hst\ observations}
\label{observation}
High spatial resolution imaging is required for the decomposition of the nuclear and host emission to accurately estimate the luminosity and stellar mass of the host galaxy. For this purpose, we observed the sample of 32 AGNs, as described above, with the \hst/WFC3 infrared channel, through the \hst\ program GO-15115 (PI: John Silverman). We selected to use the filters F125W $(1.2<z<1.44)$ and F140W $(1.44<z<1.7)$ according to the redshift of the targets so that the rest-frame spectral window is well above the 4000~\angstrom\ break. This selection further ensures that the broad \halpha\ line is not present in the bandpass so as not to contaminate the host emissions due to the broad wings of the PSF.

For each target, we obtained six separate exposures of $399$s (i.e., total exposure time $2394$s). The six exposures were dithered and combined with the {\sc astrodrizzle} software package following standard procedures, and resulted in an output pixel scale of $0\farcs{0642}$ by setting \texttt{pixfrac} parameter as $0.8$ and using a \texttt{gaussian} kernel\footnote{\label{note1}For CID255, 3/6 of the dither WFC3 images are corrupted. We achieve to analyze this sample using the same approach, taking the 3 available frames.}. In Table~\ref{tab:objlist}, we list the details of the individual observations.

Having obtained the \hst\ image, we remove the background light, arising from the sky and the detector. In this step, we adopt the {\sc photutils} by Python and model the global background light in 2D based on the \texttt{SExtractor} algorithm, which effectively accounts for gradient in the background distribution. Then, we subtract the derived  sky background light to obtain a clear image. To test the fidelity of this subtraction, we measure the surface brightness in the empty regions and verify that it is consistent with zero within the noise. Finally, we extract the postage stamp of the AGN image and PSF image to carry out the modeling process; see Section~\ref{sec:psf_library} and Section~\ref{sec:analysis}.

Multi-band information provides the SED at a more precise level. A substantial fraction (21/32) of our objects have rest-frame UV images for those in COSMOS \citep{Koekemoer2007}. Here, we utilize images taken with ACS/F814W filter. The final image is drizzled to $0\farcs{03}$ pixel scale. Given the multi-band images for our AGN, we are able to infer their host color and assess the contribution of both the young and old stellar population which ensures an accurate inference of rest-frame R-band luminosity (including a K-correction) and stellar mass \citep{Gallazzi2009}. 

\subsection{PSF library}
\label{sec:psf_library}

The knowledge of the PSF is crucial for imaging decomposition of the AGN and its host, especially when the point source contributes to the majority of the total emission. The PSF is known to vary across the detector and over time due to the effects of aberration and breathing. Simulated PSF, such as those based on {\sc TinyTim}, are usually insufficient for our purposes \citep{Mechtley2012}. Stars within the field-of-view of each observation provide a better description than the simulated PSF since they are observed simultaneously with the science targets, and reduced and analyzed in a consistent manner \citet{Kim2008, Park15}. However, we have found that such stars usually do not provide an ideal PSF for de-blending the AGN and host galaxy through extensive tests. The issues are associated with an insufficient number of bright stars near our targets, color differences, and other effects not fully understood.

To minimize the impact of such mismatches, we build a PSF library by selecting all the isolated, unsaturated PSF-stars with high S/N from our entire program. The selection consists of the following steps. First, we identify stars from the COSMOS2015 catalog \citep{Laigle2016}. However, many bright stars with an intensity similar to our AGN sample were excluded in this catalog. Therefore, we also manually select PSF-like objects as candidates from the \hst/WFC3 imaging. We then discard non-ideal PSF candidates based on their intensity, FWHM, central symmetry, and presence of nearby contaminants. In total, the PSF library contains 78 and 37 stars imaged through filters F140W and F125W, respectively. We assume that the stars in the library are representative of the possible PSFs in our program. The dispersion of PSF shapes within our library provides us with a good representation of the level of uncertainty in our measurements resulting from the image decomposition.

\begin{figure*}
\centering
{
\includegraphics[height=0.25\textwidth]{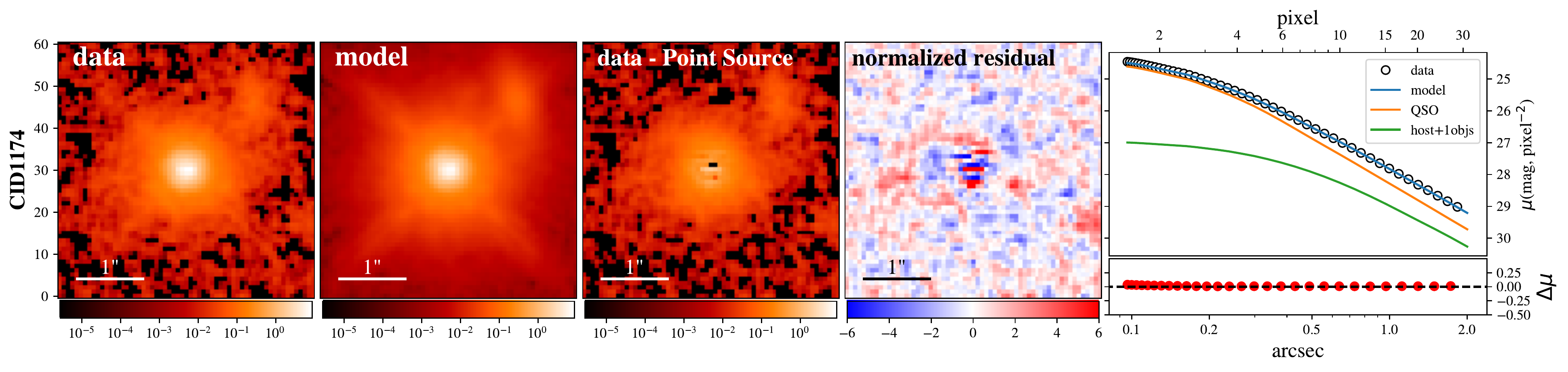}
\caption{\label{fig:AGN_decomp} AGN-host galaxy decomposition of COSMOS-CID1174 based on the \hst/WFC3 F140W image.The panels from left to right are as follows: (1) observed data, (2) best-fit model (AGN$+$host), (3) data minus the model PSF (i.e., host galaxy free of the AGN), (4) residuals divided by the variance and (5)~1-D surface brightness profiles (top) and the corresponding residual (bottom). The 1-D profiles indicate the surface brightness including the data (open circles), the best-fit model (blue line), the AGN (orange line), and the model for the extended sources (green line, i.e., host and other objects). Note that the 1-D surface brightness profiles are only for illustration purposes. The actual fitting is based on the two-dimensional images. The remaining 31 objects in the sample are shown in Appendix~\ref{sec:restsample}.
}}
\end{figure*} 

\section{AGN-host Decomposition}
\label{sec:analysis}

We simultaneously fit the two-dimensional flux distribution of the central AGN and the underlying host galaxy. Following common practice, we model the central AGN as a scaled point source and the host galaxy as a \sersic\ profile. Note that the actual morphologies of the host galaxies could be more complicated (e.g., bulge$+$disk). However, the \sersic\ model is an adequate first-order approximation of the surface brightness distribution with a flexible parameterization that provides sufficient freedom to infer the total host flux even for our high redshift sample. We simultaneously fit the nearby galaxies, that happen to be close enough to the AGN, with a \sersic\ model to account for any potential contamination from their extended profiles. The systems CID206 and ECDFS-358 have nearby objects which could not be described by \sersic\ model; thus, we mask these objects in the fitting procedure.

We use the image modeling tool \lenstronomy~\citep{Birrer2015, lenstronomy} to perform the decomposition of the host and nuclear light. \lenstronomy\ is a multi-purpose open-source gravitational lens image forward-modeling package written in Python. 
Its flexibility enables us to turn off the lensing channel and focus on the AGN and host decomposition\footnote{As a check, we compared the results from \lenstronomy\ to the commonly used galaxy modeling software {\sc Galfit} and confirmed that the results are consistent between the two.}. The main advantage of \lenstronomy\ is that it returns the full posterior distribution of each parameter (i.e., not just the best fit model) and the Laplace approximation of the uncertainties. The input ingredients to \lenstronomy\ include:
\begin{enumerate}
\item AGN imaging data. \\
-- Using aperture photometry, we find that an aperture size with radius $\sim1\farcs{}5$ sufficiently covers the AGN emission of our sample\footnote{The photometric aperture was determined as a compromise between the detection of the total flux and a minimization of the background noise while keeping the computational time low. We found that extending the radius beyond $1\farcs5$ did not add any missing light, while it increased the noise and computational time. 
}. By default, we extract an image of $61\times61$ pixels (i.e., $4''\times 4''$). If needed, a larger box size is selected to include nearby objects. 
\item Noise level map.\\
-- The origin of the noise in each pixel stems from the read noise, background noise, and Poisson noise from the astronomical sources themselves. We measure these directly from the empty regions of the data. We then calculate the effective exposure time of each pixel based on the drizzled \texttt{WHT} array maps to infer the Poisson noise level. A final noise map includes all of these sources of error. 
  
\item PSF. \\
-- The PSF is directly taken from the PSF library. Usually, a mismatch exists when subtracting the AGN as the scaled PSF, especially at the central parts. While modeling multiply imaged AGN, this mismatch can be mitigated with PSF reconstruction by the iterative method~\citep{Chen2016, Birrer2018}.  However, this approach requires multiple images which are not available in our case.  We remedy this deficiency by using a broad library which should contain sufficient information to cover all the possible PSF. 
\end{enumerate}


The host property of an AGN is determined by the following steps. First, we model the AGN and host using each PSF in the library. With the input ingredients to \lenstronomy, the posterior distribution of the parameter space is calculated and optimized by adopting the Particle Swarm Optimizer (PSO)\footnote{Note that \lenstronomy\ enables one to further infer the parametric confidence interval using MCMC. In our case, given a fixed PSF, the $1\sigma$ inference of each parameter is extremely narrow. Thus, we only take the best-fit inference using PSO for further calculations. The errors on the fit parameters are assessed by using different PSFs for each object in the sample.} \citep{PSO}.
To avoid any unphysical results, we set the upper and lower limits on the parameters: effective radius \Reff~$\in[0\farcs{}1,1\farcs{0}]$, \sersic\ index $n\in[0.3,7]$.
Then, we rank the performance of each PSF based on the $\chi^2$ value and select the top-eight PSFs as representative of the best-fit PSFs. We determine the host \sersic\ parameters (i.e., flux, \Reff, \sersic\ index) using a weighted arithmetic mean, calculated as follows:

\begin{eqnarray}
\label{eq:weights}
w_i = exp \big(- \alpha \frac{ (\chi_i ^2 - \chi_{best} ^2 )}{2 \chi_{best} ^2} \big),
\end{eqnarray} 
where the $\alpha$ is an inflation parameter\footnote{Defining $\alpha$ as the {\it inflation} parameter literally means that it has to be larger than~$1$. If $\alpha<1$, the relative likelihood between different PSFs would be too close and introducing $\alpha$ would have a side-effect.} so that when $i=8$:
\begin{eqnarray}
\label{eq:alpha}
\alpha \frac{ \chi_{i=8} ^2 - \chi_{best} ^2 }{2 \chi_{best} ^2} = 2.
\end{eqnarray} 
The goal of this recipe is to weight each PSF based on their relative goodness of fit, while ensuring at least eight are used to capture the range of systematic uncertainties. The results do not change significantly if we chose a different number of PSFs, as shown below.

Note that since each AGN was observed at a different location of the detector and at a different time, the top-eight PSFs usually vary from one AGN to another. Given the weights, the value of host properties and the root-mean-square ($\sigma$) error are calculated as:
\begin{eqnarray}
\label{eq:infer_value}
\bar{x}  =  \frac{  \sum_{i=1}^{N}   x_i * w_i  }{\Sigma w_i} ,
\end{eqnarray} 
\begin{eqnarray}
\label{eq:infer_scatter}
\sigma =   \sqrt{ \frac{  \sum_{i=1}^{N}   (x_i -  \bar{x} ) ^2 * w_i  }{\Sigma w_i} },
\end{eqnarray} 
where $N$ is the number of the ranking PSF, i.e., $N=8$.
\noindent In Figure~\ref{fig:AGN_decomp}, we demonstrate the best-fit result for COSMOS-CID1174. The adopted weights are listed in Table~\ref{tab:weight_CID1174}. 

\begin{figure}
\centering
{
\includegraphics[height=0.3\textwidth]{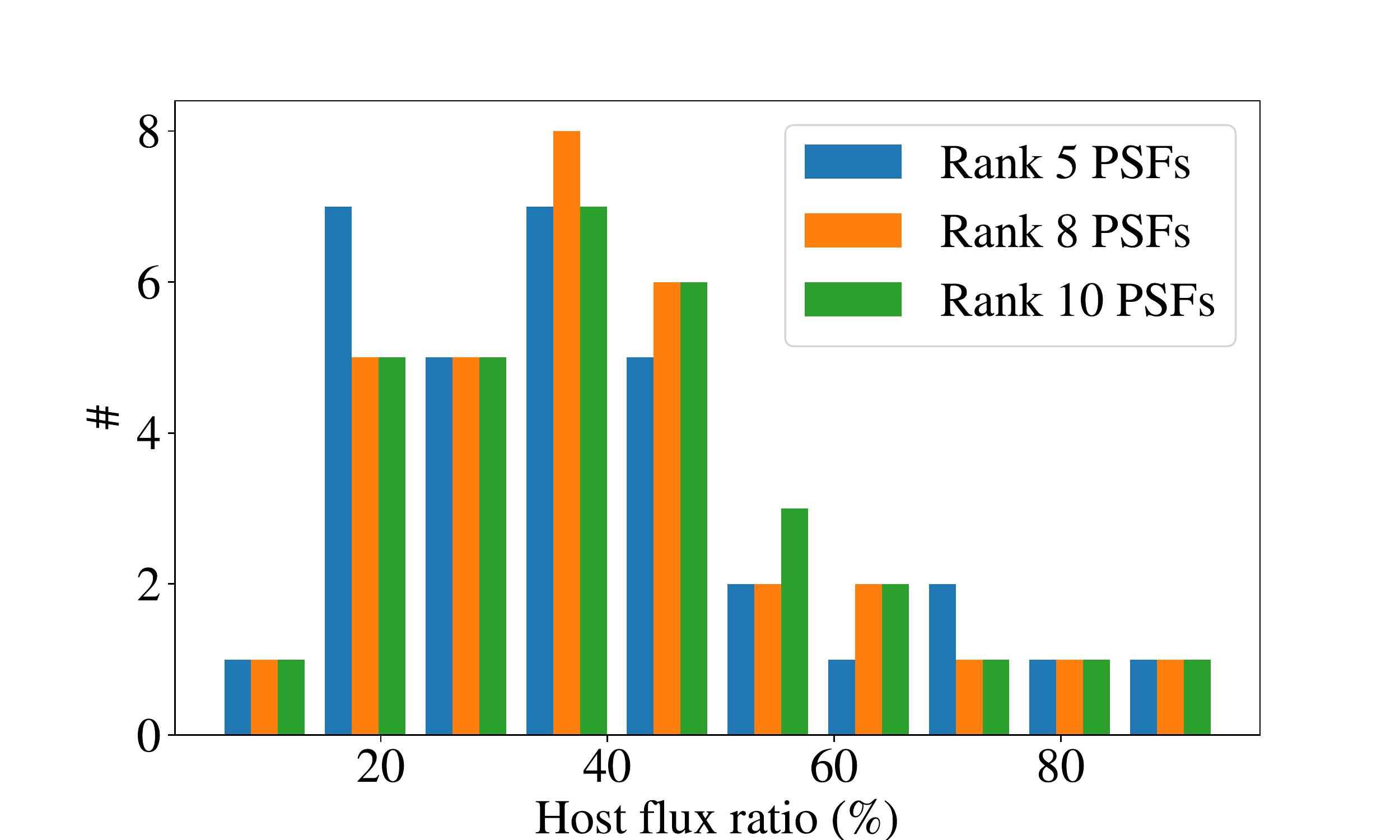}\\
\includegraphics[height=0.3\textwidth]{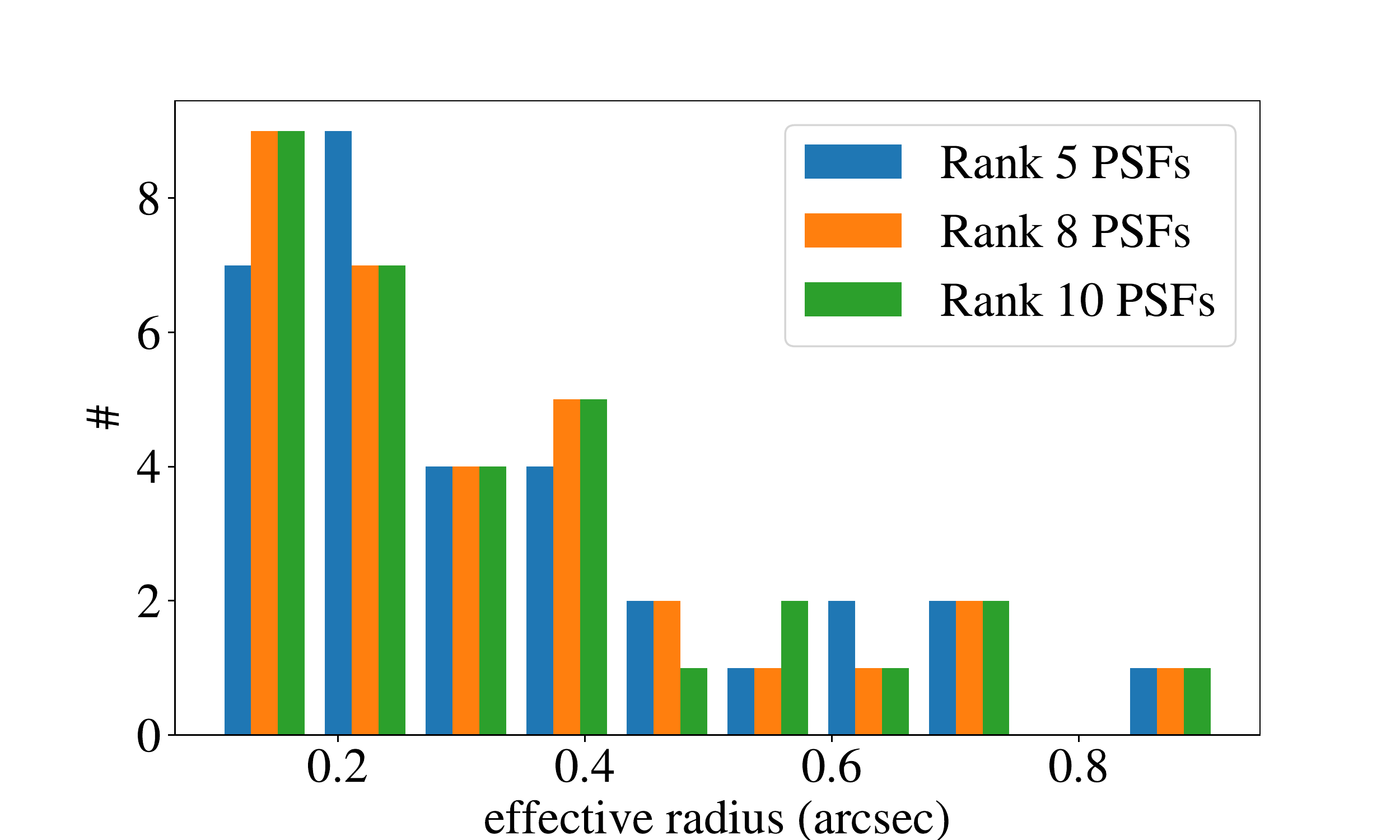}\\
}
\caption{\label{fig:hist_compare} 
Distribution of the host-to-total flux ratio and effective radius when implementing a different number of top-ranked PSFs.}
\end{figure}

We apply this approach to all AGNs to determine the global characteristics of the hosts of type-1 AGNs at these high redshifts. We measure the effective radius (\Reff), \sersic\ index, and host-to-total flux ratio and describe each of these in the following section. We recognize that these measurements are weighted by the eight top-ranked PSFs. The limited number of top-ranked PSFs may underestimate actual uncertainties. To gauge how the number of top-ranked PSF affects our results, we compare results when also using five and ten top-ranked PSFs. As shown in Figure~\ref{fig:hist_compare}, the results are consistent. For each AGN, we check the location on the detector for the top-ranked PSFs and find that they are unrelated to the AGN position on the detector. This finding underscores that position on the detector is not the main factor driving the PSF shape. Other factors, likely to be more significant, are the sub-pixel centering, the intrinsic color of the star, and the jitter and thermal status of the telescope during the observations. The complexity of the problem
highlights the necessity of decomposing the AGN using all the available PSF-stars from the entire program.

We carry out a similar analysis for 21/32 AGNs in the COSMOS field which have ACS/F184W imaging data. We assess their host flux ratio using the same approach as done for the WFC3-IR. The ACS FOV is more extensive than WFC3 one; thus we generated 174 PSFs for use. As expected, the detection of the host galaxy in the IR band is of higher significance than the UV due to the effects of dust extinction and the contrast between the (blue) AGN and (red) host. Thus, we fix the \Reff\ and \sersic\ $n$ as the value determined by IR band thus amounting to an assumption that the morphology of the galaxy is consistent between ACS and WFC3 bands. In this case, the only free parameters are the total host and AGN flux. We report the host galaxy properties in Table~\ref{tab:result_sersic}.

\section{Results}
\label{sec:result}

From our image decomposition, we detect the host galaxy in all cases at a significant level, except one case (i.e., SXDS-X763) that has a host-to-total flux ratio lower than $15\%$. \hst/WFC3 images with the AGN component removed are presented in the third panel (i.e., the ``data$-$point source'' stamps) of Figure~\ref{fig:AGN_decomp} and in Appendix~\ref{sec:restsample} for the remaining cases. While some of the galaxies have nearby neighbors, most are isolated and do not show strong signs of interaction or ongoing mergers, which indicates the fueling mechanism of AGNs may not be from major mergers for our sample. This is also relevant for model fitting with smooth \sersic\ profiles and the subsequent determination of the stellar mass. In the following subsections, we describe the properties of our ensemble of type-1 AGN host galaxies.
 
\subsection{Host galaxy properties}
\label{sec:result-hosts}
As shown in Figure~\ref{fig:hist_compare} (top panel), the host-to-total flux ratio (total$=$host$+$nuclear) of the sample spans a wide range from $10\%$ to $90\%$ with most of the sample concentrated between $20\%$ to $50\%$ (median value $37\%$). Those with flux ratios above $\sim20\%$ have a higher degree of significance with respect to the detection of the host galaxy, while the five systems (i.e., CID255, CID50, LID360, CDFS-229, SXDS-X763) which have host-to-total flux ratios lower than $20\%$ should be considered as marginal detections. 

\begin{figure}[ht]
\epsscale{1.2}
\plotone{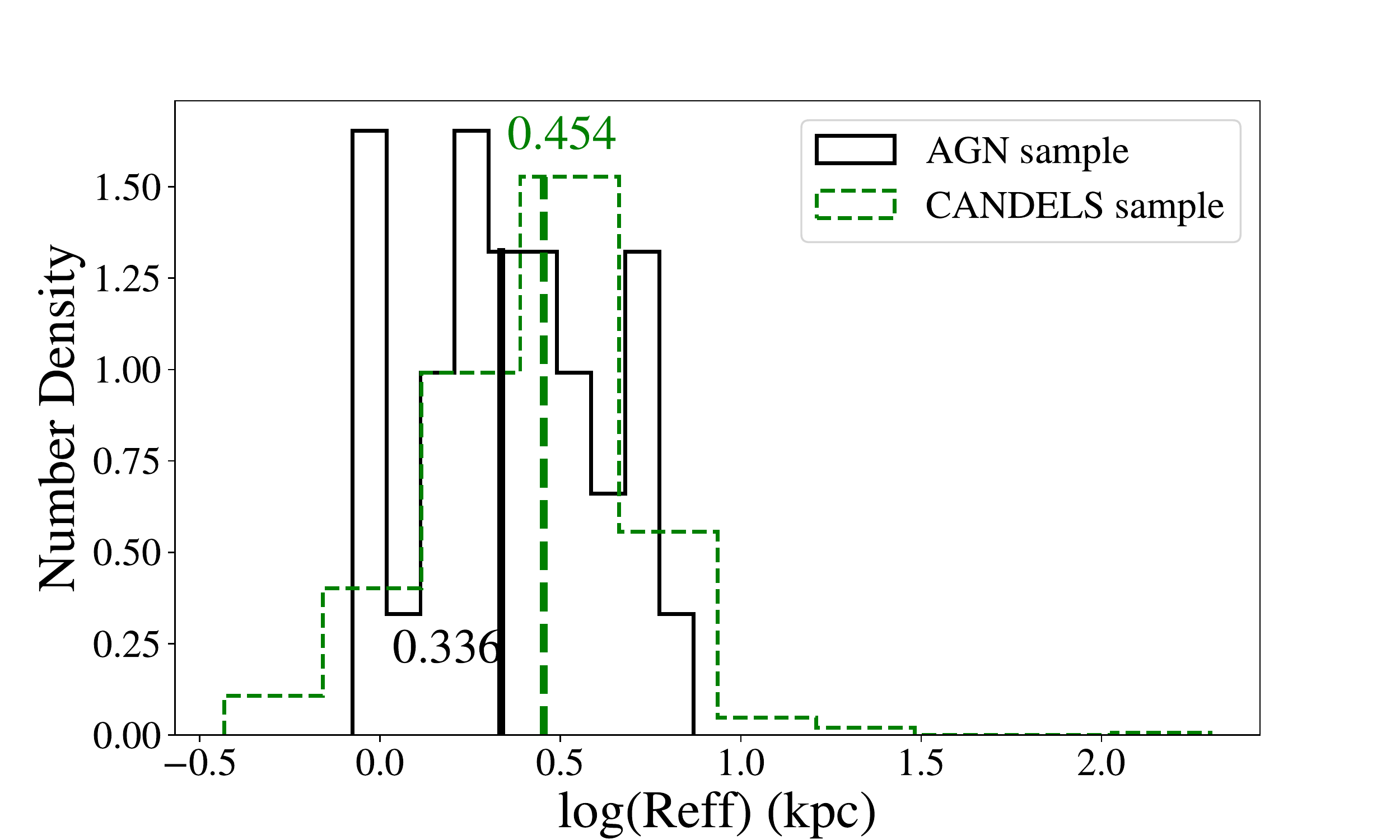}
\plotone{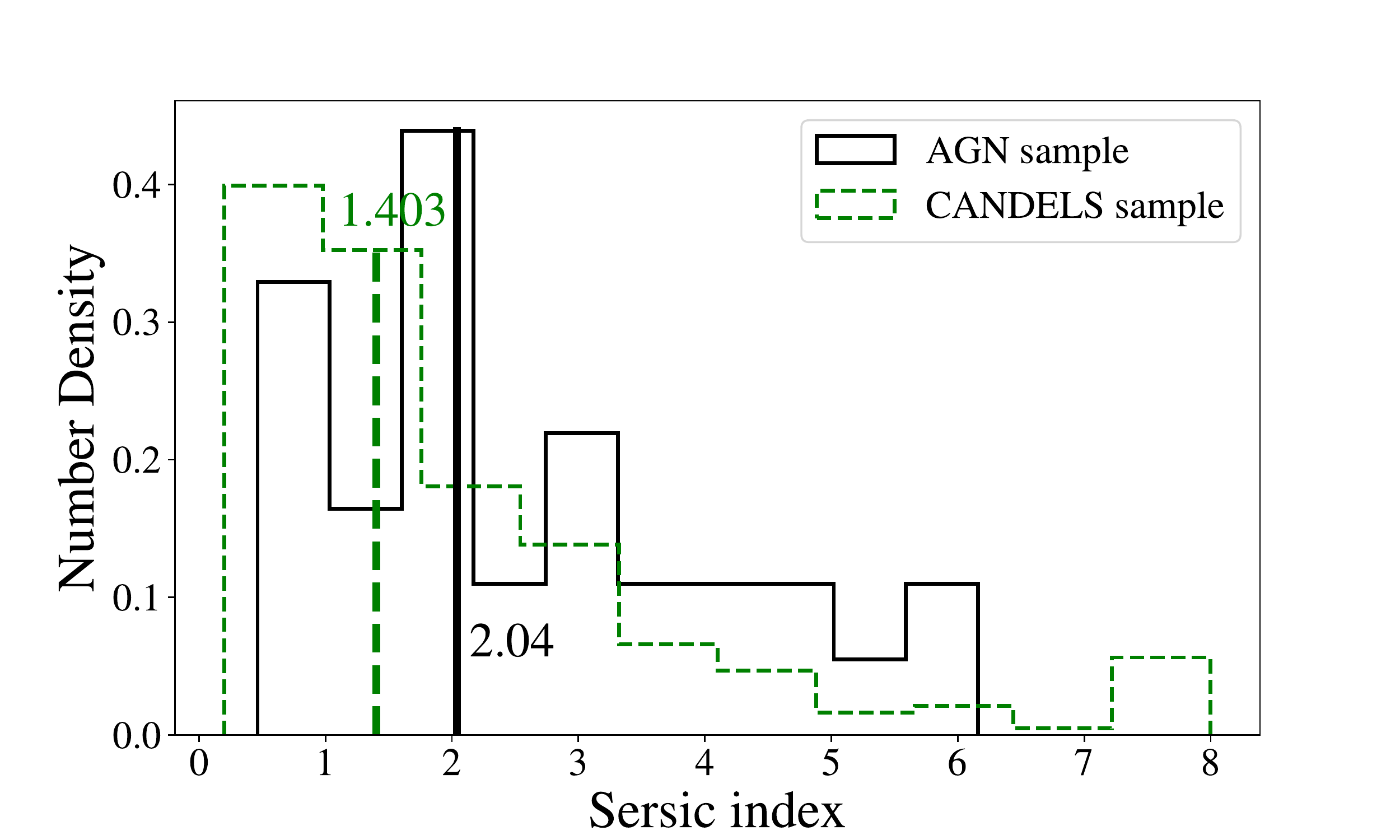}
\caption{\label{fig:hist_rn} 
The comparison of the histogram of the \Reff\ (top panel) and \sersic\ $n_s$ (bottom panel), with median value indicated. For comparison, we show the distribution of CANDELS galaxies at similar redshifts and stellar masses to our type-1 AGN sample.}
\end{figure}

The distributions of effective radius \Reff\ are shown in Figure~\ref{fig:hist_compare} (bottom panel). Based on the redshift, we calculate the physical scale of the radius for each object in kpc assuming a standard cosmology. We plot them together with measured \sersic\ index in Figure~\ref{fig:hist_rn}. These values are distributed within the allowed range and not concentrated on either the upper or lower bound. The \Reff\ of our sample are between $\sim1-7$~kpc (peaked at $\sim2.2$~kpc). Nearly half ($15/32$) of our systems have \sersic\ index $n<2$, indicating they have a significant disk component, likely in addition to the presence of a bulge. Five systems have a high \sersic\ index ($n>4.5$) for the first run; we use $n\in[1,4]$ as prior to re-fit these systems and find that the changes on the inference of their host luminosity are very limited ($<0.03$ dex). 
Seven systems have effective radius \Reff$<0\farcs18$, i.e., smaller than 3 pixels. In order to check whether this affects our conclusions,  we re-fit them with a \Reff\ $\in[0\farcs{}2,1\farcs{0}]$ prior and find that the inferred host flux barely changes ($<1\%$). In particular, the
inferred \Reff\ for three systems (CID543, XID2202, and SXDS-X1136) have hit the lower limit (i.e., $0\farcs{}1$), and thus the scatter on the parameter is formally zero. This means the size inference of these three systems should really be considered an upper limit, rather than a measurement. However, the inference of their host luminosities and stellar masses are still reliable given the consistency of the re-fitting results and the small scatter in the host flux ($<10\%$).

We compare the morphology of AGNs host galaxies to inactive galaxies from the CANDELS survey~\citep{Grogin2011, Koekemoer2011}. We identify 4401 inactive galaxies within a redshift range ($1.2<z<1.7$), comparable to our AGN sample, whose \sersic\ measurements are provided by \citet{VDwel++2012} using \galfit. Their stellar masses are derived based on the 3-D-\hst\ spectroscopic survey~\citep{Momcheva2016, Brammer2012} and comparable to our AGN hosts ($9.5< \log (M_* /M_{\odot})< 11.5$; Section~\ref{sec:mm}). We compare the histogram of the inferred \Reff\ and \sersic\ index to the inactive galaxies in Figure~\ref{fig:hist_rn}, where we find no significant difference between their distributions and median value. We also test whether the distributions can be drawn from the same parent population using a Kolmogorov-Smirnov test and determining the p-value to be 0.42 and 0.04 for \Reff\ and $n$, respectively. We conclude that the host galaxies of our AGN sample are representative of the overall population of galaxies, with a significant disk component, at comparable luminosity and stellar mass at the same redshift. In Section~\ref{sec:bh_bulge}, we use this information to infer the likely bulge masses, hence the \mbh\ -- \bmass\ relation.

\subsection{Rest-frame colors}

\begin{figure*}
\centering
\begin{tabular}{c c}
{\includegraphics[height=0.4\textwidth]{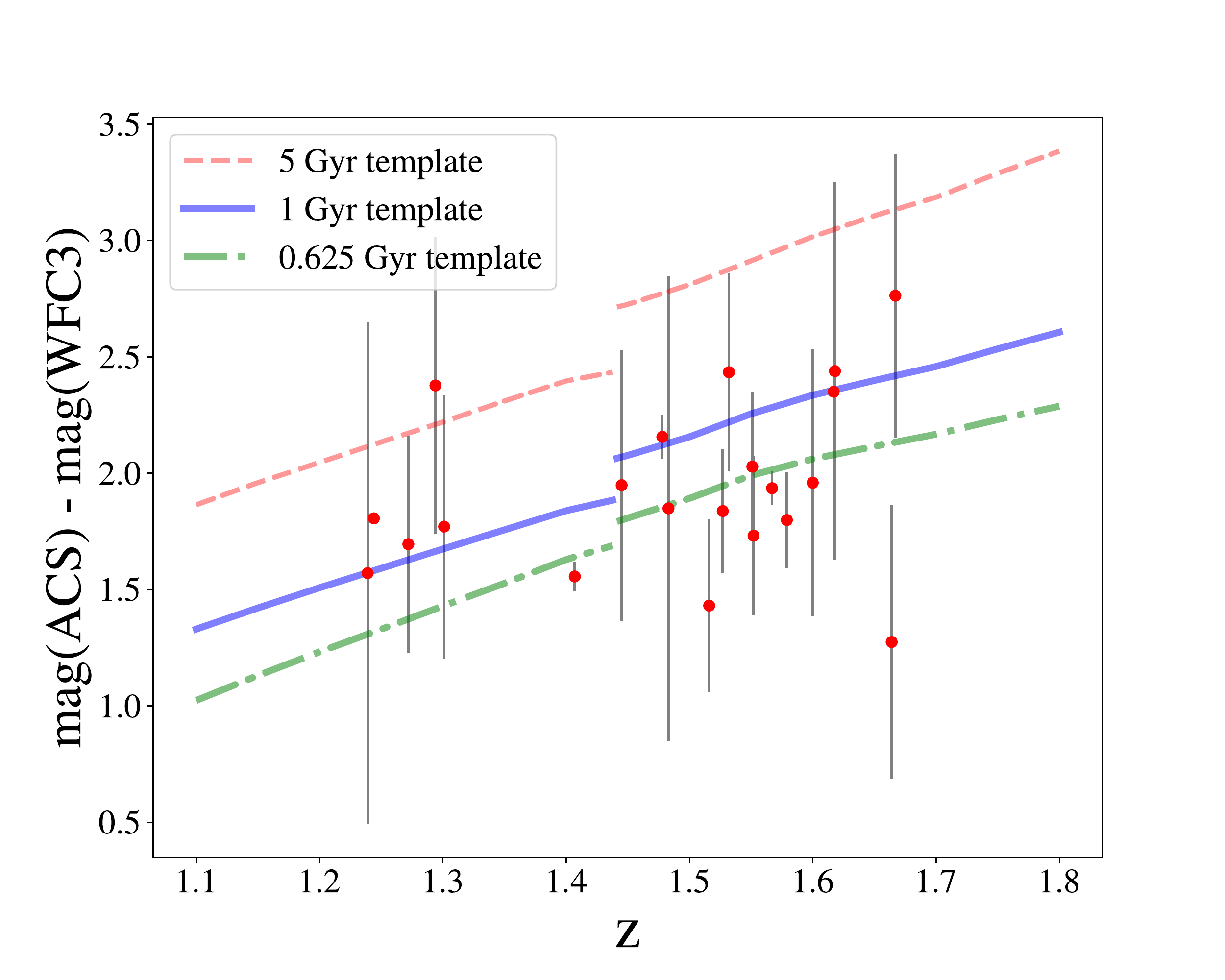}}&
{\includegraphics[height=0.4\textwidth]{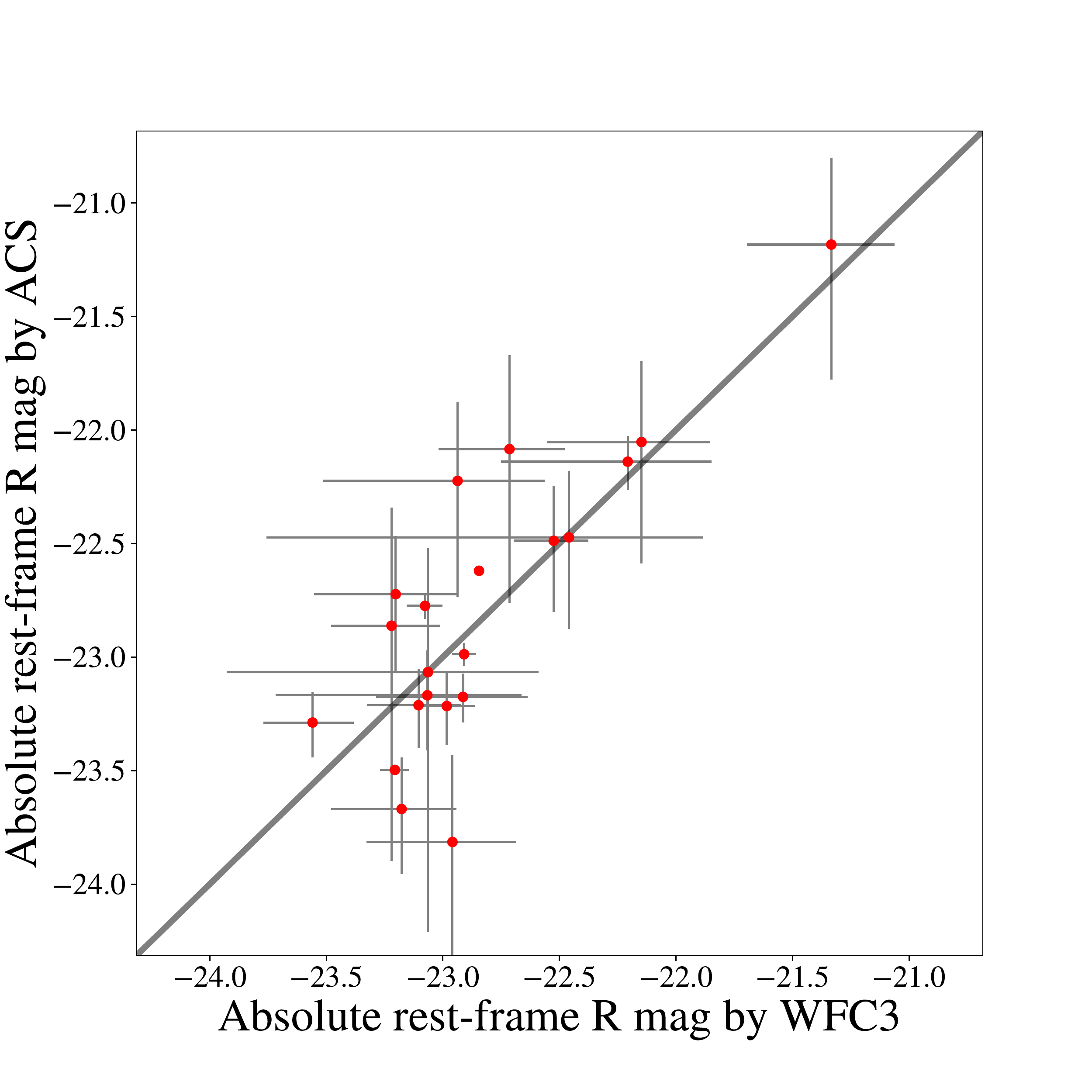}}\\
\end{tabular}
\caption{\label{fig:compare_temp} 
{\it Left}: The observed color as a function of redshift, based on $21/32$ AGNs which have both WFC3 and ACS imaging data. Note that the filter  combinations WFC3/F125W and WFC3/F140W are adopted for galaxies at $z<1.44$ and $z>1.44$, respectively. 
We also plot three predicted models including $5$~Gyr, $1$~Gyr and $0.625$~Gyr. Clearly, the $5$~Gyr stellar population is ruled out by our data, and we adopt the $1$~Gyr and $0.625$~Gyr for galaxies at $z<1.44$ and $z>1.44$, respectively.
 {\it Right}: Comparison of the inferred rest-frame R band magnitude, based on the adopted stellar populations.
}
\end{figure*}

For $21/32$ AGNs, we have multi-band host magnitudes that enable us to select the appropriate stellar population templates to determine rest-frame luminosities and stellar masses for the overall sample.  We find that the $1$~Gyr and the $0.625$~Gyr stellar population with solar metallicity and a Chabrier IMF~\citep{Bruzual2003} provide the excellent matches to the observed colors of our sample at $z<1.44$ and $z>1.44$, respectively  (see Figure~\ref{fig:compare_temp}). To minimize the uncertainty associated with these corrections, we use these two templates to interpolate to the rest-frame R band which is very close to the observed wavelengths. We note that the choice is not unique, and other combinations of ages and metallicities and star formation history could match the observed colors and would provide very similar R-band magnitudes, and also stellar masses \citep{Bell2000, Bell2001}. 

\subsection{\mbh-\lhost\ relation}\label{sec:ml}

Adopting the $1$~Gyr and $0.625$~Gyr stellar population, we perform a K-correction to derive the rest-frame R-band magnitude of our sample, based on the host inference in the  WFC3 band. As mentioned above, since the WFC3 filter is already close to the rest-frame R-band, we expect the $M_R$ uncertainty introduced by this K-correction to be within $0.05$~mag. We derive the rest-frame R band luminosity from $\log L_R/L_{R, \odot} = 0.4\times(M_{R, \odot}-M_R)$, where $M_{R, \odot}=4.61$~\citep{Blanton07}. $L_R$ ranges between $\log (L_R/L_{\odot,R})  \in [9.5, 11.5]$ with individual values listed in Table~\ref{tab:result_sersic}. 

\begin{figure}
\centering
{\includegraphics[width=0.5\textwidth]{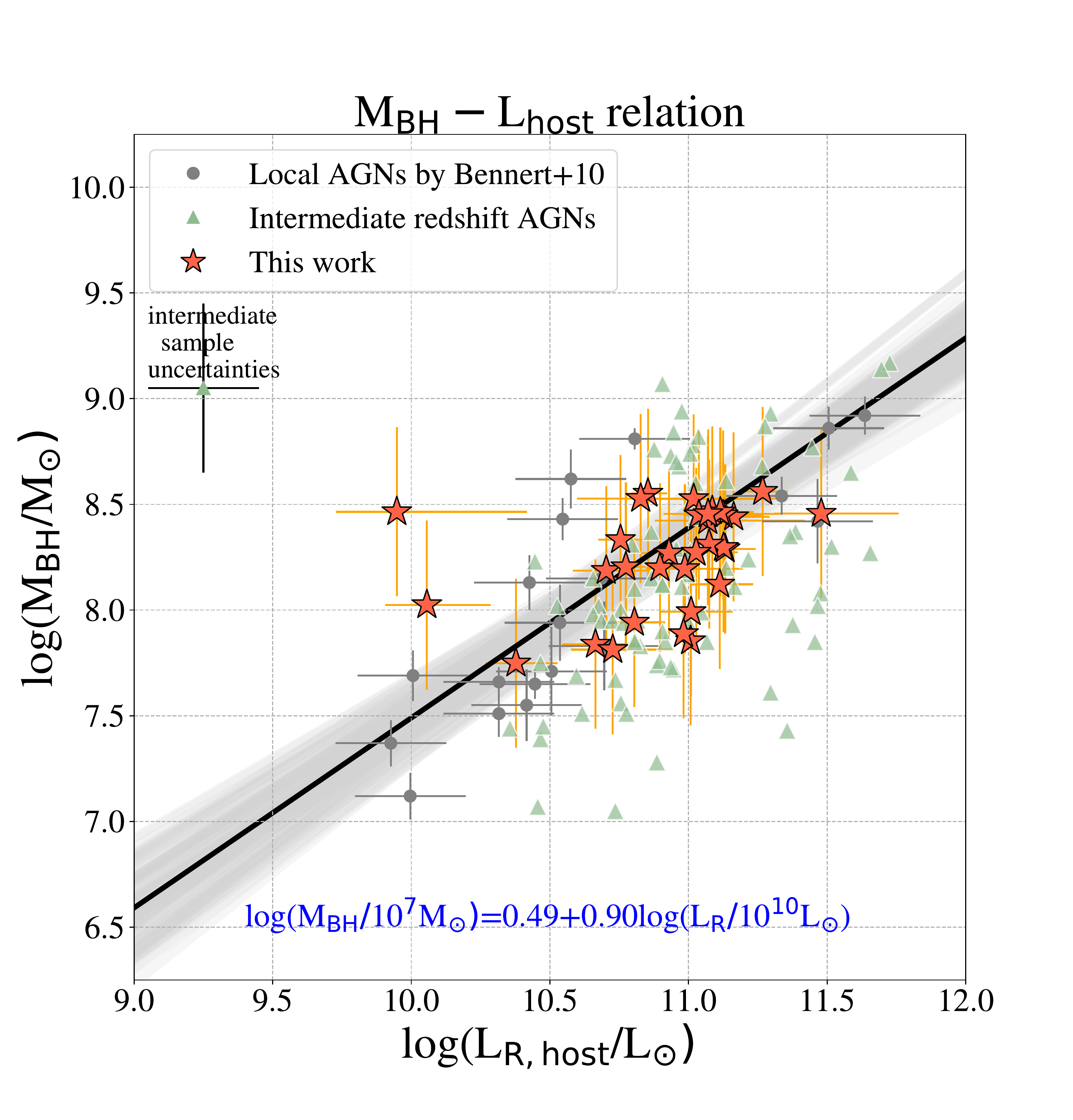}}
\caption{\label{fig:ML} 
The {\it observed} black hole mass vs. R-band luminosity relation. The black line and the blue colored equation indicates the best-fit result of the local sample as given in Equation~\ref{eq:MLlocal}, with $1\sigma$ confidence interval indicated by the gray region. The values for the comparison samples at low and intermediate redshift are listed in \citet{Ding2017b}, Table 1 and 2.
}
\end{figure}

We show the relation between \mbh\ and \lhost\ in Figure~\ref{fig:ML} with comparison samples. For reference, we fit the local data with a linear relation,

\begin{eqnarray}
\label{eq:MLlocal}
\log \big( \frac{\mathcal M_{\rm BH}}{10^{7}M_{\odot}})= \alpha_0 + \beta_0 \log(\frac{L_R}{10^{10}L_{\odot}}),
\end {eqnarray}

\noindent which enables a direct comparison between our high-$z$ sample and the local relation. The distribution of our data appears to be in good agreement with the local relation and the other AGN samples at lower redshift. Therefore, the observational data indicate that the relation between black hole mass and the host luminosity are similar at different periods of the universe. In Appendix~\ref{sec:ml-ev}, we explore how the high-$z$  sample would evolve in this plane solely with the luminosity evolution of the host galaxy as done in past studies \citep[e.g., ][]{Ding2017b}. 

We note that there are outliers that deviate from the distribution of the overall sample, such as SXDS-X763. We suspect that this source may have abnormal host properties. Indeed, the value of \Reff\ has a large uncertainty ($\sim 75\%$), and the host-to-total flux ratio is the lowest of the high-$z$ sample ($<10\%$). 

\subsection{\mbh-\smass\ relation}\label{sec:mm}

Using the NIR imaging with \hst, we take the host luminosity along with color information to estimate the stellar mass content of each host galaxy based on the mass-to-light ratio of the adopted stellar populations. The uncertainty level associated with the stellar mass is expected to be of order $0.1$~dex (changing the IMF would affect all our stellar masses systematically). We find $M_*$ ranges between $\log (M_*/M_{\odot}) \in [9.7, 11.3]$. These values are listed in Table~\ref{tab:result_sersic}. 

\begin{figure}
\centering
{
\includegraphics[width=0.5\textwidth]{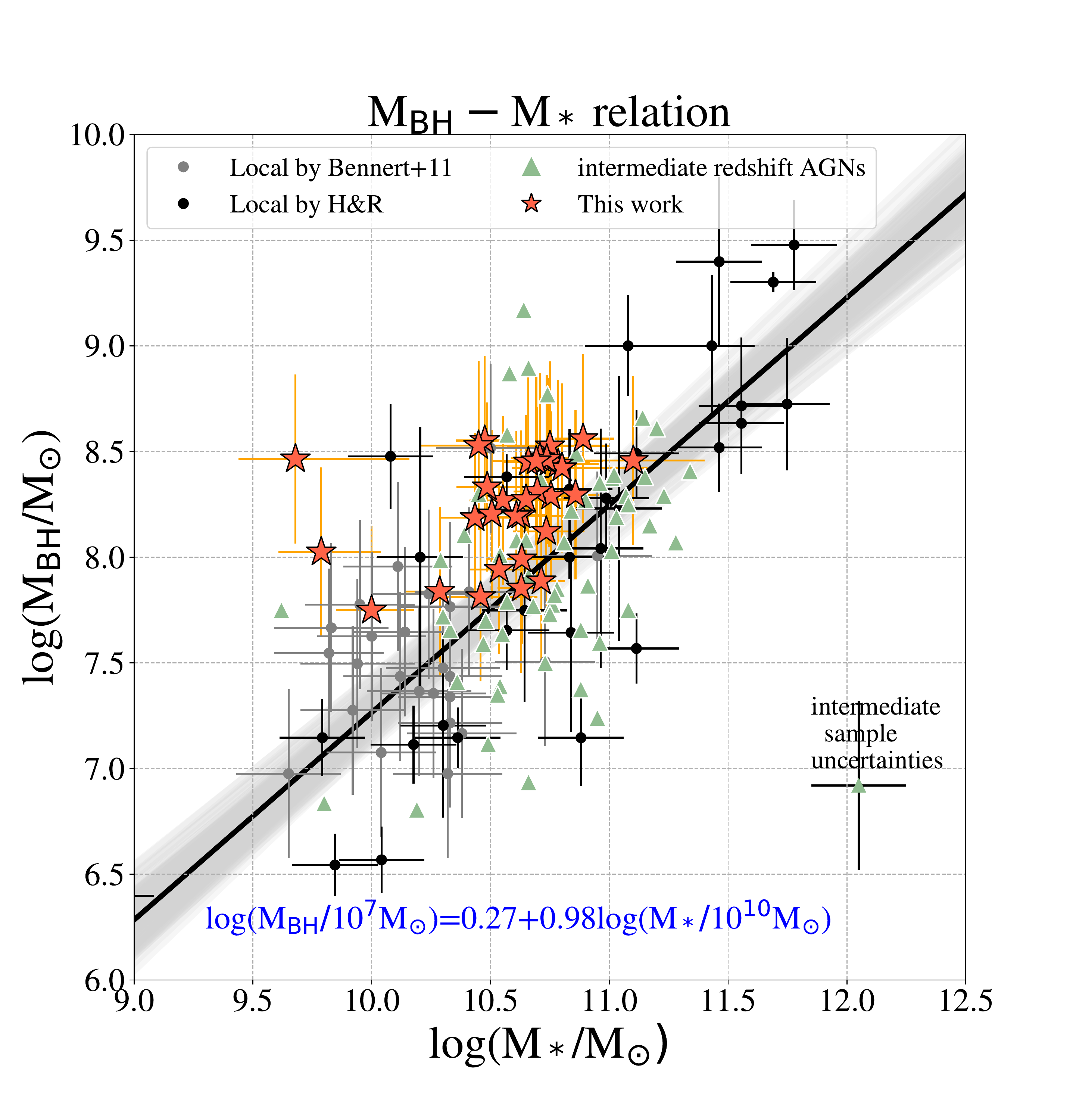}
}
\caption{\label{fig:MM} 
Black hole mass vs. stellar mass relation ( \mbh -- \smass\ ). The best-fit local relation is shown as described in Figure~\ref{fig:ML}.
}
\end{figure}

\begin{figure*}
\centering
\begin{tabular}{c c}
{\includegraphics[width=0.5\textwidth]{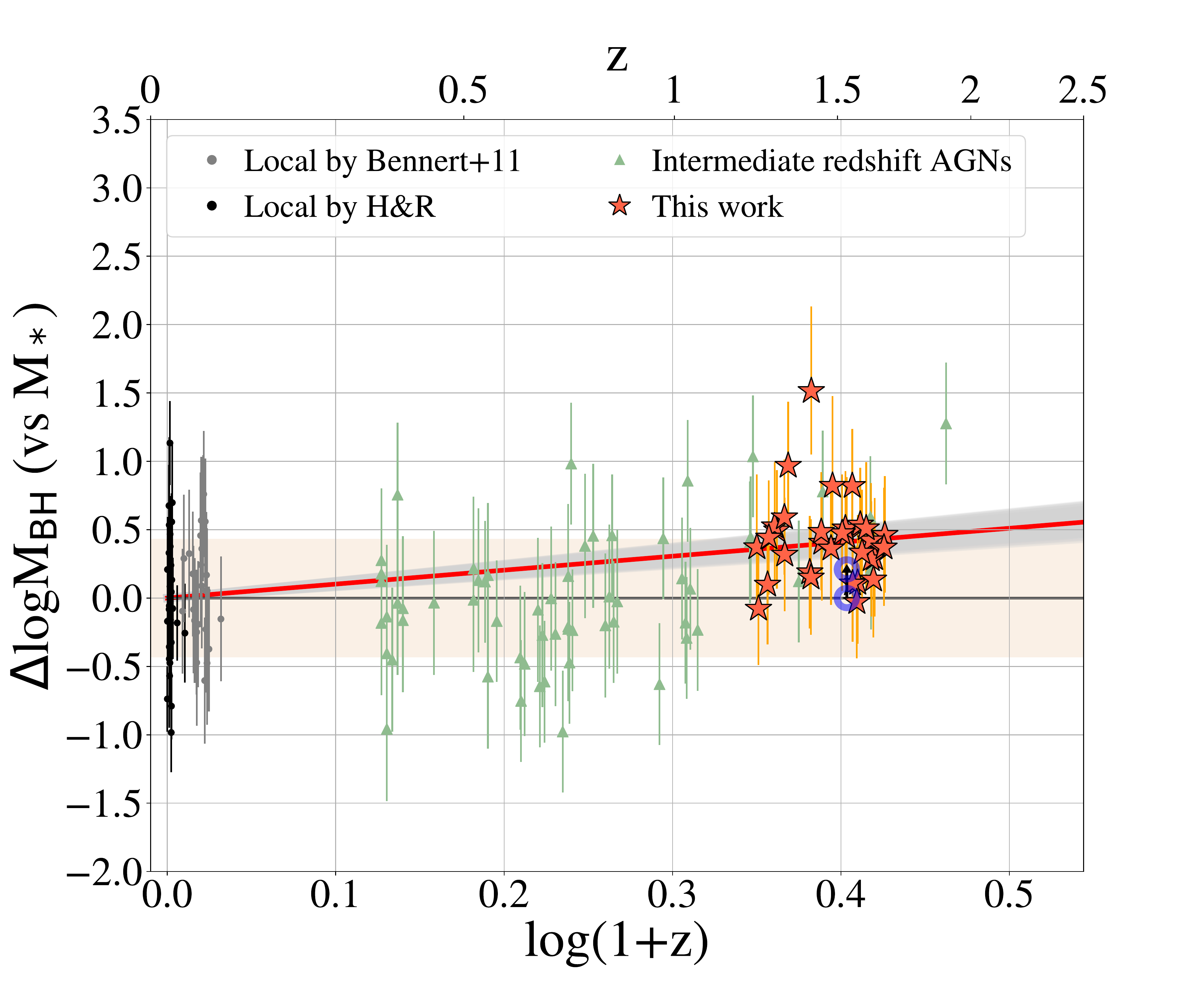}}&
{\includegraphics[width=0.5\textwidth]{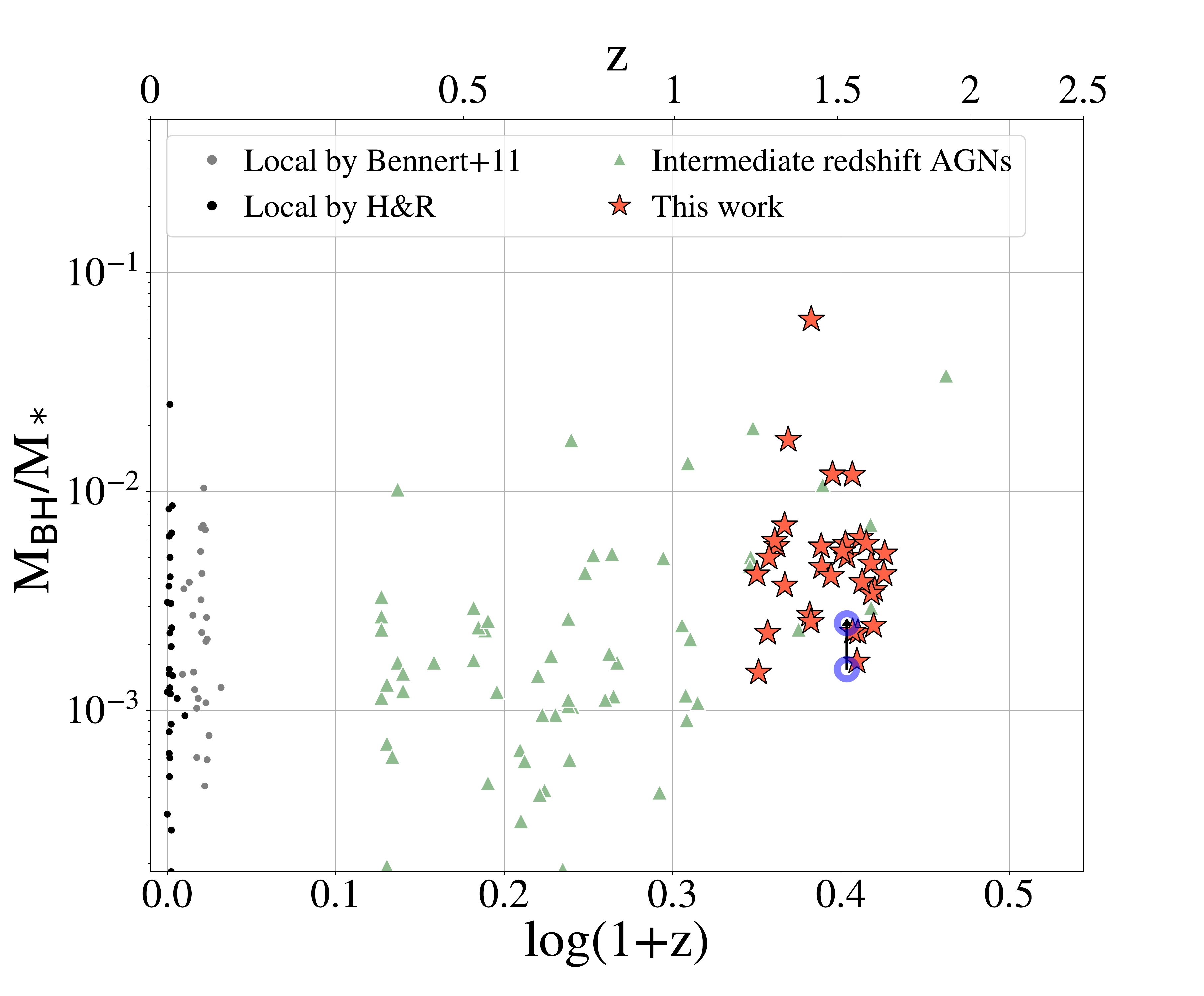}}\\
\end{tabular}
\caption{\label{fig:MM-vz} 
 {\it Left}: Offset in  $\log($\mbh$)$ (vs. \smass) as a function of redshift. The orange band is the intrinsic scatter of local linear relation. The red line and gray band are the best-fit and $1\sigma$ offset fitting by Equation~\ref{eq:offset} using our 32 high-$z$ AGNs, with slope value $\gamma  = 1.03 \pm 0.25$. {\it Right}: \mbh/\smass\ as a function of redshift. In both panels, we use blue open circles to show the expected bias of the median value for $\Delta \log$\mbh. We use framework of \citet{Schulze2011,Schulze2014} to show that even with no evolution selection effects would shift the expectation higher towards the measured values as indicated by the small arrow (Section~\ref{sec:sf_framework}.) The measurements of the local and intermediate redshift samples are re-calibrated using self-consistent recipes and listed in Table~\ref{tab:comp_sample}.}
\end{figure*}

In Figure~\ref{fig:MM}, we plot the our \mbh -- \smass\ measurements and find that the distributions in this plane for the four samples are similar. However, there is a slight shift of the high-$z$ data towards higher black hole masses at a fixed host mass. In Figure~\ref{fig:MM-vz}, we plot the mass ratio as a function of redshift ({\it right} panel) and the mass ratio difference ($\Delta \log \mathcal M_{\rm BH} = \log \big( \frac{\mathcal M_{\rm  BH}}{10^{7}M_{\odot}}) -\alpha-\beta\log(\frac{M_*}{10^{10}M_{\odot}})$ which is considered as the difference between the high redshift data and the best-fit local relation; {\it left} panel). For our high-$z$ sample, we find the averaged $\Delta \log \mathcal M_{\rm BH}$ to be $0.43\pm0.06$, a factor of $~\sim2.7$ higher than the local mass ratio as indicated by the top blue circle in Figure~\ref{fig:MM-vz}.

As with Equation~\ref{eq:MLlocal}, we fit the \mbh-\smass\ data with a linear relation as:
\begin{eqnarray}
\label{eq:MMlocal}
\log \big( \frac{\mathcal M_{\rm BH}}{10^{7}M_{\odot}})= \alpha_1 + \beta_1 \log(\frac{M_*}{10^{10}M_{\odot}}),
\end {eqnarray}

\noindent and an evolution term parameterized in the following form:

\begin{eqnarray}
\label{eq:offset}
\Delta \log \mathcal M_{\rm BH}= \gamma \log (1 + z).
\end{eqnarray} 
Performing such a fit based on the {\it observed} data (i.e., no correction for selection effects), we obtain $\gamma  = 1.03 \pm 0.25$. Even with the exclusion of the few outliers, the slope is significantly non-zero with $\gamma  = 0.89 \pm 0.27$. The best-fit {\it observed} evolution model is shown in Figure~\ref{fig:MM-vz}.

As shown in these panels, the scatter of the high-$z$ correlation presents a similarity to the local relation. To quantitively make this comparison, we investigate the intrinsic scatter of our sample. Note that the {\it observed} high-$z$ sample has effects from both the measurement uncertainty and the narrow selection window, and thus one needs to take them both into account to extract the intrinsic scatter. Based on our forward modeling framework, we find that our AGN sample has intrinsic scatter as $0.25$~dex on the vertical axis, taking into account the measurement uncertainty and the selection function. We will return to this topic in a forthcoming paper (Ding et al. in prep.), where we focus on the scatter of the sample, as a diagnostic of AGN-host galaxy feedback mechanism. The fact that the intrinsic scatter for the high-$z$ sample is no larger than the local one, which is $\sim0.35$~dex \citep{Gul++09}, may pose a challenge for explanations of scaling relations where random mergers are the origin of the correlation \citep{Peng2007,Jahnke2011}, indicating that there may likely be a connection between the SMBHs and their host galaxies during their formation.

\subsection{Taking into account the selection function}
\label{select_eff}

\begin{figure}
\centering
{
\includegraphics[width=0.5\textwidth]{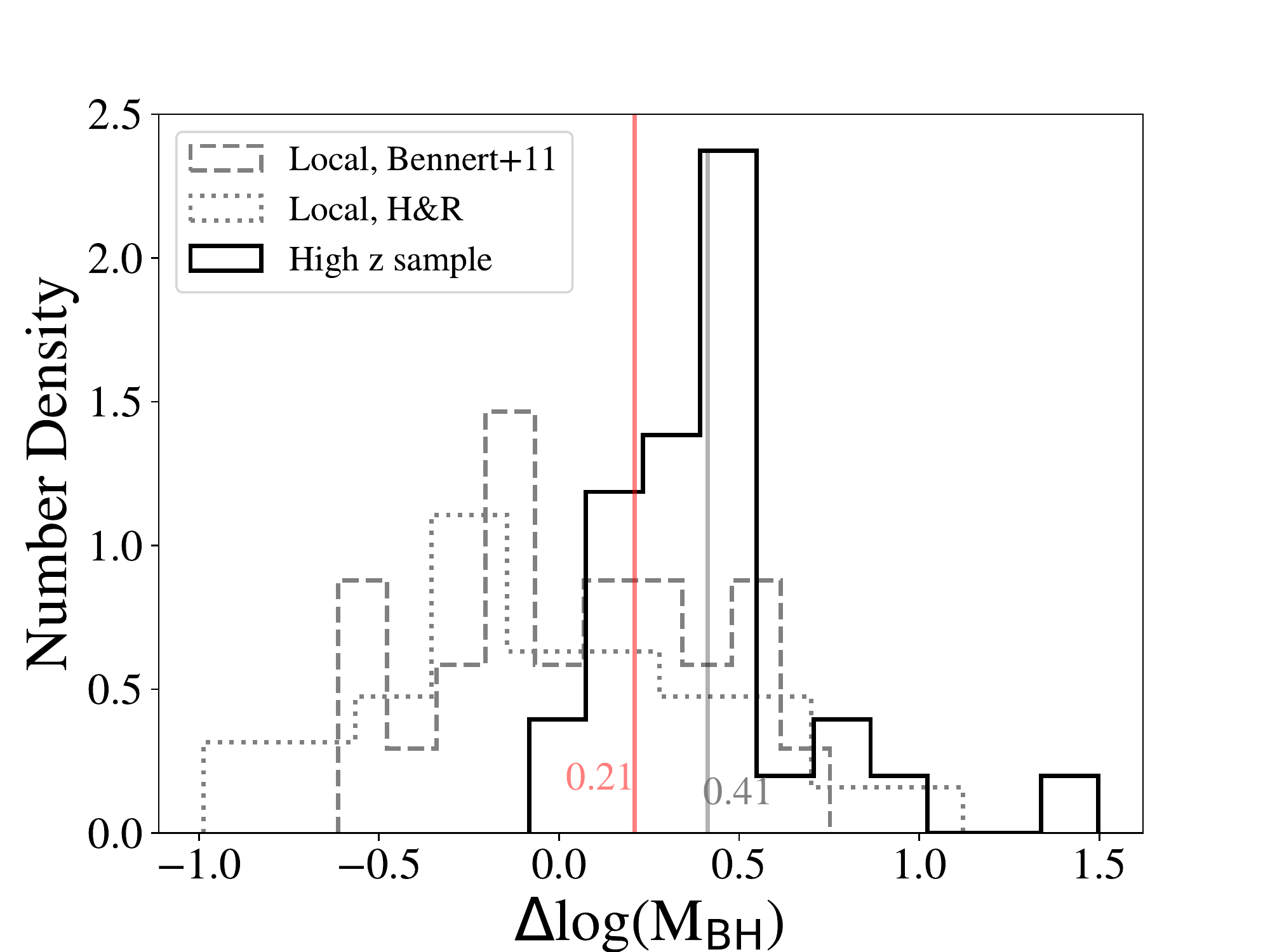}
}
\caption{\label{fig:offset_hist} 
Histogram of the $\log($\mbh$)$ offset for the local samples and our 32 high-$z$ sample. The gray vertical line shows the mean value of the offset for the high-$z$ sample, and the red vertical line shows the shift the expectation by the selection effect, even if there were no evolution. The median value of the offset is very close to the mean value ($+0.30$).
}
\end{figure}

In Section~\ref{sec:sf_framework}, we used the Bayesian framework introduced by \citet{Schulze2011} and find that in the case of no intrinsic evolution of the correlations we would expect to measure an offset corresponding to a $+0.21$~dex bias in the inferred $\Delta \log$\mbh, owing to our selection function. Interestingly, this bias could account for the majority of the observed offset as obtained in the \mbh-\smass\ correlations in the last section. 
To demonstrate, we use blue open circles to show the expected bias raised by this selection effect in Figure~\ref{fig:MM-vz}. We also show, in Figure~\ref{fig:offset_hist}, the histogram $\Delta \log$\mbh\ of the high-$z$ sample and compare it to the local ones. 
The plots show that the correlations between \mbh\ and host galaxy total stellar mass or luminosity could be consistent with those of the local samples given the uncertainties and selection function.

To further evaluate the ``true" underlying evolution and its uncertainty, corrected for selection effects, taking into account the actual observations, we adopt an independent method based on the approach introduced by \citet{Tre++07} and developed by \citet{Ben++10}, \citet{Park15} and \citet{Ding2017b}. The method parametrizes the evolution of the correlations between \mbh\ and host properties as an offset from the local one $\gamma$, and an intrinsic scatter \sint, which can be a free parameter or tied to the local relation. It then imposes a selection function in \mbh\ and calculates the intrinsic parameters of the model given the observations. In practice, we start from the local black hole mass function and the evolution model, generate mock samples using a Monte Carlo approach, apply the selection function and compare with the data to generate the likelihood. To illustrate the importance of the intrinsic scatter in the selection effects, we adopt both uniform (flat) prior and lognormal prior for \sint. Note that this method assumes a narrow Eddington ratio distribution, which is different from the one by \citet{Schulze2011}, as described in Section~\ref{sec:sf_framework}. Also, this method adopts the local black hole mass function rather than the high-$z$ BH mass function. These differences have a second-order effect and could be responsible for the different magnitude of the selection effect. On the other hand, this method probes the importance of the scatter at high-$z$, which is complementary. 

\begin{figure*}
\centering
\begin{tabular}{c c}
\subfloat[\mbh-\smass, flat prior]
{\includegraphics[width=0.5\textwidth]{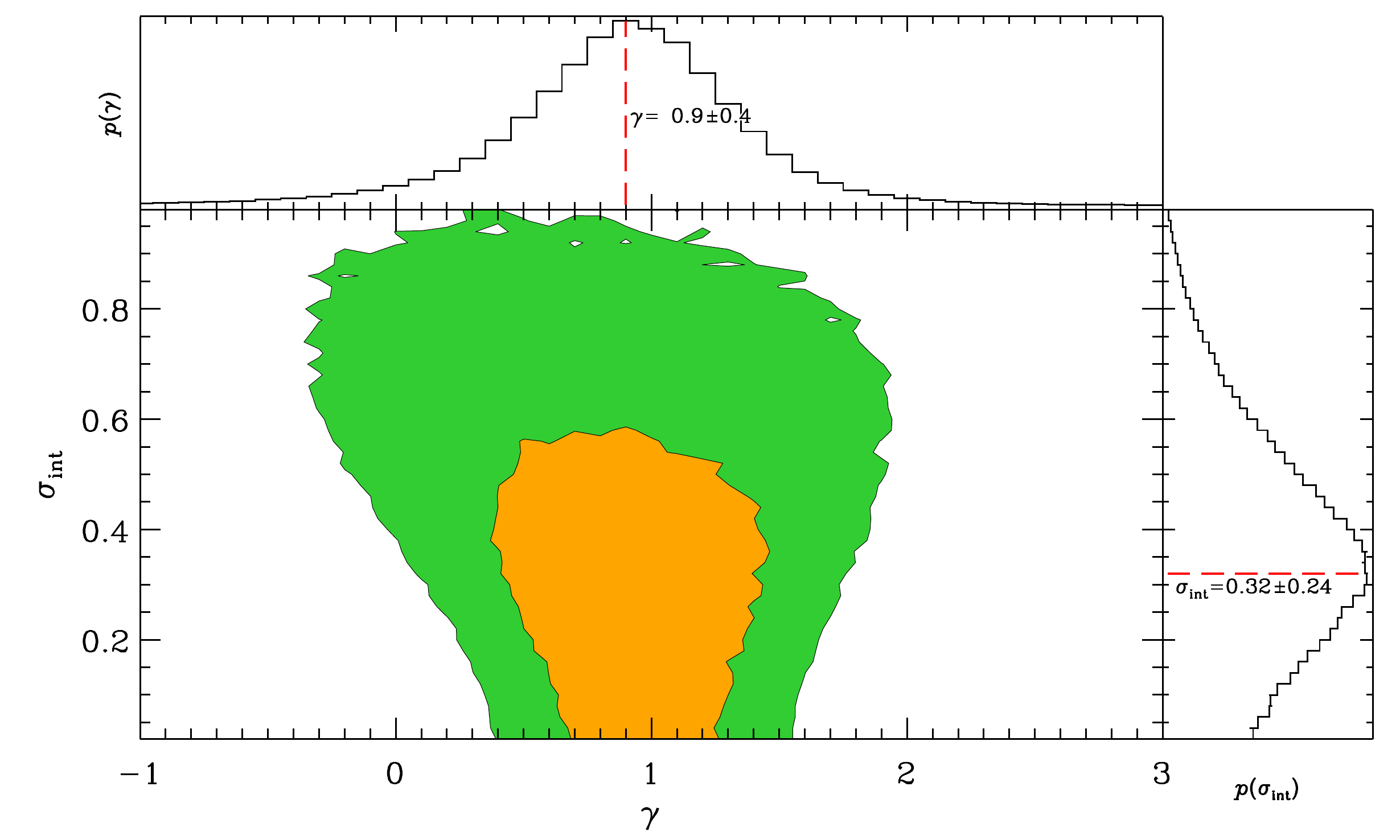}}&
\subfloat[\mbh-\smass, lognormal prior]
{\includegraphics[width=0.5\textwidth]{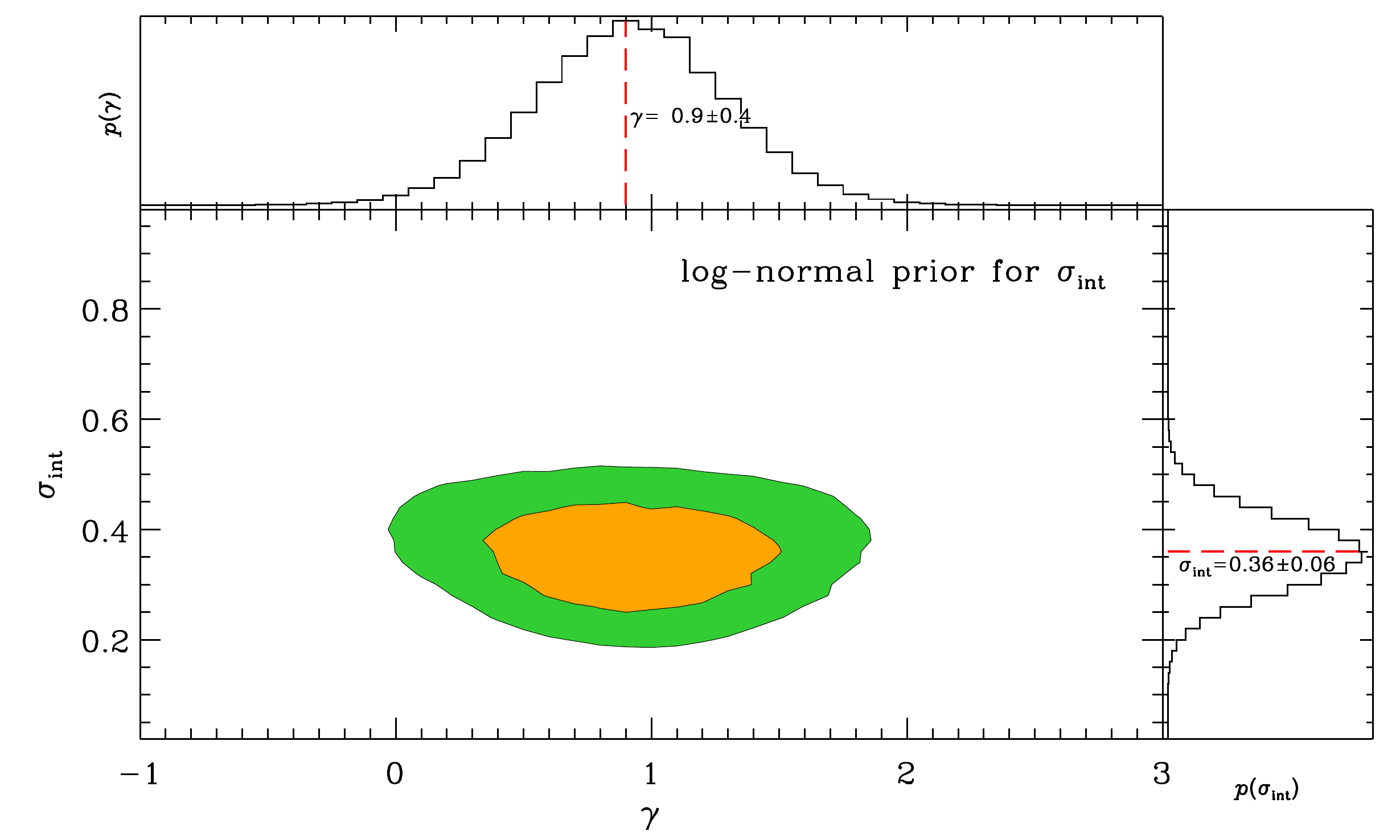}}\\
\end{tabular}
\caption{\label{fig:select_effect} 
Constraining the evolution factor $\gamma$ of \mbh-\smass\ relation (Equation~\ref{eq:offset}), with intrinsic scatter \sint, using a Monte Carlo simulation. The adopted sample includes our 32 AGNs and the intermediate redshift AGNs, using flat prior of \sint\ ({\it left} panel) and lognormal prior ({\it right} panel). The colored regions indicate the 68 and 95\% confidence regions.
}
\end{figure*}

Combining the 32 AGNs together with the intermediate redshift sample, we present the inferred  $\gamma$ and \sint\ in the two-dimensional planes in Figure~\ref{fig:select_effect}. The plots show that the inferred evolution is uncertain and depends crucially on the intrinsic scatter, especially when applying a luminosity evolution for the host (see Appendix~\ref{sec:ml-ev}). Assuming the lognormal prior \sint, to mimic the assumption of the method discussed in Section~\ref{sec:sf_framework}, one sees that the best estimate of $\gamma$ is positive, but the 95\% confidence intervals extend to zero. Thus one cannot conclude that evolution is significantly detected in our data. 
When relaxing the prior on \sint\ we find that the scatter is consistent with being as low as in the local samples, with a one-sided interval including zero and $1\sigma$ upper limit at around $0.5$~dex. 
We also study the selection effect by only considering the new 32 AGNs, resulting in a higher evolutionary trend with a higher value of $\gamma$. We show all the result of the $\gamma$ and \sint\ in Table~\ref{table:gamma_sf}.

A simple check using our prior estimate of the bias yields a similar result. The mean offset of the sample of 32 objects from the local relationship is $0.43\pm0.06$ dex, which would correspond to $\gamma=1.08\pm0.15$. However, after correcting for the selection bias ($0.21$~dex, see Section~\ref{sec:sf_framework}), the offset reduces to $0.22\pm0.06$ and thus $\gamma=0.55\pm0.15$, marginally positive, but not conclusively inconsistent with the local value given the error bars and uncertainty in the correction. We also note that the distribution of offsets shown in Figure~\ref{fig:offset_hist} displays a positive asymmetric tail. Whereas we expect the negative tail to be suppressed by our selection function, and thus it should be accounted for in our treatment.
If one computes the median offset instead of the mean, the offset is almost completely consistent with the expected bias. Larger samples with different selection function are needed to establish whether the positive tail is real or due to small sample statistics. 

\subsection{\mbh-\bmass\ relation}\label{sec:bh_bulge}

The local sample of inactive galaxies used in this analysis is mainly comprised of bulge-dominated galaxies, and the entire local \smass\ we adopted are their bulge masses. That is, in previous sections, we are comparing the \mbh-\smass$_{\rm ,total}$ relations in the distant universe to the \mbh-\smass$_{\rm ,bulge}$ relations locally. Considering that a significant stellar component of the high-$z$ AGNs have a disk component (i.e., AGN hosts with fitted \sersic\ index close to 1),  \smass$_{\rm , bulge}$ must be smaller than \smass$_{\rm ,total}$. Given that the structures of our AGN hosts are similar to inactive galaxies at equivalent redshifts and stellar masses (Section~\ref{sec:result-hosts}), we expect their B/T ratios to follow a similar distribution that can be estimated using a single-\sersic\ index \citep{Bruce2014}. We can then infer the bulge stellar mass of our AGN hosts based on the inferred B/T ratios. 

\begin{figure}
\centering
{
\includegraphics[width=0.5\textwidth]{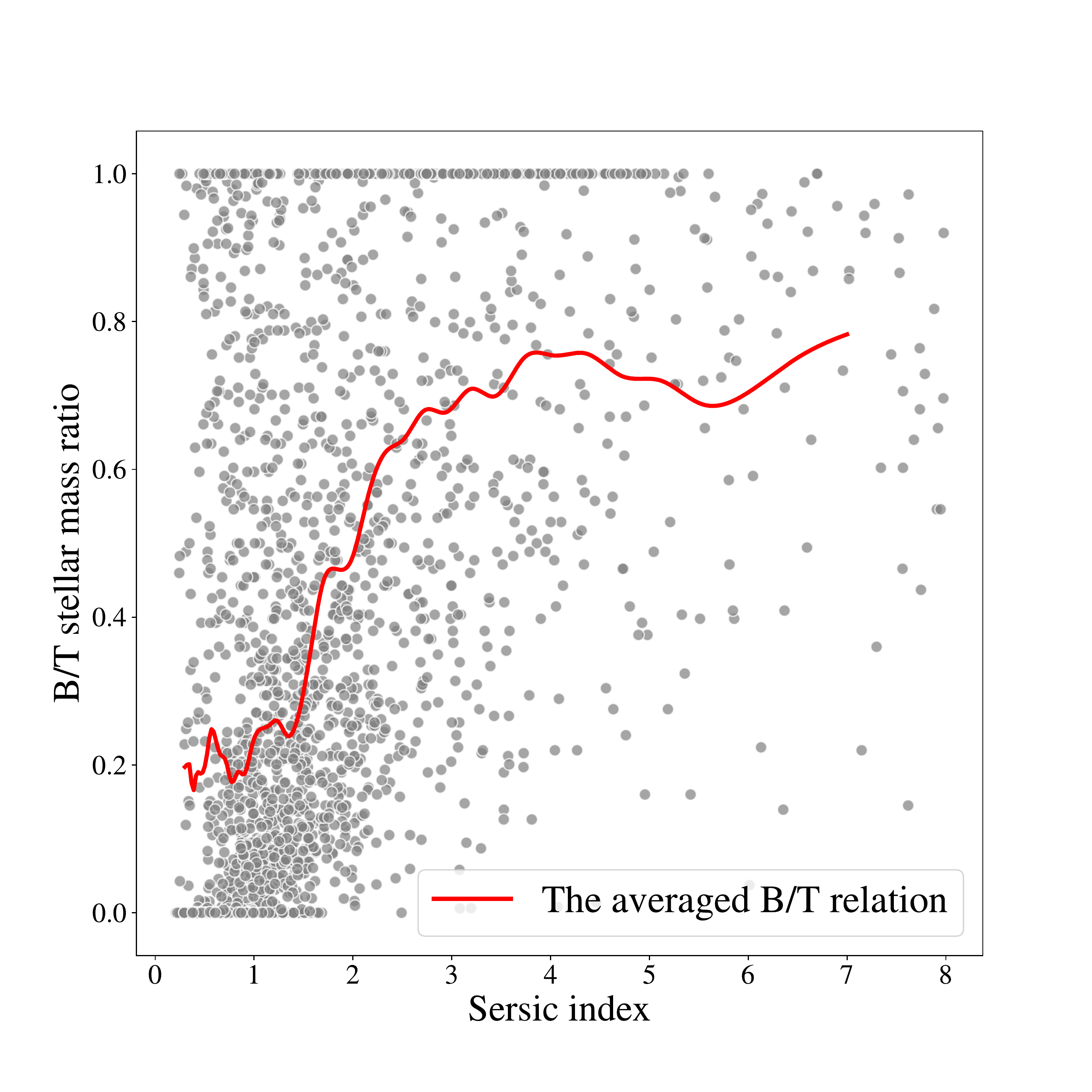}
}
\caption{\label{fig:BT-n_relation} 
Relation between bulge-to-total stellar mass ratio and single-\sersic\ index using CANDELS inactive galaxies with $\log (M_*/M_{\odot})\in~[9.0, 11.5]$. The red curve is the averaged B/T ratio at different \sersic\ index. Note that this curve is only for demonstrating the expected B/T value at different \sersic\ index. In this work we adopt the Monte Carlo approach to randomly sample the B/T to account for the scatter.
}
\end{figure}

First, we establish a relation between the single-\sersic\ index and the B/T ratio using inactive galaxies from CANDELS at similar redshifts and stellar masses (Figure~\ref{fig:BT-n_relation}; $\log (M_*/M_{\odot})\in~ [9.0, 11.5]$) using the B/T measurements of \citet{Dimauro2018}. In the figure, the red line indicates the average B/T ratio at a given \sersic\ index. We then implement a Monte Carlo approach by randomly sampling a Gaussian \sersic\ distribution for each AGN in our sample based on our measurements with 1$\sigma$ errors (Table~\ref{tab:result_sersic}). Next, we randomly sample the associated B/T ratio for a given \sersic\ index. To avoid the case of an unphysical faint bulge flux, we set the lower limit for B/T ratio as $0.1$. In each realization, we compare the \smass$_{\rm, bulge}$ to that of the \mbh\ and estimate the offset and $\gamma$.  We use 10,000 realizations to make sure the distribution of random samples is stable. Note that the Monte Carlo approach would take the scatter into account, including the \sersic\ index and its relation with the B/T ratio. We list the resulting bulge masses \bmass\ in Table~\ref{tab:result_sersic}.

\begin{figure*}
\centering
\begin{tabular}{c c}
{\includegraphics[height=0.39\textwidth]{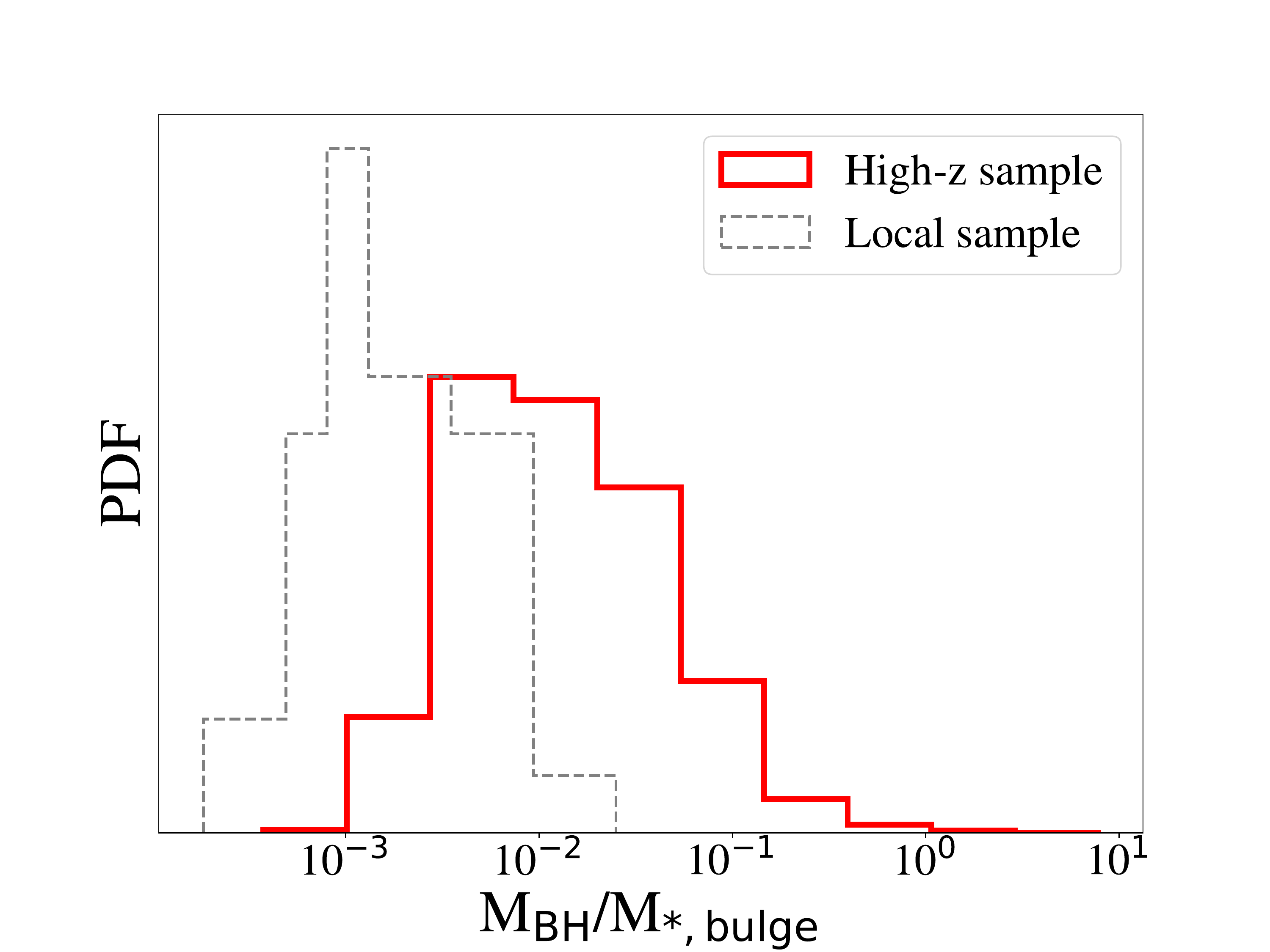}}&
{\includegraphics[height=0.39\textwidth]{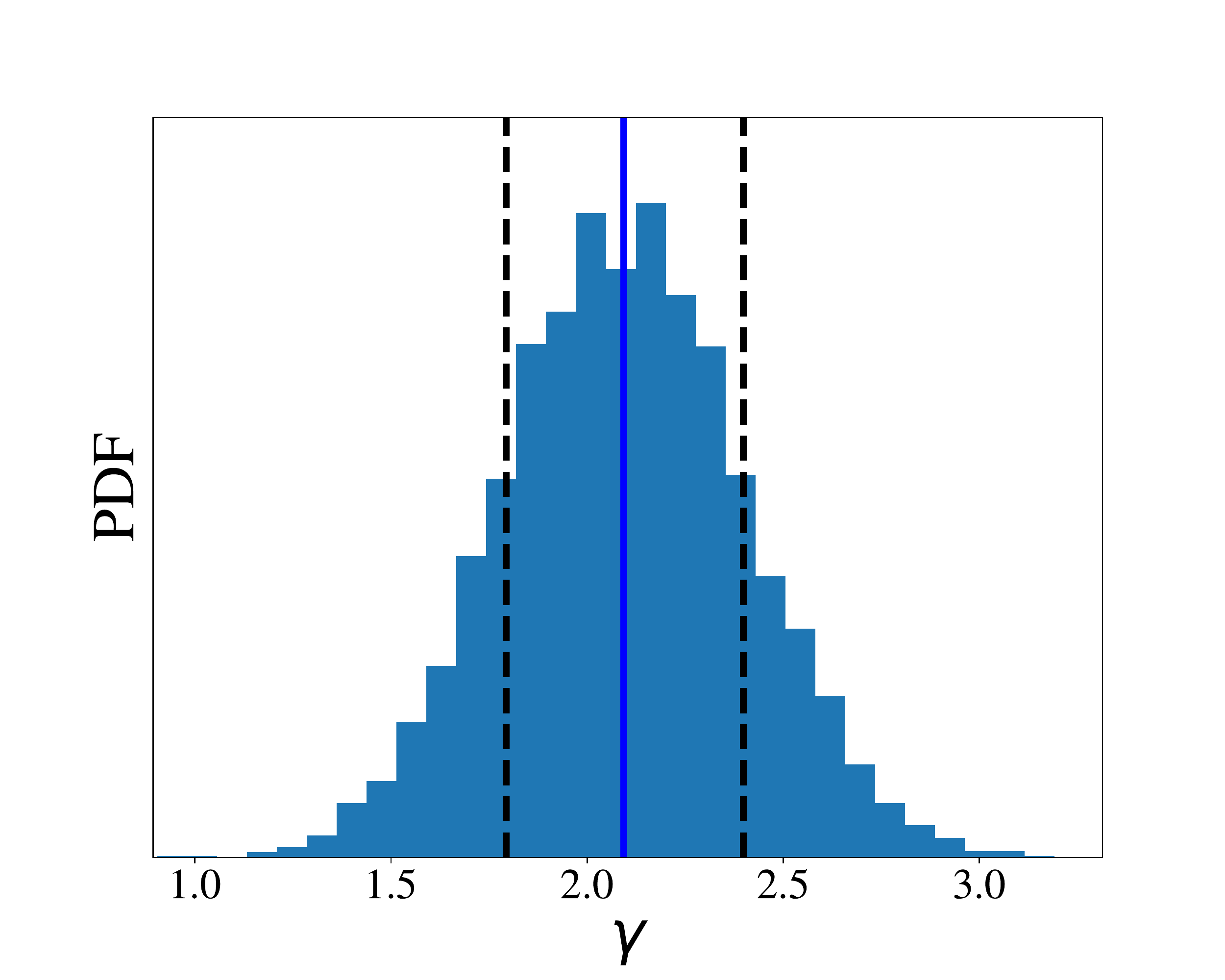}}\\
\end{tabular}
\caption{\label{fig:gamma_hist} 
Illustration of the distribution from the $10,000$ time realizations. {\it Left}: \mbh/\smass\ distribution for 32 high-$z$ sample, compared to the local one. {\it Right}: $\gamma$ distribution, with median value (blue line) and $1\sigma$ region (dashed line) .
}
\end{figure*}

\begin{figure*}
\centering
\begin{tabular}{c c}
{\includegraphics[width=0.5\textwidth]{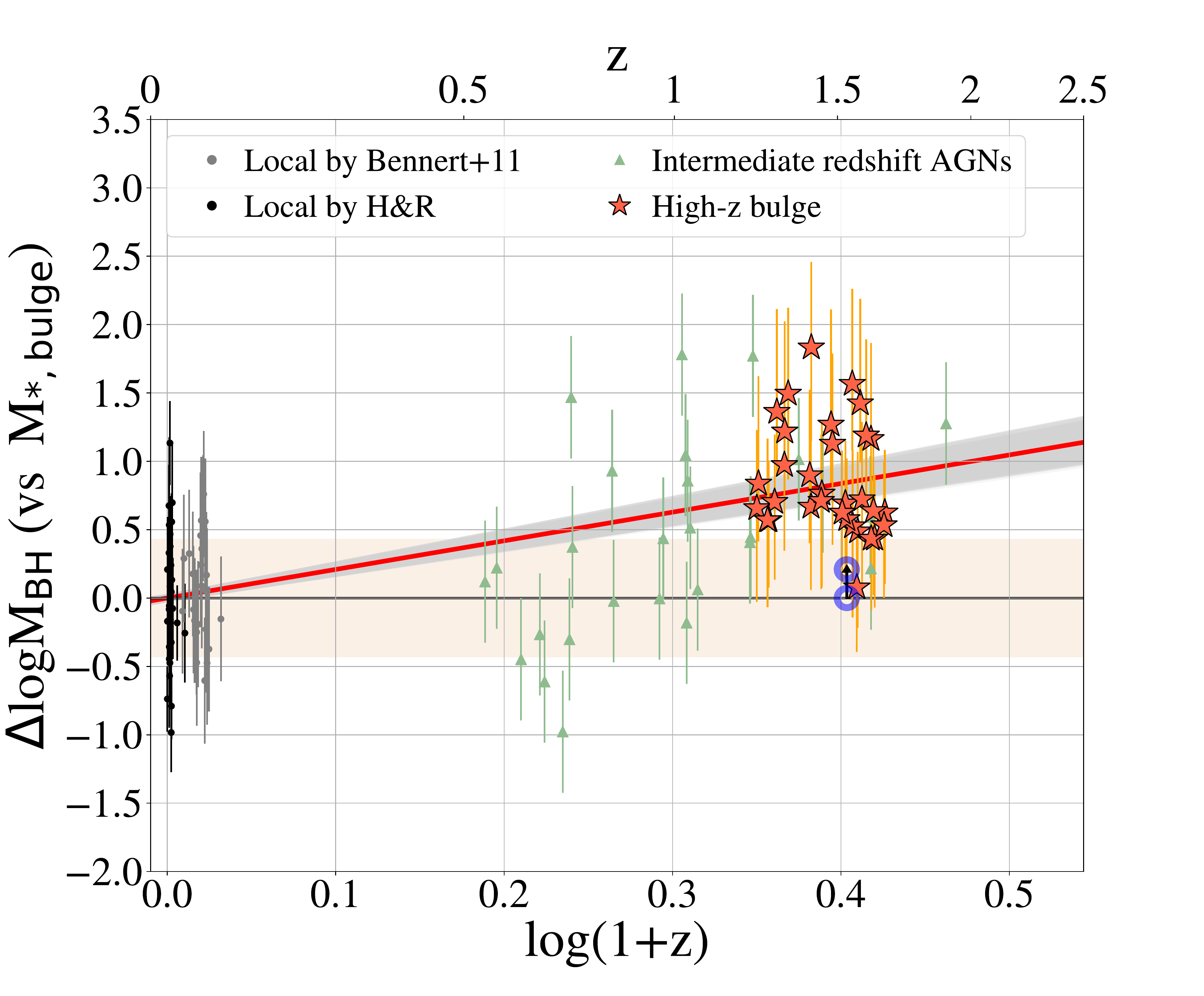}}&
{\includegraphics[width=0.5\textwidth]{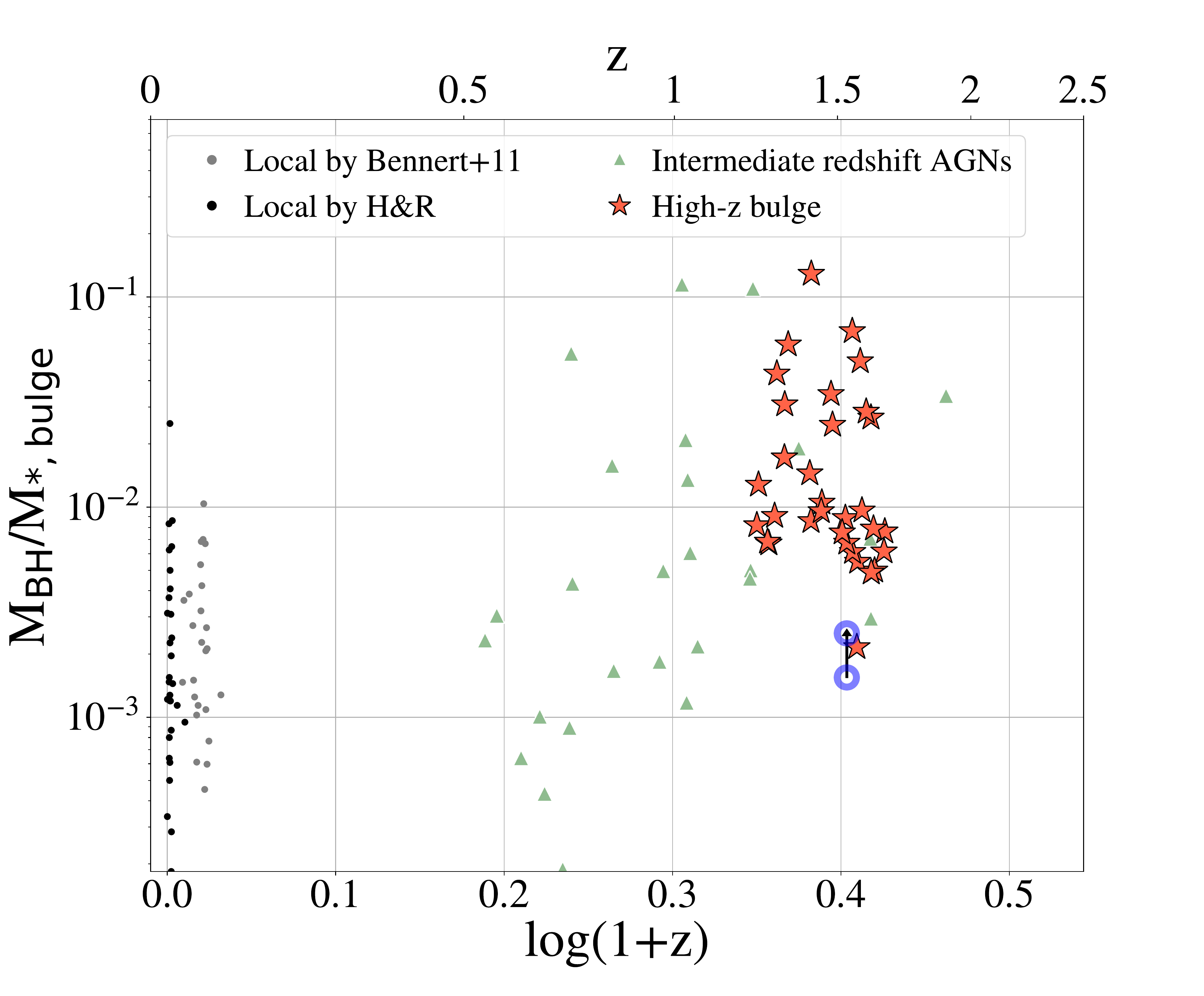}}\\
\end{tabular}
\caption{\label{fig:MM_bulge-vz}
Same as Figure~\ref{fig:MM-vz}, but for \mbh-\bmass\ relations. The intermediate sample consists of 27 objects from \citet{Bennert11} and \citet{SS13}, listed in Table~\ref{tab:comp_sample}. \citet{Cisternas2011} do not provide information to infer their bulge mass.}
\end{figure*}

As expected, we find a stronger offset of
$\Delta \log \mathcal M_{\rm BH}^{bulge} = 0.87\pm0.07$~dex in \mbh/\smass$_{\rm, bulge}$ with respect to the
local relation, which corresponds to an evolutionary trend with
$\gamma = 2.09\pm0.30$. Since selection effects are independent of B/T in
our framework, we can apply our prior estimate of the correction for selection effects by removing $0.21$~dex, leaving still a significant offset of $0.67\pm0.07$~dex. We show the histogram of  \mbh/\smass\ and $\gamma$ inferred by
the Monte Carlo approach in Figure~\ref{fig:gamma_hist} and illustrate the offset between bulge and BH
as a function of redshift in Figure~\ref{fig:MM_bulge-vz}.

Even with the uncertainty associated with our estimate of the
bulge-to-total ratio, it is clear that there is a significant evolution
in the \mbh\ -- \bmass\ relation, considering the
marginal evidence for evolution in the \mbh-\smass$_{\rm, total}$
relation and that our galaxies are not pure bulges. This finding
confirms with higher fidelity the conclusion by \citet{Bennert11}
based on optical data of a smaller sample of 11 objects in the same
redshift range \citep[see][for similar results]{SS13, Jah++09, Cisternas2011}. We list the inferred $\gamma$ using the different sample combinations in this section in
Table~\ref{table:diff_sample_gam}.

\section{Final technical remarks}
\label{sec:dis}

\subsection{Systematic errors}\label{sec:sysm_err}

In this work, we use state-of-the-art techniques to measure the host galaxy flux, separating it from that of the point source. The fidelity of the host galaxy magnitude is high as demonstrated by the robustness with respect to the choice of PSF, which is the primary source of uncertainty. However, we introduce some uncertainty by adopting two common simple stellar population templates to derive the rest-frame R-band luminosity and stellar mass for the full population. In principle, if we had more information on the color, we could improve on this source of uncertainty by adopting a specific stellar population model for each target. However, considering that the host magnitude in the \hst/ACS band is faint (host-to-total flux ratio $< 30\%$), the individual color that we measure is insufficient to further discriminate between templates, and thus we adopt the common templates for simplicity and to facilitate reproduction of our work.

In any case, we stress that the uncertainty in the scaling relations is dominated by the single epoch black hole mass estimates (i.e., $0.4$~dex), and not by uncertainties in the photometry or stellar mass estimates (i.e., typically $\sim0.15$~dex). Thus, the choice of template is a subdominant source of error. Note that the large uncertainties of \mbh\ is not immediately apparent from the figures, since only the AGNs that have {\it measured} \mbh\ that fall into our selection window (i.e., $\log($\mbh$/M_{\odot})\in[7.5, 8.56]$) have been selected. A posteriori, the fact that the intrinsic scatter we observe is not larger than the one in the local universe is consistent with this statement.

\subsection{Systematic errors related to the choice of local anchor}
\label{sec:local_sys}

To compare our high-$z$ sample to the local relation, we adopt the local measurements by \citet{Ben++10, Bennert++2011} and \citet{H+R04}. In the literature~\citep{Kormendy13, Bentz2018}, other analyses of the local sample are available, which could also be considered as the local anchor. However, from the point of view of our differential evolutionary measurement, we stress once again the importance of selecting local anchors that have been measured and calibrated in a similar way to the distant high-$z$ samples. It is crucial that the black hole masses be obtained in the same of self-consistent manner, and it is also critical that the host/AGN decomposition be measured in the most similar way possible.

Therefore, for these reason, the B10 and HR04 samples are the most appropriate for our goal, in spite of their limitations. For instance, B10 did not consider morphological features like strong bars in the decomposition. This choice might affect the inference of the bulge/disk properties in absolute terms, but it is the most similar decomposition to what is possible at high-$z$ where these features cannot be resolved. Also, the HR04 sample does not use color information to estimate the \smass~of their sample, rather than use Jeans modeling. This gives rise to systematic uncertainties related to the choice of the initial mass function for example, which need to be kept in mind. We refer the reader to \citet{Kormendy13} for more discussion of systematic issues in local samples.

The issue of a consistent calibration of black hole masses is perhaps the most subtle and thorny.
Not all the local correlations are mutually consistent, especially when considering different galaxy properties like luminosity, stellar mass, and velocity dispersion, and therefore a single self-consistent calibration is not possible. To illustrate the problem, we compare our local \mbh-\smass\ sample to that published by \citet{Kormendy13} (hereafter K13). We find that the \mbh-\smass\ measurements by K13 have a global $\sim+0.3$~dex offset relative to our local sample. Note that a $+0.3$~dex difference would almost erase the $\Delta\log$\mbh\ for the high redshift sample as described in the last section, if it were independent of the black hole mass calibration. However, the \mbh-\sigstar\ relations by K13 are also offset by  $\sim+0.3$~dex with respect to those given by \citet{Woo2010}, which is the baseline for our adopted black hole mass estimators. That is, if we were to adopt the K13 normalization instead, the \mbh\ of all the AGN sample should be increased by $\sim$0.3~dex at the same time. As a result, adopting a different local anchor would only globally affect the level of the \mbh\ to local and high-$z$ sample at the same time, with the result of leaving $\Delta\log$\mbh\ unchanged. To illustrate this point further, we adopt the K13 local ellipticals and classical bulges as the $z\sim0$ anchor and compare to the recalibrated \mbh\ of our high-$z$ AGNs sample. The inference of  $\gamma$ changes to $0.74\pm0.31$ which is consistent at $1\sigma$ level with the value (i.e., $\gamma=1.03\pm0.25$ ) based on our chosen anchor.

\citet{Bentz2018}, given the superior resolution of their data in the local universe, decomposed their host using more components than bulge$+$disk (e.g., Bar(s), Barlens, Ring(s)). This approach is impossible at high-$z$ given current data, and therefore their sample should not be used as local anchor, for consistency.
Just for illustration, when adopting the \citet{Bentz2018} as the local anchor, we obtain $\gamma = 2.12\pm 0.32$ using the 32 AGNs, indicating still a positive evolution, although as expected the amount of evolution is different for the reasons given above.

In conclusion, whereas the uncertainty in the local anchor does not affect our measurement of evolution, it is important to keep in mind this issue in evolutionary studies and verify that the measurements are as self-consistent as possible in the local and distant universe. In order to further reduce the uncertainty, one should perform a self-consistent measurement with identical techniques and data quality both in the local and distant universe. This effort is currently in progress and when completed should enable us to completely eliminate this source of uncertainty \citep{Bennert11,Harris2012, Bennert2015}.

\section{Conclusions} \label{sec:sum}

We studied the evolution of the correlations between the supermassive black hole and their host galaxies using new measurements of 32 X-ray selected AGNs at $1.2<z<1.7$. Near-infrared spectroscopic observations with Subaru/FMOS of the \halpha\ emission line are available that provide reliable BH masses. To obtain the properties of the AGN host galaxies, we performed an image decomposition using state-of-the-art techniques on \hst/WFC3 IR data to obtain high-resolution imaging of the AGN and its host. This required us to collect PSF-stars across all the fields to build a library of PSFs. Using state-of-the-art image modeling tool \lenstronomy, we decomposed the AGN image in the 2D plane taking each PSF in the library. We obtained the host properties (i.e., host flux, effective radius, \sersic\ index) using a weighted arithmetic mean based on the inference from the eight top-ranked PSFs. With additional \hst/ACS image in the optical, we identified the appropriate stellar populations and derived the rest-frame R-band luminosity (\lhost) and stellar mass (\smass) of the host with the corresponding mass-to-light ratio.
We then determine the mass relations at high-$z$ and establish any signs of evolution by combining our high-$z$ measurements with local and intermediate redshift samples from the literature~\citep{Park15, Bennert11, SS13, Cisternas2011}. 

We find the average ``observed" ratio between the mass of a SMBH and either their total host luminosity or total stellar mass is larger at $z\sim1.5$, by $0.41$~dex (a factor of $\sim2.6$) and  $0.43$~dex (a factor of $\sim2.7$), respectively, as compared to that in the local universe. 
However, taking into account uncertainties and the bias due to selection effect could bias, the offset to the local universe is only marginally significant.
Even with the remaining uncertainties in the corrections for selection, we are confident that any evolution in the \mbh/\smass\ relation is at the most a factor of two with the case of no evolution entirely plausible.

Considering the bulge component \citep{Bennert11, Woo++08} separately, we find that our high-$z$ AGN sample has significantly lower mass/luminosity bulges at fixed black hole mass than in the local universe. Comparing to their \mbh, we find evolution with the ratio evolving as $(1+z)^{2.09\pm0.30}$, which is significant even when accounting for selection effects. We caution that the evolution of the bulge component is more uncertain than our measurement of the total galaxy components, since it is based on an estimate of the B/T ratio from the measured \sersic\ index based on a matched sample of inactive galaxies. Nevertheless, since the signal is stronger than for the integrated quantities, the measurement is more significant even given larger uncertainties.
Thus, we can conclude that the BHs $8-10$~Gyrs ago reside in bulges which are less massive/luminous than today \citep[see also][]{Bennert11,Park15}, even though a proper measure of the bulge mass at high-$z$ may need to wait for the next generation ground-based 20-30m class telescopes.

We summarize several limitations of this work. First, due to finite resolution of \hst, we model the host as a single \sersic\ profile. The residuals do not show evidence of additional components, indicating that our choice is appropriate given the noise and resolution of the data. However, we cannot exclude that more features would be revealed by higher quality data.

Second, we adopt simple stellar population templates to derive the host luminosity and stellar mass for the sample, which introduces some uncertainty in the inference of the host properties. However, this source of uncertainty is subdominant with respect to that associated with single epoch black hole mass estimates (see Section~\ref{sec:sysm_err} for detail). 

Third, for a few AGN hosts, their inferred host effective radius hits the lower limit, and thus the inference of their \Reff\ should be considered as the upper limit. However, the inference of the host luminosity and stellar mass is still reliable in these cases, as shown by our test using different lower limit boundaries (Section~\ref{sec:result-hosts}). 

Fourth and last, the choice of local anchor affects our inferred evolution at some level.  We adopted as our baseline the local samples for which measurements have been made with a procedure that is most similar to the one we have applied at high-$z$, but other choices are possible. We investigated the impact of different choices for the local anchor in detail and we found that even though the numerical values can change somewhat, the inferred evolution is always positive  (see Section~\ref{sec:local_sys}).

Given that the \smass$_{\rm ,total}$-\mbh\ relation is closer to or consistent with the local relation, the inferred evolution of the \bmass-\mbh\ correlations is qualitatively consistent with a scenario where the assembly of the black hole predates that of the bulge, which is processed by the transfer of stellar mass from the disk via mergers or secular processes \citep{Jah++09,Bennert++2011,SS13}. The stellar mass needed to build the bulge appears to be present in the disk component at high-$z$.

Of further interest, the scatter in the \mbh-\smass\ ratios at high-$z$ is similar to the local one. This was unexpected since the measurements of the masses of both the black holes and their hosts should have more uncertainty than local estimates. If true, this may indicate that cosmic averaging of initially unrelated states \citep{Peng2007} are not driving the relation between SMBHs and their host galaxies, since then a larger scatter at higher redshifts would be expected. Thus, there is likely a connection between the two yet to be fully understood; contenders include AGN feedback or links to common gas reservoirs.

Finally, the forthcoming launch of the {\it James Webb Space Telescope} ({\it JWST}) and the first light of adaptive-optics assisted extremely large telescopes may provide high-quality imaging data of AGNs at higher redshift (up to $z\sim7$) and thus trace the evolution of correlations at the more distant universe. Although, galaxies, as we know, are smaller at high-$z$ thus may remain challenging to detect under the glare of a luminous AGN. In the lower redshift Universe, wide-area surveys with Subaru/HSC, LSST, and WFIRST offer much promise to build samples for studying these mass ratios and dependencies on other factors (e.g., environment).  

\section*{Acknowledgments}

Based in part on observations made with the NASA/ESA Hubble Space Telescope, obtained at the Space Telescope Science Institute, which is operated by the Association of Universities for Research in Astronomy, Inc., under NASA contract NAS 5-26555. These observations are associated with programs \#15115. Support for this work was provided by NASA through grant number HST-GO-15115 from the Space Telescope Science Institute, which is operated by AURA, Inc., under NASA contract NAS 5-26555.

The authors thank the anonymous referee for helpful suggestions and comments which improved this paper. We acknowledge the importance of the conference ``Particle Astrophysics and Cosmology Including Fundamental InteraCtions (PACIFIC-2013)" in preparing for this program. We fully appreciate input from Luis Ho, Renyue Cen, Hyewon Suh, Takahiro Morishita and Peter Williams. XD, SB and TT acknowledge support by the Packard Foundation through a Packard Research fellowship to TT. TT acknowledges support by NSF through grant AST-1412315. VNB gratefully acknowledges assistance from a NASA grant associated with HST proposal GO 15215. JS is supported by JSPS KAKENHI Grant Number JP18H01251 and the World Premier International Research Center Initiative (WPI), MEXT, Japan. A.S. is supported by the EACOA fellowship.


This work has made use of \lenstronomy~\citep{lenstronomy}, {\sc Astropy}~\citep{Astropy}, {\sc photutils}~\citep{photutils}, {\sc TOPCAT}~\citep{TOPCAT}, {\sc Matplotlib}~\citep{Matplotlib} 
and standard Python libraries.

\bibliographystyle{apj.bst}

\newpage

\appendix

\section{A. AGN - host galaxy decomposition for the remaining 31 AGNs}\label{sec:restsample}
The AGNs decomposition for the other 31 objects of our sample, presented in the same way as Figure~\ref{fig:AGN_decomp}.

\begin{figure}[ht]
\centering
{
\includegraphics[height=0.25\textwidth]{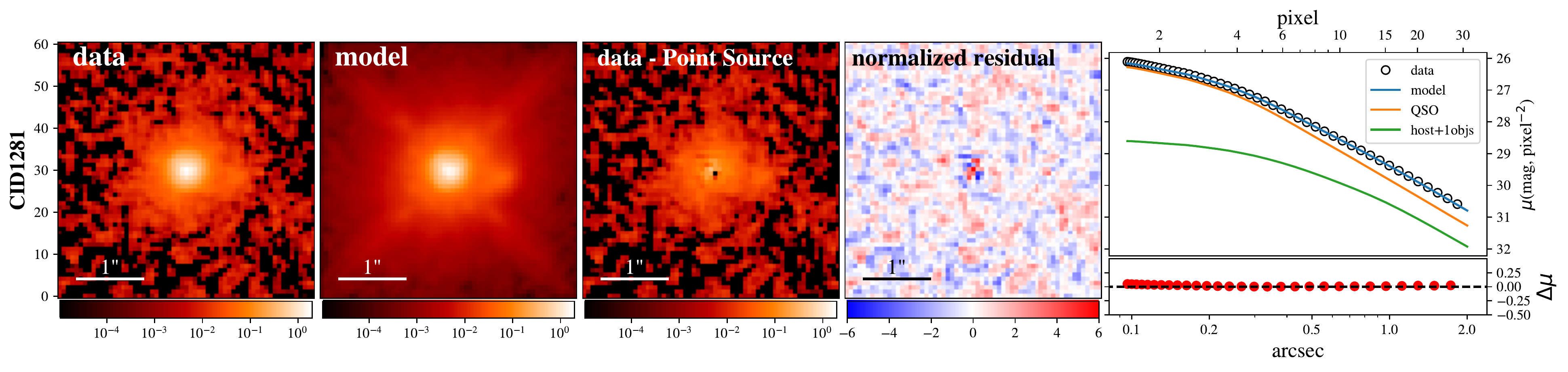}
\includegraphics[height=0.25\textwidth]{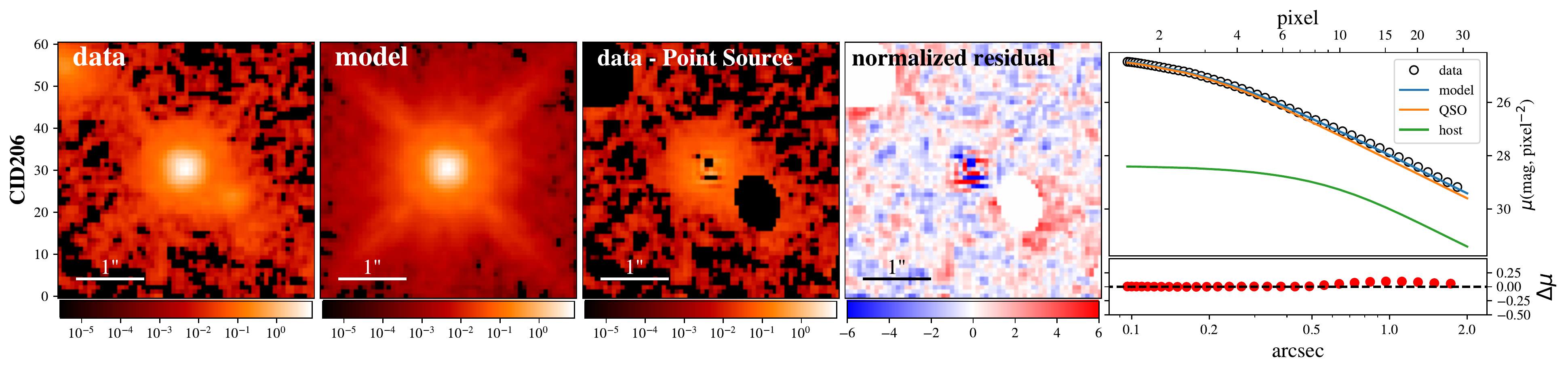}
\includegraphics[height=0.25\textwidth]{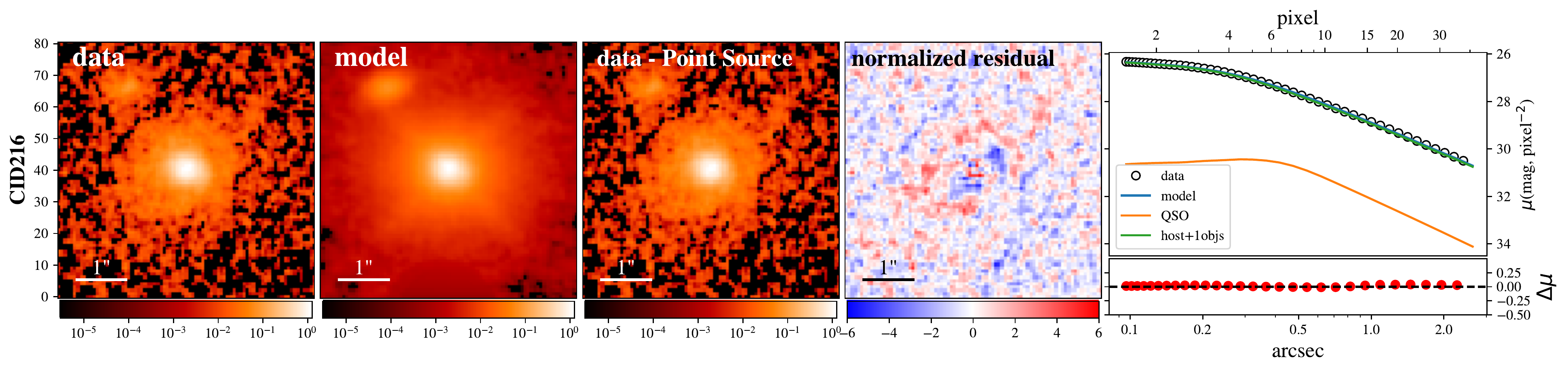}
\includegraphics[height=0.25\textwidth]{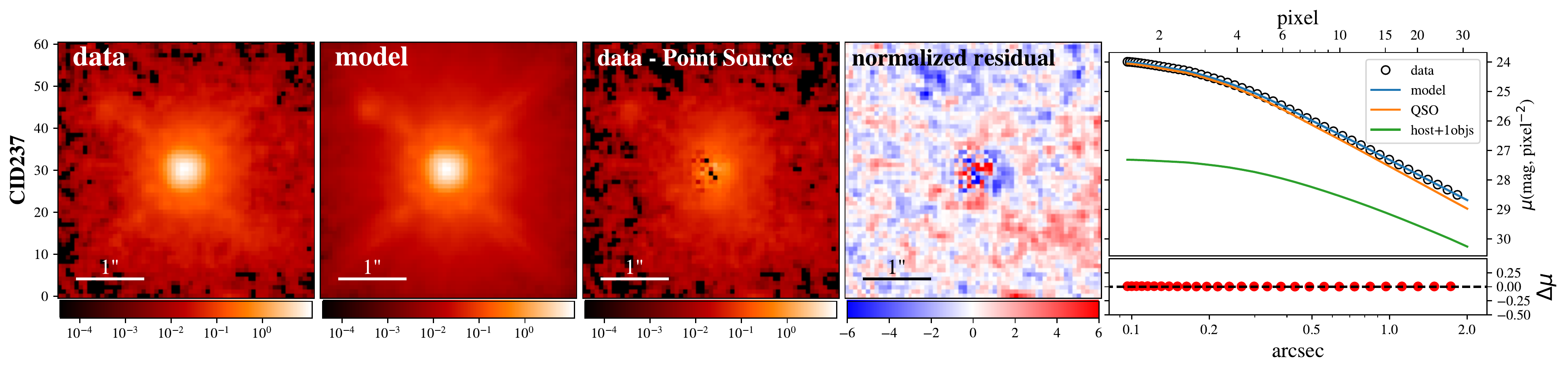}
}
\end{figure} 

\begin{figure}
\centering
{
\includegraphics[height=0.25\textwidth]{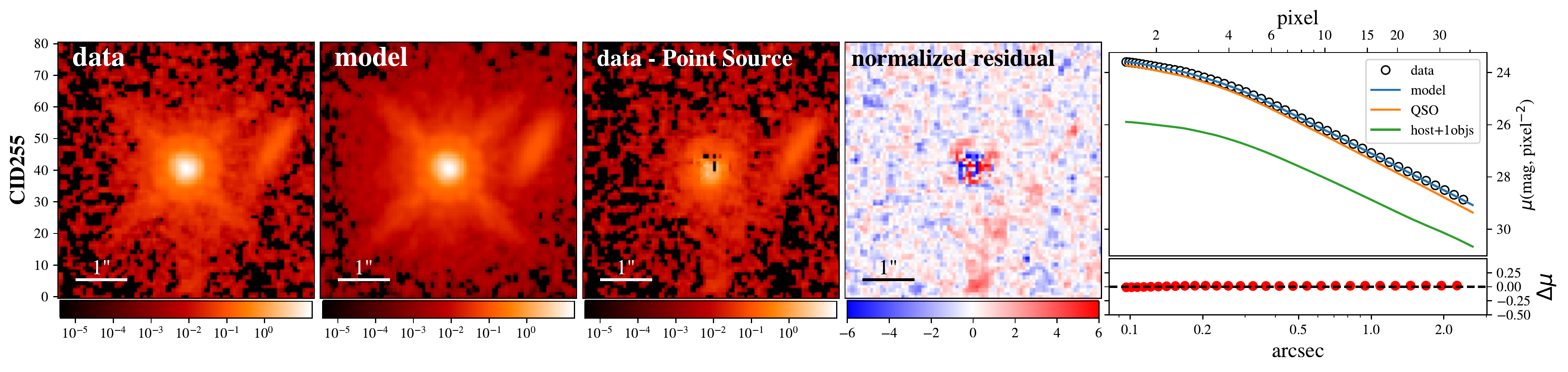}
\includegraphics[height=0.25\textwidth]{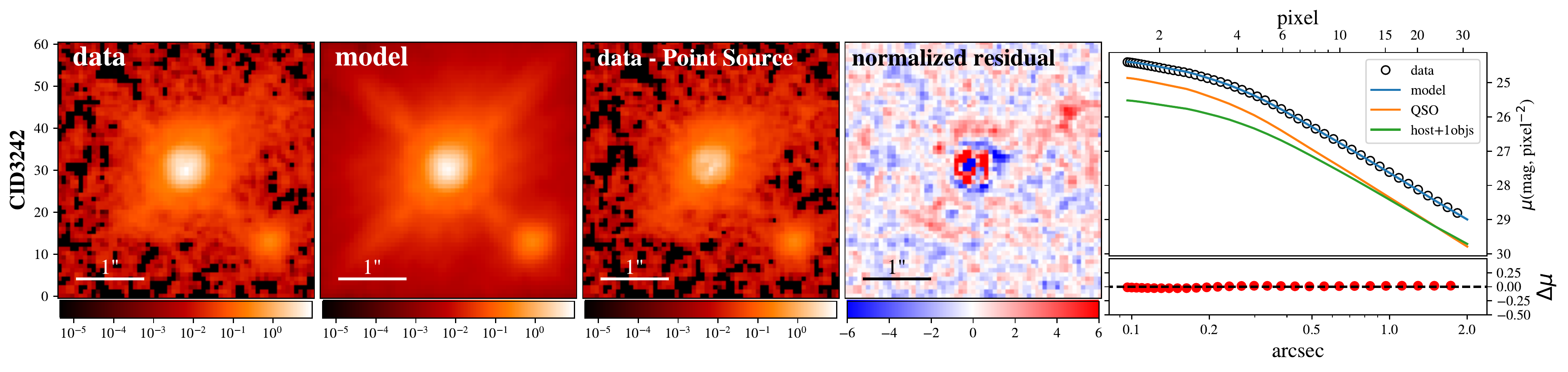}
\includegraphics[height=0.25\textwidth]{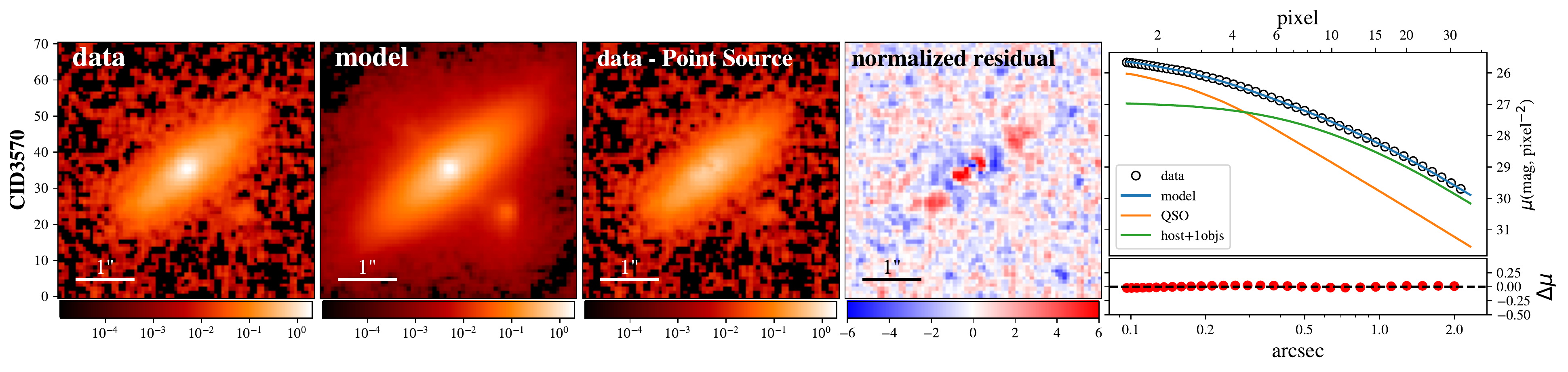}
\includegraphics[height=0.25\textwidth]{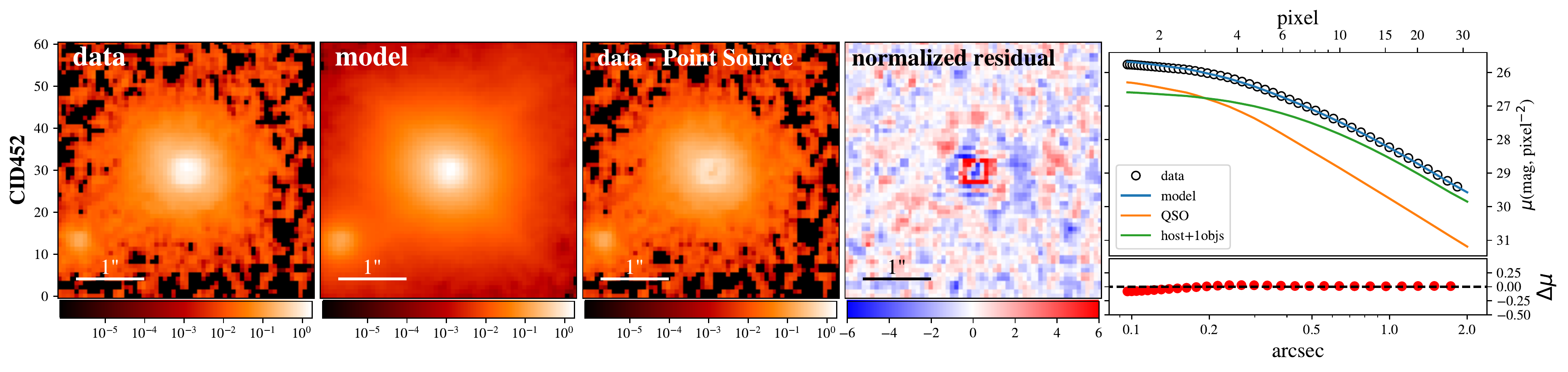}
\includegraphics[height=0.25\textwidth]{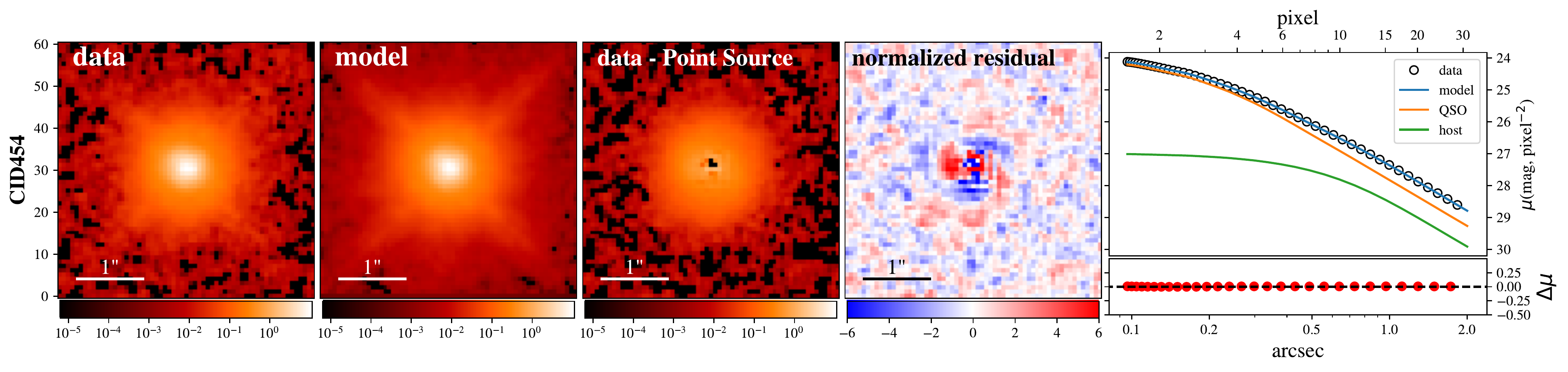}
}
\end{figure} 

\begin{figure*}
\centering
{
\includegraphics[height=0.25\textwidth]{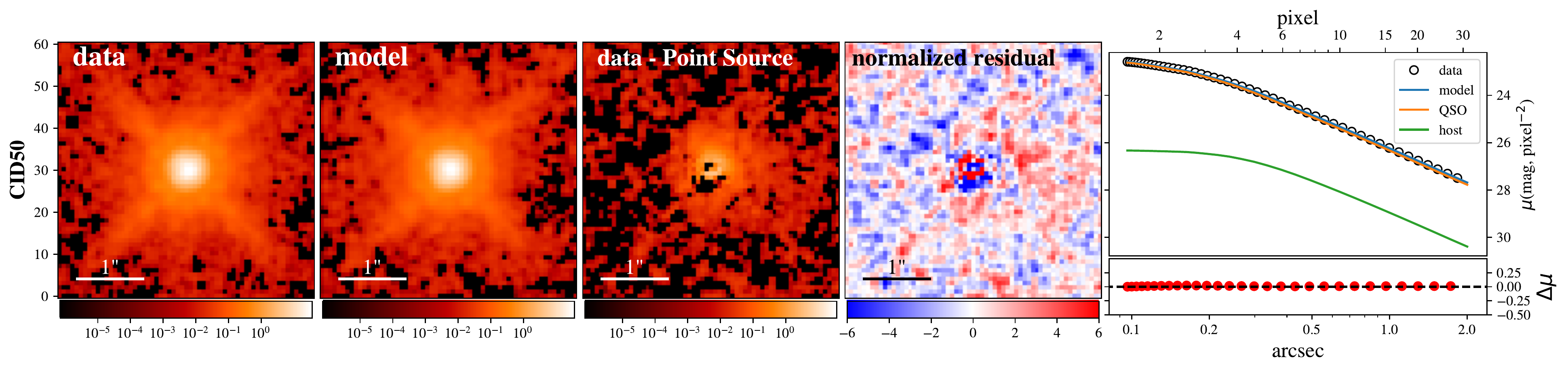}
\includegraphics[height=0.25\textwidth]{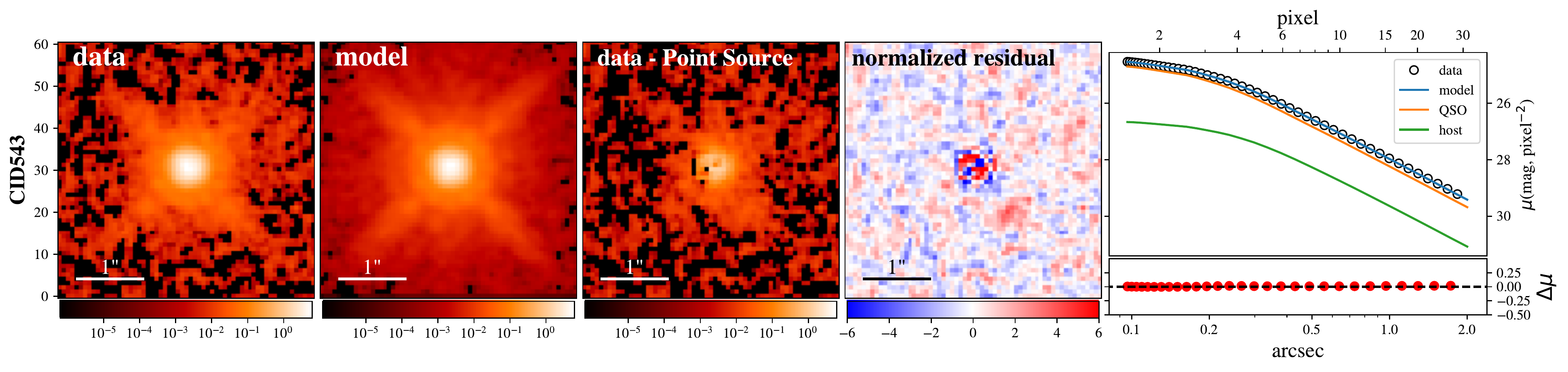}
\includegraphics[height=0.25\textwidth]{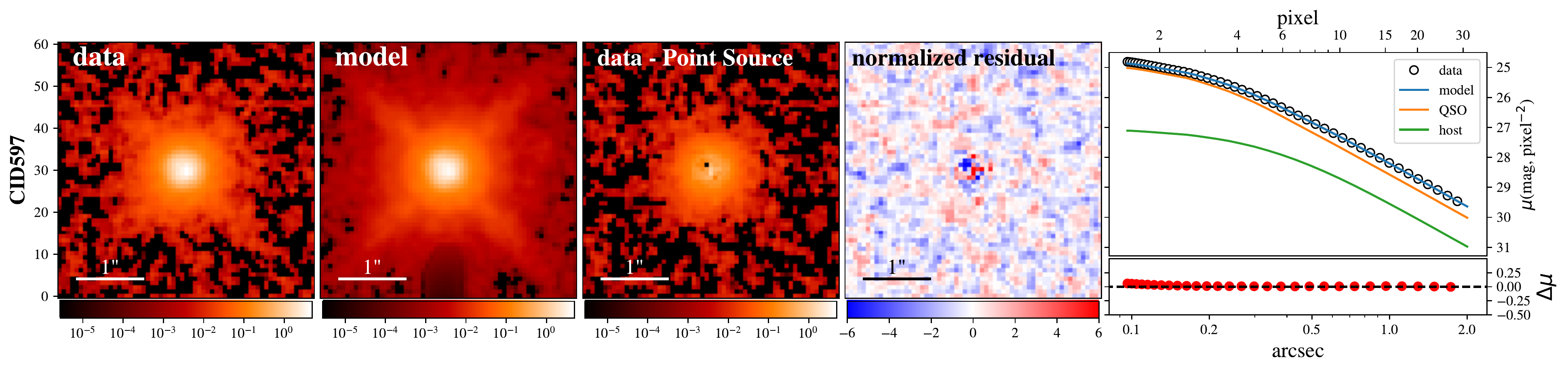}
\includegraphics[height=0.25\textwidth]{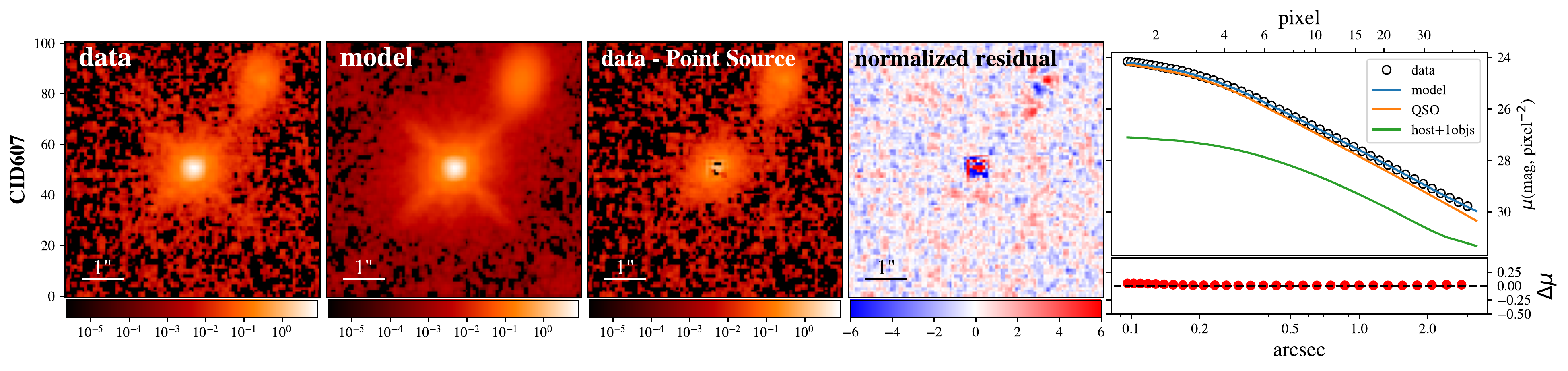}
\includegraphics[height=0.25\textwidth]{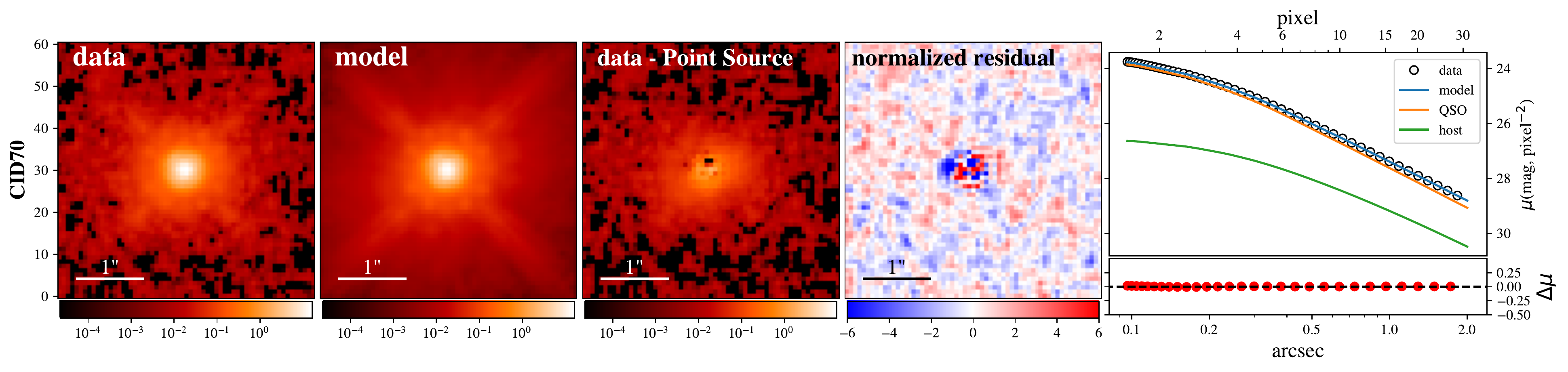}
}
\end{figure*} 

\begin{figure*}
\centering
{
\includegraphics[height=0.25\textwidth]{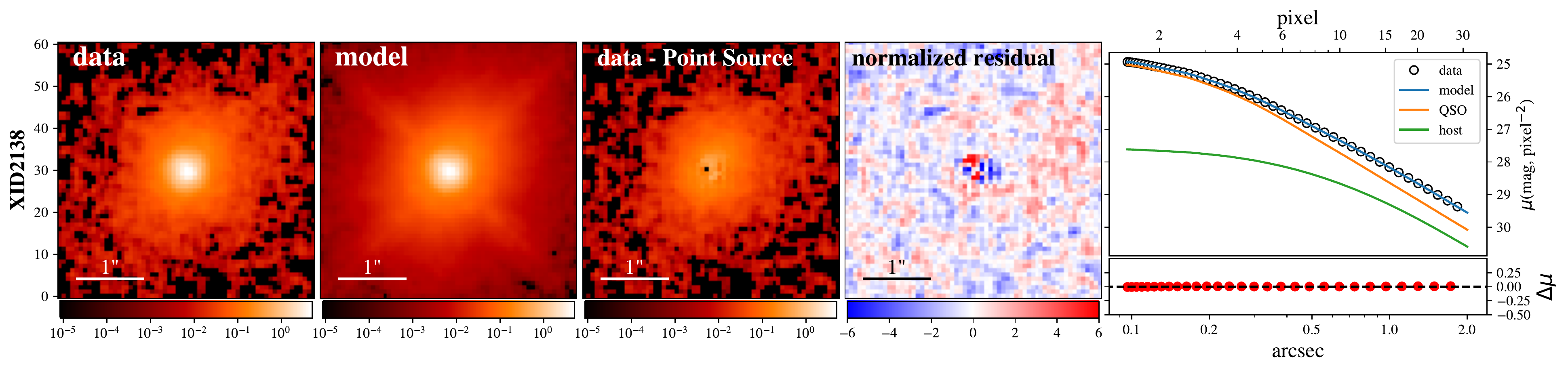}
\includegraphics[height=0.25\textwidth]{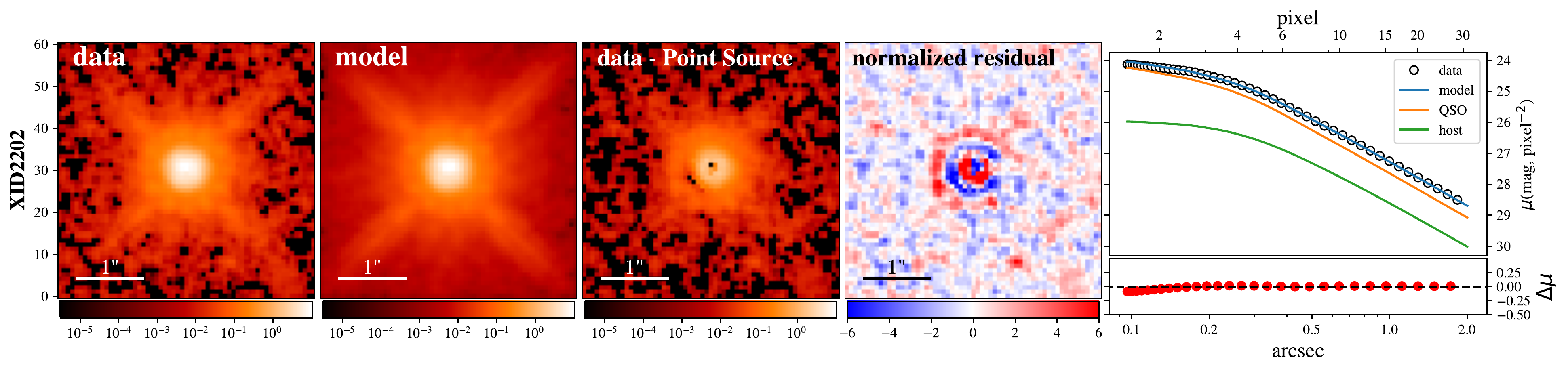}
\includegraphics[height=0.25\textwidth]{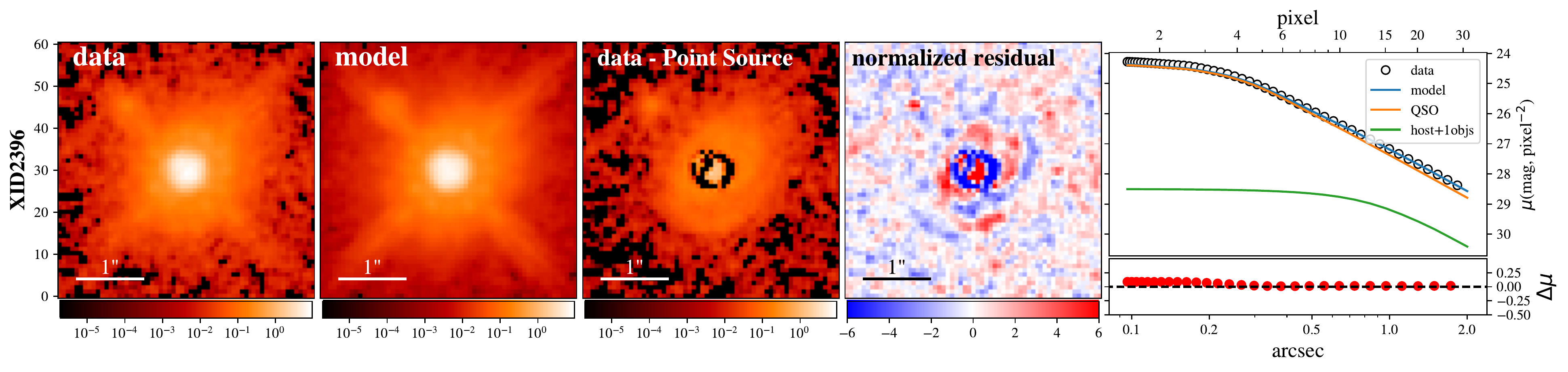}
\includegraphics[height=0.25\textwidth]{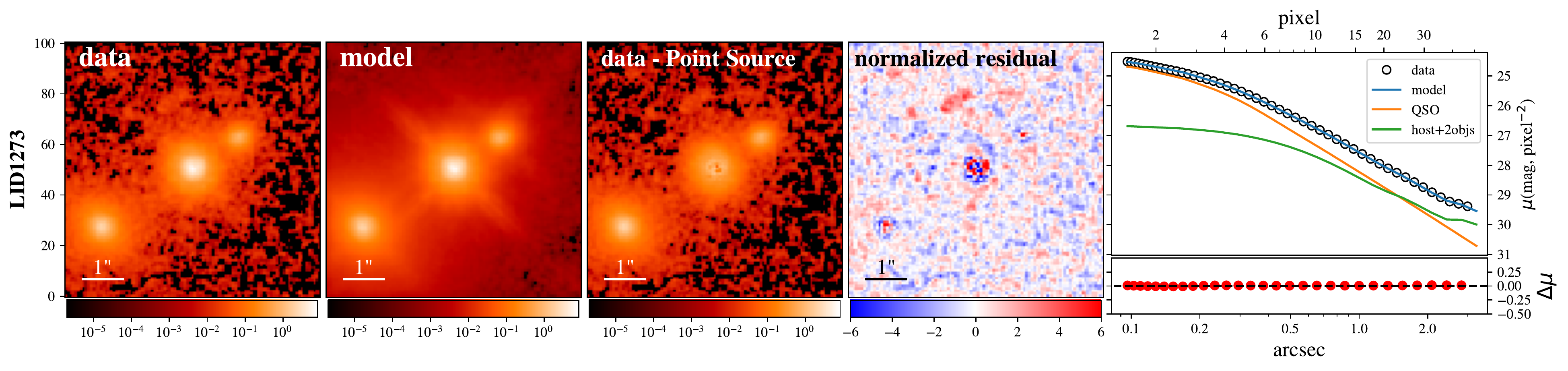}
\includegraphics[height=0.25\textwidth]{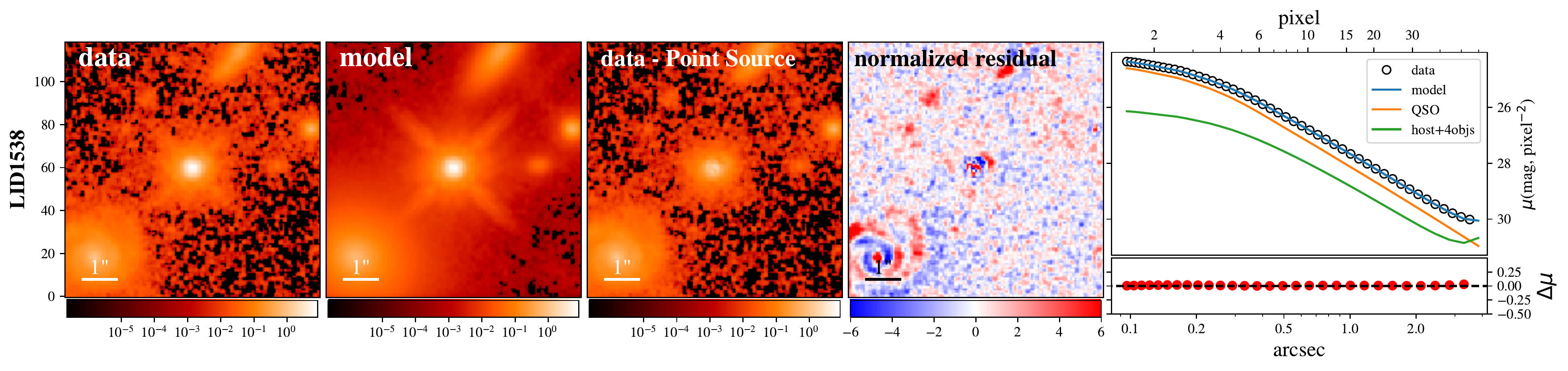}
}
\end{figure*} 

\begin{figure*}
\centering
{
\includegraphics[height=0.25\textwidth]{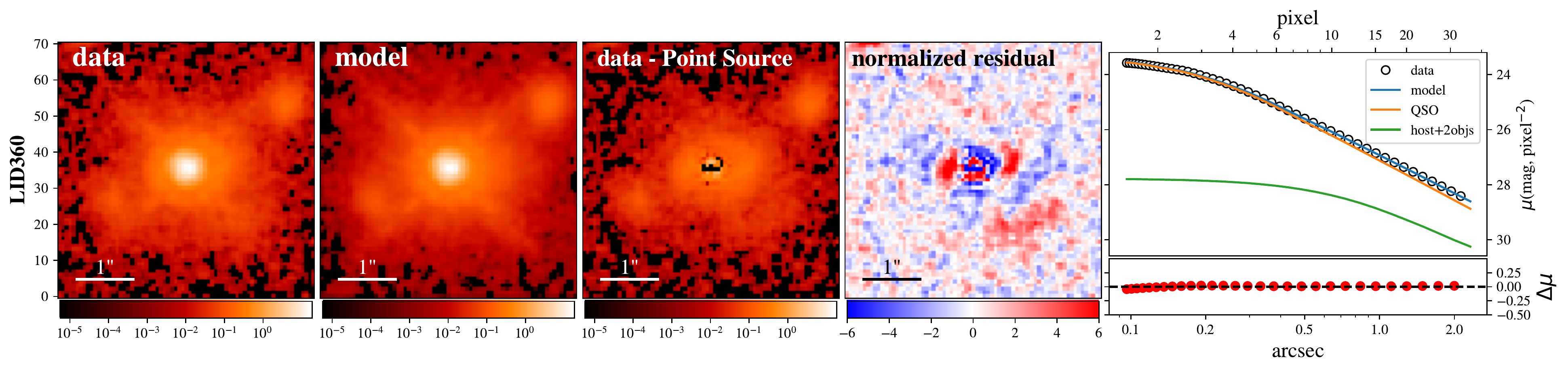}
\includegraphics[height=0.25\textwidth]{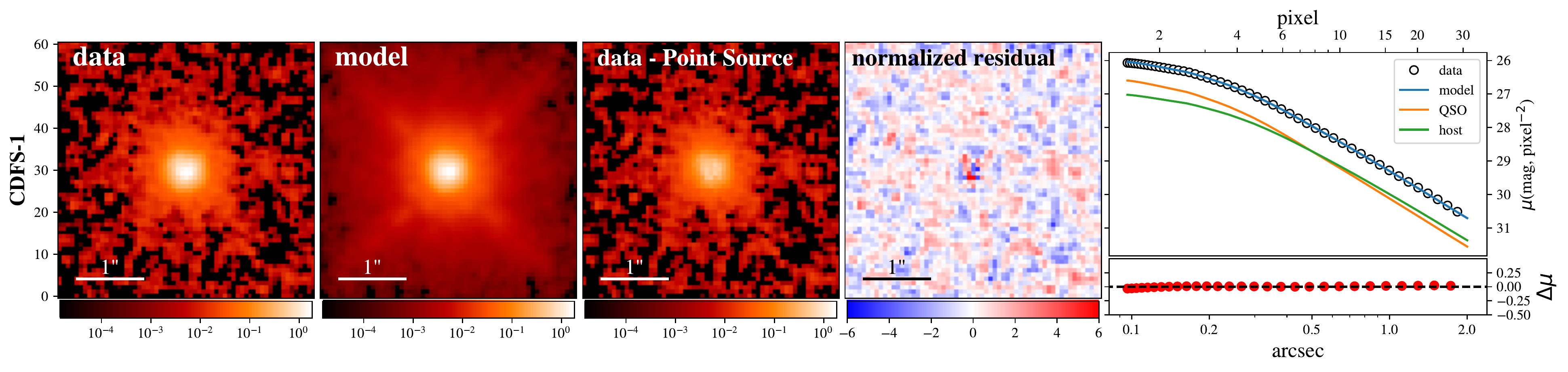}
\includegraphics[height=0.25\textwidth]{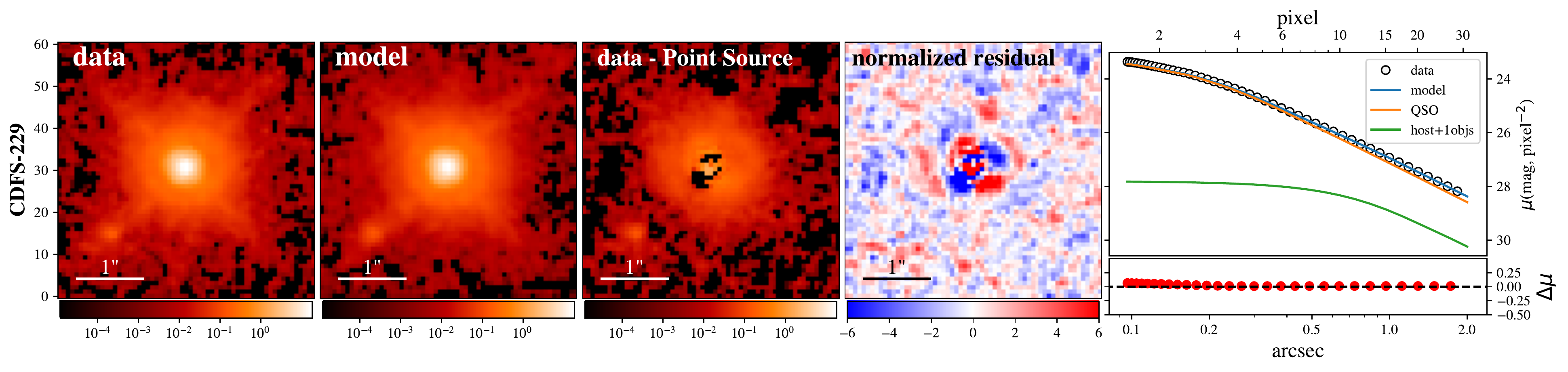}
\includegraphics[height=0.25\textwidth]{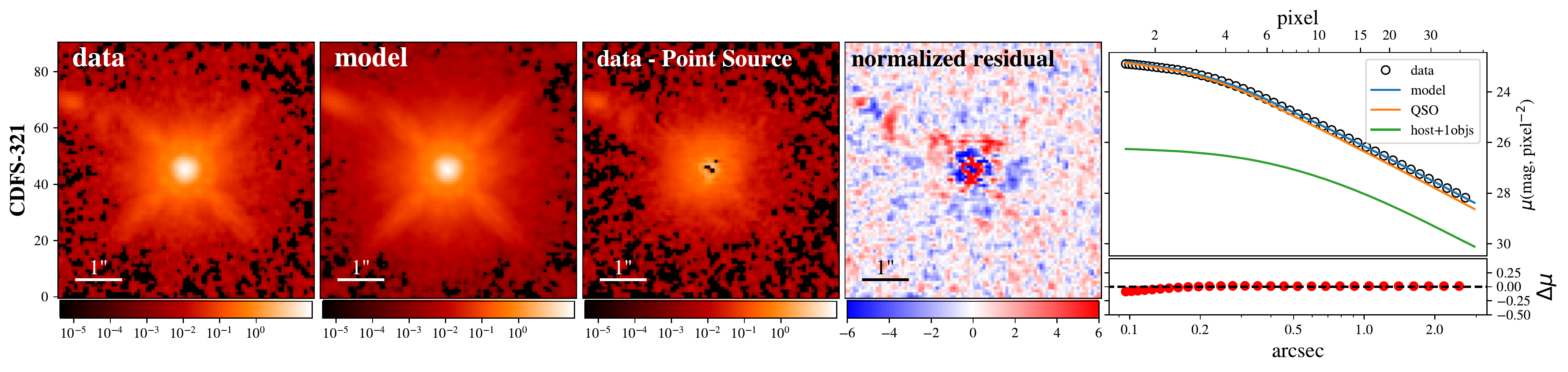}
\includegraphics[height=0.25\textwidth]{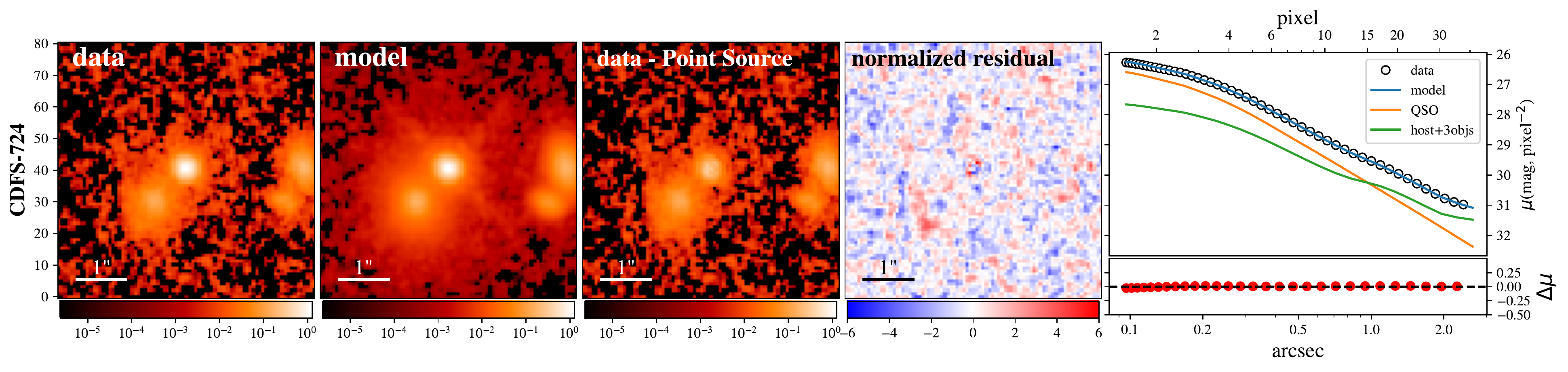}
}
\end{figure*} 

\begin{figure*}
\centering
{
\includegraphics[height=0.25\textwidth]{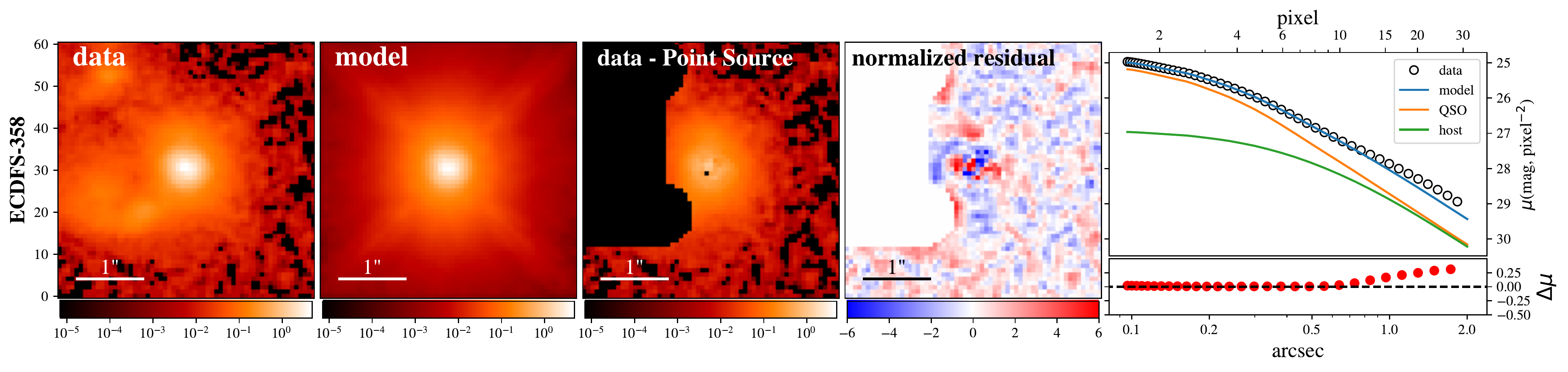}
\includegraphics[height=0.25\textwidth]{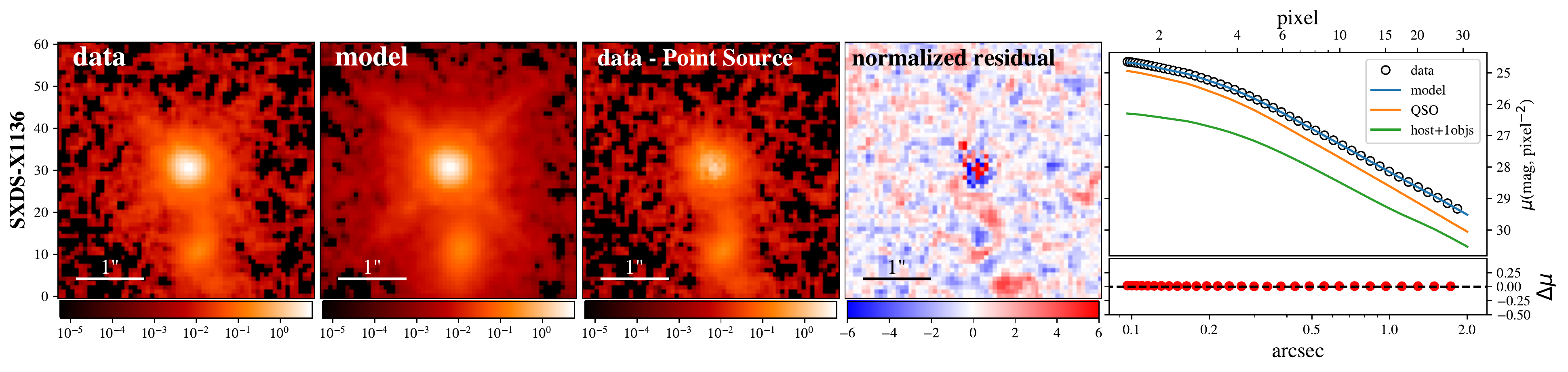}
\includegraphics[height=0.25\textwidth]{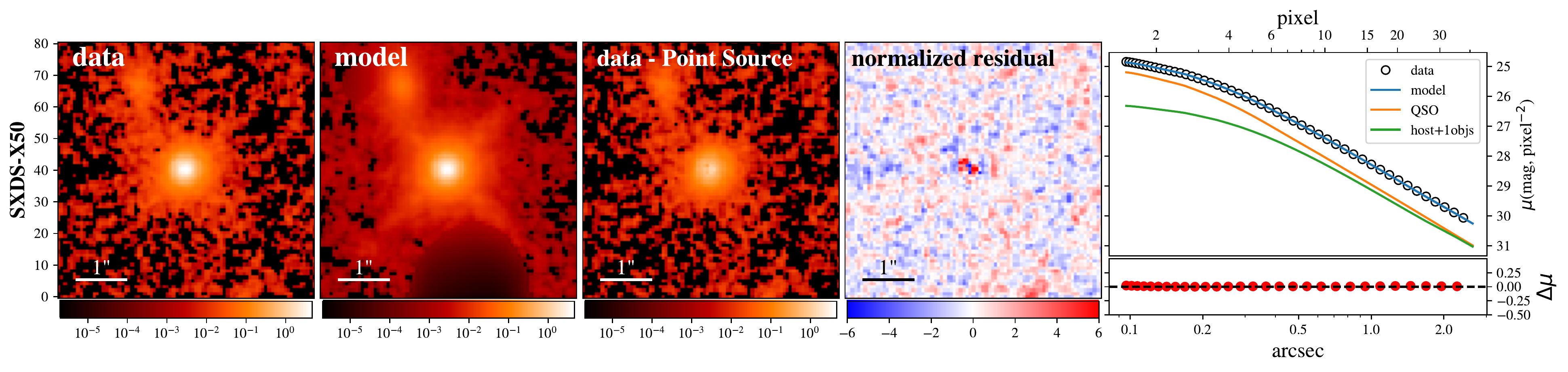}
\includegraphics[height=0.25\textwidth]{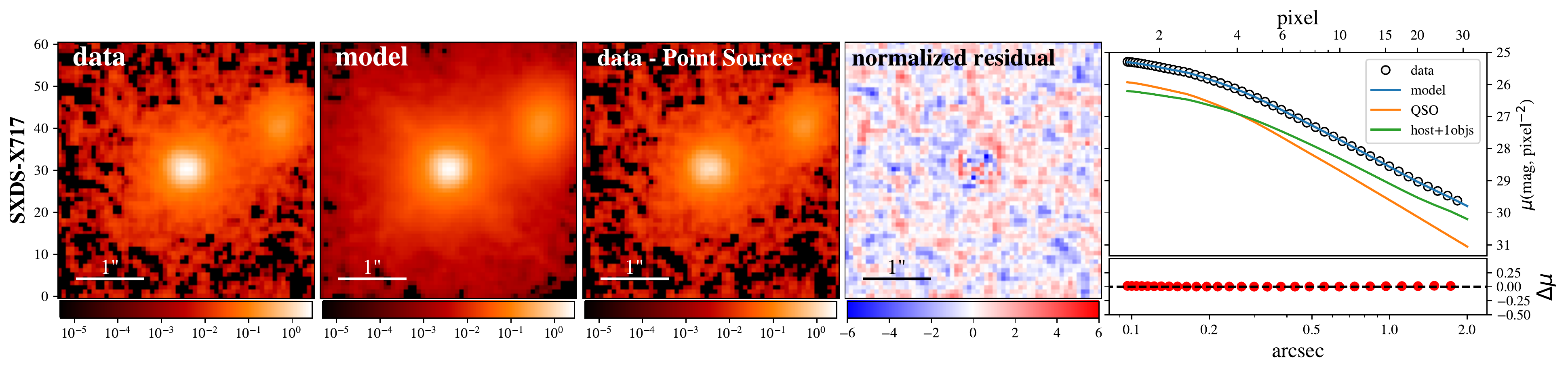}
\includegraphics[height=0.25\textwidth]{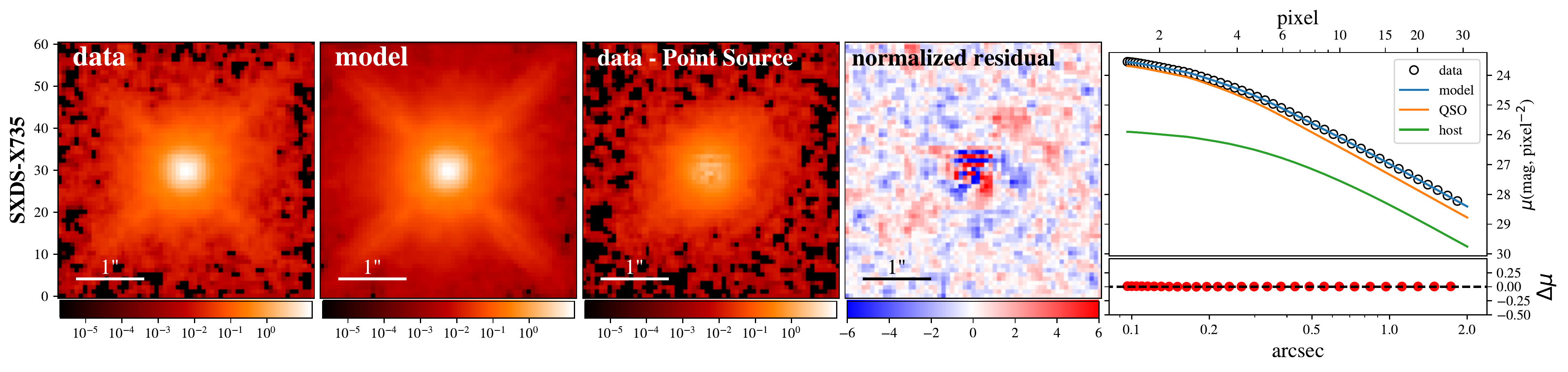}
}
\end{figure*} 

\begin{figure*}
\centering
{
\includegraphics[height=0.25\textwidth]{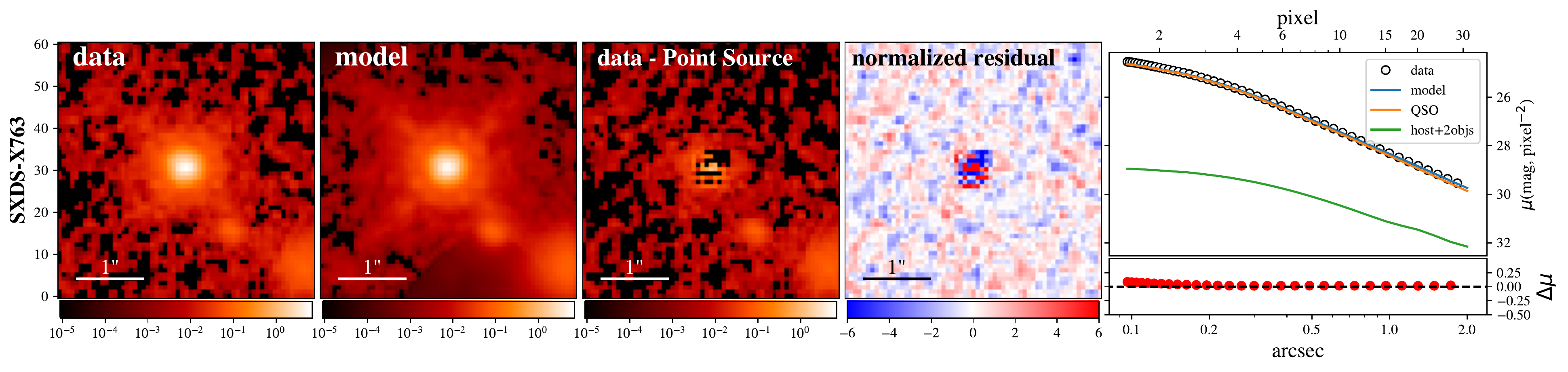}
\includegraphics[height=0.25\textwidth]{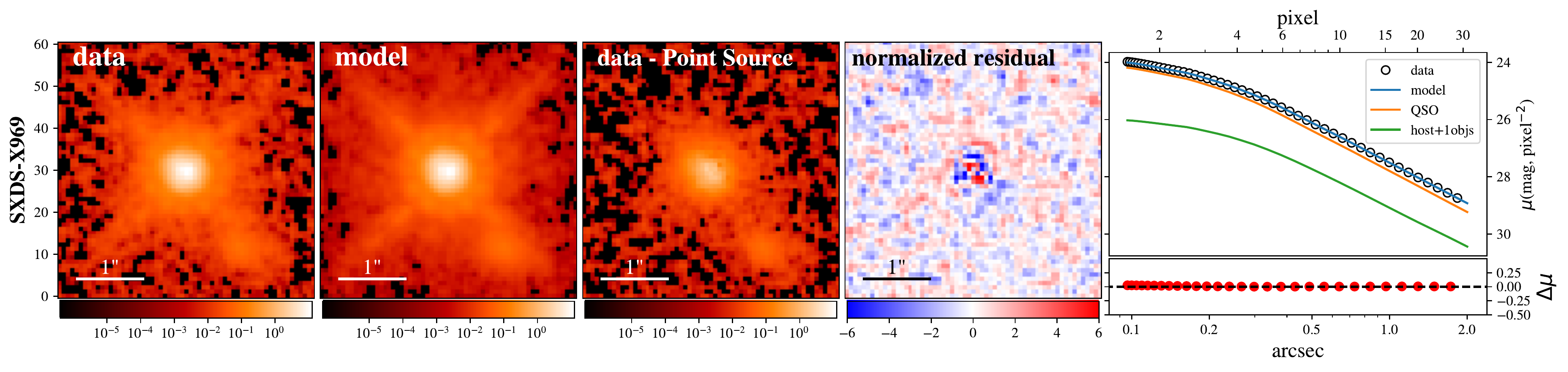}
}
\end{figure*} 

\clearpage
\section{B. AGN host galaxy luminosity with a correction for passive evolution}\label{sec:ml-ev}
In the passive evolution scenario, we expect the galaxy luminosity to fade over time. Thus, we transfer the \lhost\ for distant samples at today so as to compare the \mbh-\lhost\ relation to the local in the equivalent frame.
We consider this scenario following \citet{Ding2017b} by parametrizing the luminosity evolution with the functional form as
$d{\rm mag}_{\rm R}\sim~d\log(1+z)$, i.e.,
\begin{eqnarray}
\label{eq:L_relation}
\log(L_{R,0})=\log(L_{R}) - 1.48 \log (1+z).
\end{eqnarray} 
This formalism is more accurate to fit a broad range redshift comparing to a single slope as $d$mag$/dz$. We refer the interested readers to \citet[][section 5.4]{Ding2017b} for more details.

\begin{figure*}[ht]
\centering
\begin{tabular}{c c}
\subfloat[\mbh-\lhost\ relation, evolution-corrected]
{\includegraphics[height=0.45\textwidth]{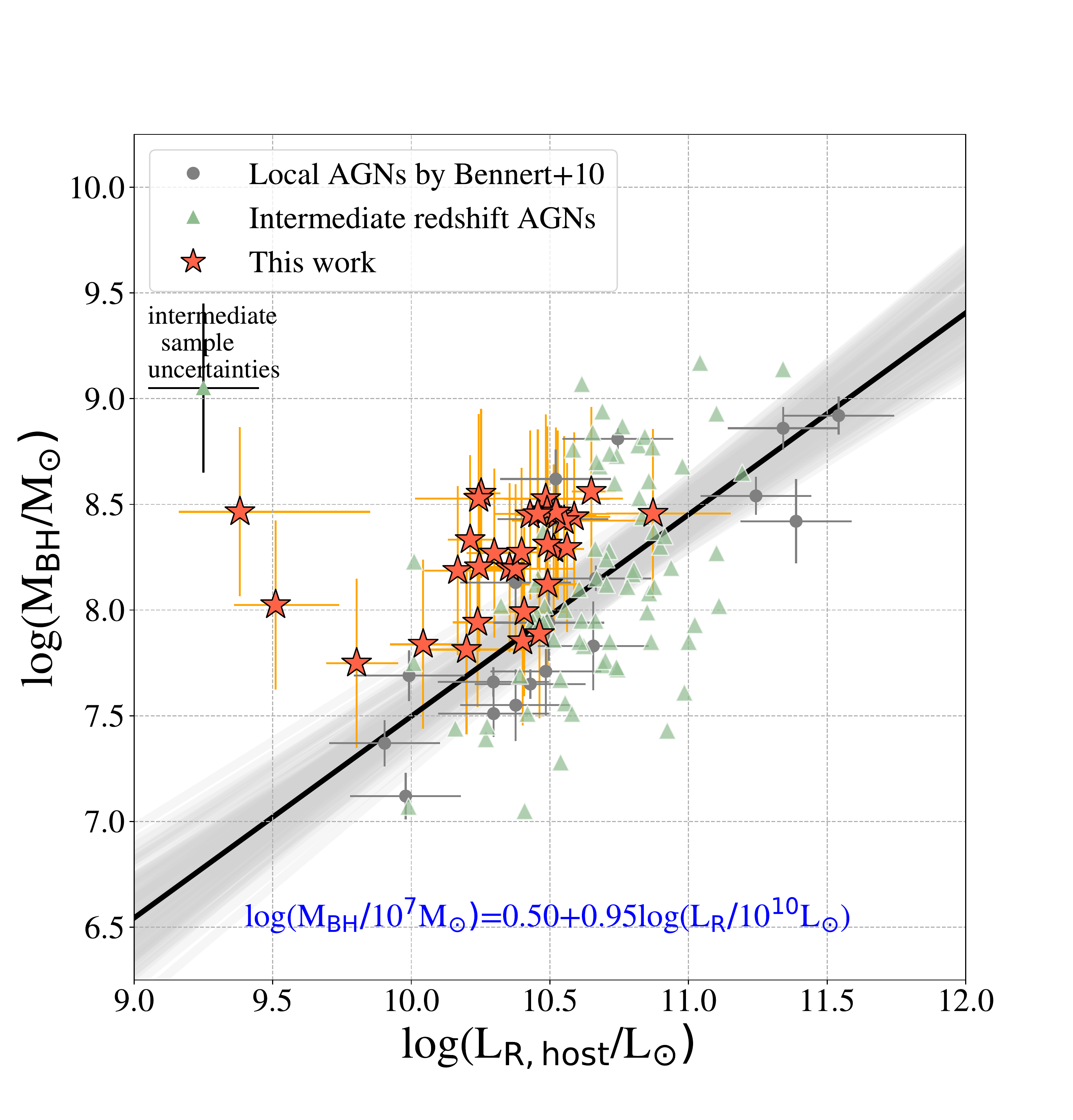}}&
\subfloat[offset in  $\log($\mbh$)$ (VS. \lhost) as a function of redshift]
{\includegraphics[height=0.45\textwidth]{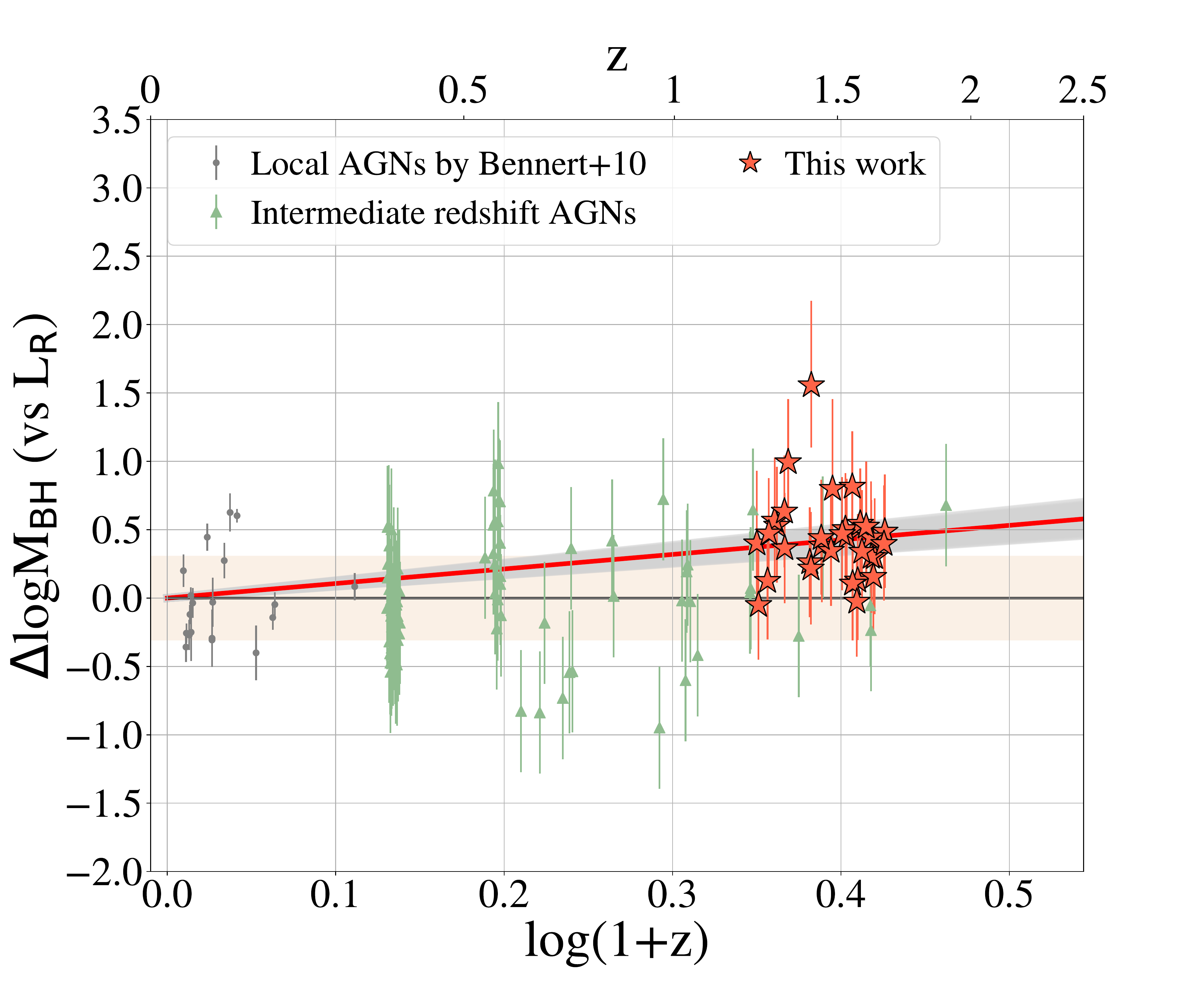}}\\
\subfloat[\mbh-\lhost, flat prior]
{\includegraphics[width=0.5\textwidth]{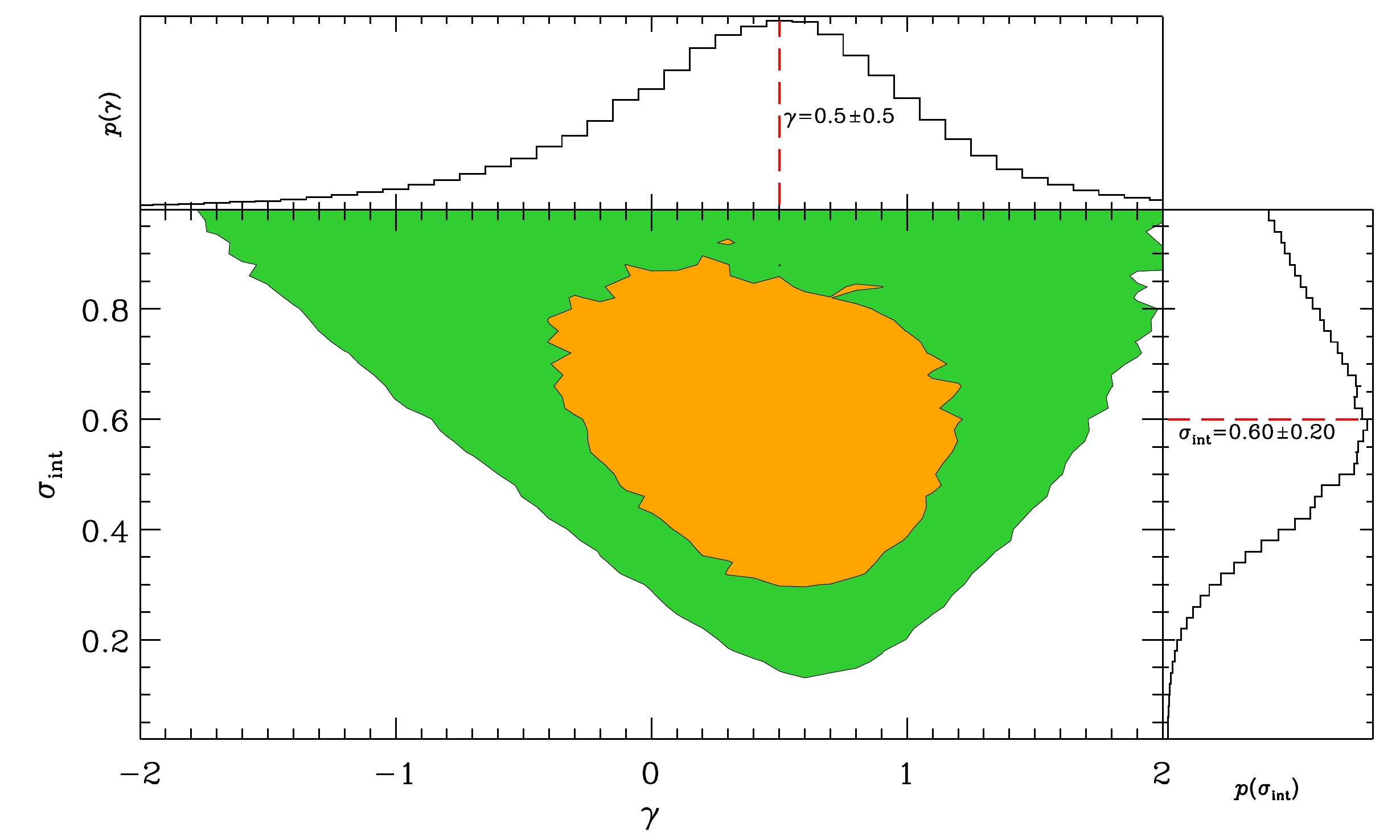}}&
\subfloat[\mbh-\lhost, lognormal prior]
{\includegraphics[width=0.5\textwidth]{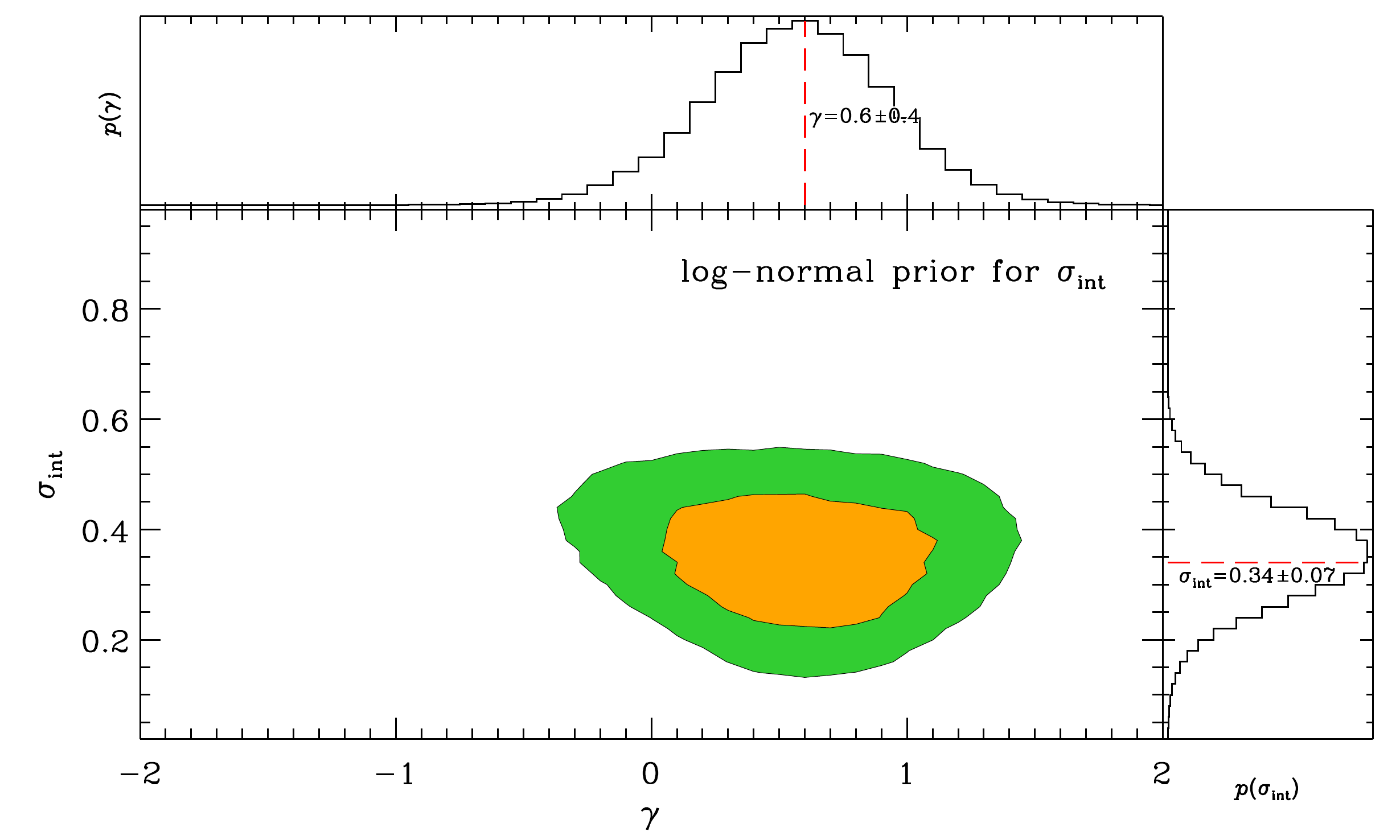}}\\
\end{tabular}
\caption{\label{fig:ML-vz} 
Same as previous figures but for \mbh-\lhost\ relation, considering the passive evolution correction for host galaxy luminosity.}
\end{figure*}

Having transferred the \lhost\ to today, we find that, as showing in Figure~\ref{fig:ML-vz}-(a), at fixed mass, the BH in the more distant universe tends to reside in less luminous hosts, which is consistent to the \mbh-\smass\ relation. Based on the 32 AGN systems, we fit the offset as a function of redshift in form as Equation~\ref{eq:offset} and obtain $\gamma = 1.07\pm0.23$, as showing in Figure~\ref{fig:ML-vz}-(b). We further consider the selection effect using the previous approach as introduced in Section~\ref{select_eff}, and obtain $\gamma = 0.5\pm0.5$ and $\gamma = 0.6\pm0.4$, with flat and lognormal prior, respectively, as showing in Figure~\ref{fig:ML-vz}-(c), (d).

The inferred $\gamma$ here could have large systematics given the following limitations. First, the passive evolution is based on a simplified correction; after all, we are not clear exactly how the host evolves to $z=0$. Moreover, we only consider the evolution of the host galaxy and assume the \mbh\ does not change much. 

\section{C. The \mbh-\smass\ comparison sample}\label{sec:comp_sample_value}
We use the self-consistent recipes to re-calibrate the measurements of our comparison sample. We list the values of all the comparison \mbh-\smass\ sample in the Table~\ref{tab:comp_sample}.

\begin{deluxetable*}{llcccc}
\tablecolumns{7}
\tablewidth{0pt}
\tablecaption{Details of Observation} 
\tablehead{ 
\colhead{Object ID} &
\colhead{$z$} & 
\colhead{WFC3/Filter} &
\colhead{$RA$}&
\colhead{$DEC$}&
\colhead{Observing date}
\\ 
\colhead{(1)} &
\colhead{(2)} &
\colhead{(3)} &
\colhead{(4)} &
\colhead{(5)} &
\colhead{(6)} 
} 
\startdata
COSMOS-CID1174 & 1.552 & F140W & 150.2789 & 1.9595 & 2017-10-26 \\
COSMOS-CID1281 & 1.445 & F140W & 150.4160 & 2.5258 & 2018-11-26 \\
COSMOS-CID206 & 1.483 & F140W & 149.8371 & 2.0088 & 2017-10-23 \\
COSMOS-CID216 & 1.567 & F140W & 149.7918 & 1.8729 & 2017-10-23 \\
COSMOS-CID237 & 1.618 & F140W & 149.9916 & 1.7243 & 2018-06-03 \\
COSMOS-CID255\footnoteref{note1} & 1.664 & F140W & 150.1017 & 1.8483 & 2019-03-16 \\
COSMOS-CID3242 & 1.532 & F140W & 149.7113 & 2.1452 & 2017-10-26 \\
COSMOS-CID3570 & 1.244 & F125W & 149.6411 & 2.1076 & 2017-10-27 \\
COSMOS-CID452 & 1.407 & F125W & 150.0045 & 2.2371 & 2017-10-25 \\
COSMOS-CID454 & 1.478 & F140W & 149.8681 & 2.3307 & 2018-02-26 \\
COSMOS-CID50 & 1.239 & F125W & 150.2080 & 2.0833 & 2017-10-23 \\
COSMOS-CID543 & 1.301 & F125W & 150.4519 & 2.1448 & 2018-04-30 \\
COSMOS-CID597 & 1.272 & F125W & 150.5262 & 2.2449 & 2018-11-25 \\
COSMOS-CID607 & 1.294 & F125W & 150.6097 & 2.3231 & 2017-10-25 \\
COSMOS-CID70 & 1.667 & F140W & 150.4051 & 2.2701 & 2017-10-27 \\
COSMOS-LID1273 & 1.617 & F140W & 150.0565 & 1.6275 & 2017-10-31 \\
COSMOS-LID1538 & 1.527 & F140W & 150.6215 & 2.1588 & 2018-05-01 \\
COSMOS-LID360 & 1.579 & F140W & 150.1251 & 2.8617 & 2017-10-30 \\
COSMOS-XID2138 & 1.551 & F140W & 149.7036 & 2.5781 & 2017-11-01 \\
COSMOS-XID2202 & 1.516 & F140W & 150.6530 & 1.9969 & 2017-11-05 \\
COSMOS-XID2396 & 1.600 & F140W & 149.4779 & 2.6425 & 2017-11-12\\
CDFS-1 & 1.630 & F140W & 52.8990 & -27.8600 & 2018-04-03 \\
CDFS-229 & 1.326 & F125W & 53.0680 & -27.6580 & 2018-04-04 \\ 
CDFS-321 & 1.570 & F140W & 53.0486 & -27.6239 & 2018-08-18\\ 
CDFS-724 & 1.337 & F125W & 53.2870 & -27.6940 & 2018-04-05 \\ 
ECDFS-358 & 1.626 & F140W & 53.0850 & -28.0370 & 2018-02-09\\ 
SXDS-X1136 & 1.325 & F125W & 34.8925 & -5.1498 & 2018-01-29\\ 
SXDS-X50 & 1.411 & F125W & 34.0267 & -5.0602 & 2018-03-01\\ 
SXDS-X717 & 1.276 & F125W & 34.5400 & -5.0334 & 2018-07-02 \\ 
SXDS-X735 & 1.447 & F140W & 34.5581 & -4.8781 & 2017-11-14 \\ 
SXDS-X763 & 1.412 & F125W & 34.5849 & -4.7864 & 2018-07-03 \\ 
SXDS-X969 & 1.585 & F140W & 34.7594 & -5.4291 & 2018-07-02 \\ 
\enddata
\label{tab:objlist}
\tablecomments{
Column 1: Object field and ID.
Column 2: Spectroscopic redshift.
Column 3: WFC3 filter. Note that the targets from the COSMOS field also have ACS imaging.
Column 4 and 5: J2000 $RA$ and $DEC$ coordinates.
Column 6: The observing start date.
The total exposure time of each target is 2394s.
}
\end{deluxetable*}

\begin{deluxetable*}
{@{\extracolsep{0pt}}lcccccccc}   
\tablecolumns{9}
\tablewidth{0pt}
\tablecaption{AGN properties} 
\tablehead
{ 
\colhead{Target ID}&
  \multicolumn{4}{c}{\halpha ~emission line}&
  \multicolumn{4}{c}{\hbeta ~emission line} \\
  \cline{2-5}  \cline{6-9} \\
\colhead{}& 
\colhead{FWHM(\halpha)}& \colhead{$\log( \rm L _{H\alpha}$)}& \colhead{$\log$\mbh}& Eddington ratio &
\colhead{FWHM(\hbeta)}& \colhead{$\log( \rm L _{\lambda5100})$}& \colhead{$\log$\mbh}& Eddington ratio \\
\colhead{}& 
\colhead{(\kms)}& \colhead{(${\rm erg~s^{-1}}$)}& 
\colhead{(M$_{\odot}$)}& (log$( L_{\rm Bol}/L_{\rm Edd})$)&
\colhead{(\kms)}& 
\colhead{(${\rm erg~s^{-1}}$)}&\colhead{(M$_{\odot}$)} & (log$(L_{\rm Bol}/L_{\rm Edd})$)\\
\colhead{(1)}& 
\colhead{(2)}& \colhead{(3)}& 
\colhead{(4)}& \colhead{(5)}& 
\colhead{(6)}&\colhead{(7)}&
\colhead{(8)}& \colhead{(9)}
}
\startdata 
CID1174 & 1906 & 43.43 & 7.99 & -0.47 & 5898 & 44.76 & 8.83& -1.34 \\
CID1281 & 1619 & 43.24 & 7.75 & -0.41 & \nodata & \nodata & \nodata & \nodata \\
CID206 & 3334 & 43.48 & 8.53 & -0.97 & \nodata & \nodata & \nodata& \nodata \\
CID216 & 2230 & 42.85 & 7.85 & -0.90 & \nodata & \nodata & \nodata& \nodata \\
CID237 & 2112 & 43.86 & 8.29 & -0.36 & \nodata & \nodata & \nodata& \nodata \\
CID255 & 1932 & 43.99 & 8.27 & -0.22 & 3709 & 45.37 & 8.73& -0.60 \\
CID3242 & 2543 & 43.83 & 8.45 & -0.55 & 3775 & 45.10 & 8.61& -0.75 \\
CID3570 & 1959 & 43.16 & 7.89 & -0.63 & \nodata & \nodata & \nodata& \nodata \\
CID452 & 3458 & 42.92 & 8.30 & -1.26 & 3127 & 44.63 & 8.22& -0.88 \\
CID454 & 2824 & 43.34 & 8.31 & -0.88 & \nodata & \nodata & \nodata& \nodata \\
CID50 & 2340 & 43.94 & 8.42 & -0.42 & 1939 & 45.33 & 8.15& -0.06 \\
CID543 & 2189 & 43.57 & 8.19 & -0.53 & \nodata & \nodata & \nodata& \nodata \\
CID597 & 1656 & 43.33 & 7.81 & -0.39 & \nodata & \nodata & \nodata& \nodata \\
CID607 & 3009 & 43.67 & 8.53 & -0.78 & 4242 & 44.78 & 8.56& -1.04 \\
CID70 & 2480 & 43.51 & 8.27 & -0.68 & 3982 & 45.16 & 8.69& -0.77 \\
LID1273 & 3224 & 43.61 & 8.56 & -0.87 & \nodata & \nodata & \nodata& \nodata \\
LID1538 & 2941 & 43.60 & 8.47 & -0.79 & \nodata & \nodata & \nodata& \nodata \\
LID360 & 2482 & 43.88 & 8.45 & -0.50 & 2869 & 45.09 & 8.37& -0.52 \\
XID2138 & 3186 & 43.61 & 8.55 & -0.86 & 2945 & 44.81 & 8.25& -0.71 \\
XID2202 & 2973 & 43.56 & 8.46 & -0.82 & \nodata & \nodata & \nodata& \nodata \\
XID2396 & 2271 & 44.06 & 8.46 & -0.33 & 2658 & 45.50 & 8.51& -0.24 \\
CDFS-1 & 2000 & 43.02 & 7.83 & -0.03 & \nodata & \nodata & \nodata& \nodata \\
CDFS-229 & 2190 & 43.60 & 8.20 & -0.60 & \nodata & \nodata & \nodata& \nodata \\
CDFS-321 & 2442 & 43.93 & 8.46 & -0.46 & \nodata & \nodata & \nodata& \nodata \\
CDFS-724 & 2541 & 42.95 & 8.03 & -1.15 & \nodata & \nodata & \nodata& \nodata \\
ECDFS-358 & 2237 & 43.40 & 8.12 & -0.64 & \nodata & \nodata & \nodata& \nodata \\
SXDS-X1136 & 2760 & 43.43 & 8.33 & -0.81 & 6761 & 44.71 & 8.93& -1.49 \\
SXDS-X50 & 1817 & 43.42 & 7.94 & -0.43 & \nodata & \nodata & \nodata& \nodata \\
SXDS-X717 & 2931 & 43.05 & 8.20 & -1.05 & \nodata & \nodata & \nodata& \nodata \\
SXDS-X735 & 2702 & 43.70 & 8.44 & -0.67 & 3520 & 45.07 & 8.54& -0.70 \\
SXDS-X763 & 2961 & 43.57 & 8.47 & -0.81 & 4509 & 44.51 & 8.47& -1.29 \\
SXDS-X969 & 2296 & 43.50 & 8.20 & -0.61 & 1696 & 45.05 & 7.90& -0.08 \\
\enddata
\label{tab:result_mbh}
\tablecomments{
Column 1: Object ID.
Column 2-5: \halpha\ emission line width (FWHM), \halpha\ luminosity, inferred \mbh\ and Eddington ratio. 
Column 6-9: \hbeta\ emission line BH properties. 
The typical uncertainty level for FWHM is $15\%$, and for $\log( \rm L _{\lambda})$ is in the range $0.01\sim0.2$~dex, respectively. The inferred uncertainty level for \mbh\ are assumed as 0.4 dex.
}
\end{deluxetable*}

\begin{deluxetable*}{ccccccc}
\tablecolumns{7}
\tablewidth{0pt}
\tablecaption{Host galaxy measurements of CID1174} 
\tablehead{ 
\colhead{PSF rank} &
\colhead{total $\chi ^2$} & 
\colhead{weights $w_i$} &
\colhead{host flux (counts)} &
\colhead{host flux ratio}&
\colhead{\Reff (arcsec)}&
\colhead{\sersic\ $n$}
 \\ 
\colhead{(1)} &
\colhead{(2)} &
\colhead{(3)} &
\colhead{(4)} &
\colhead{(5)} &
\colhead{(6)} &
\colhead{(7)} 
} 
\startdata
1 & $8584.429$ & $1.000$ & $82.2$ & $35\%$ & $0\farcs{}345$ & $1.1$ \\
2 & $8646.711$ & $0.920$ & $99.1$ & $42\%$ & $0\farcs{}298$ & $1.9$ \\
3 & $8816.947$ & $0.734$ & $76.7$ & $33\%$ & $0\farcs{}365$ & $1.1$ \\
4 & $9304.841$ & $0.383$ & $128.6$ & $55\%$ & $0\farcs{}231$ & $2.8$ \\
5 & $9652.575$ & $0.241$ & $187.5$ & $79\%$ & $0\farcs{}116$ & $6.2$ \\
6 & $9917.101$ & $0.170$ & $100.2$ & $42\%$ & $0\farcs{}287$ & $2.1$ \\
7 & $10018.324$ & $0.148$ & $75.1$ & $32\%$ & $0\farcs{}365$ & $1.2$ \\
8 & $10087.456$ & $0.135$ & $79.8$ & $34\%$ & $0\farcs{}358$ & $1.2$ \\
\hline\\
Weighted value & & & $97.322\pm28.336$ & $42\%\pm12\%$& $0\farcs{}309\pm0\farcs{}065$  & $1.9\pm1.3$  \\
\enddata
\label{tab:weight_CID1174}
\tablecomments{
Column~1: Rank of the PSF from the library.
Column~2: Total $\chi ^2$ for the corresponding PSF.
Column~3: Weights for the inference.
Column~4-7: Fitted value for the host flux, host/total flux ratio, effective radius, and \sersic\ index.
For this sample, the inflation parameter $\alpha$ calculated by Equation~\ref{eq:alpha} is 16.671.
}
\end{deluxetable*}

\tabcolsep=0.03cm
\begin{deluxetable*}
{@{\extracolsep{2pt}}lccccccccccc}   
\tablecolumns{10}
\tablewidth{0pt}
\tablecaption{Host galaxy properties} 
\tablehead
{ 
\colhead{Target ID}&
  \multicolumn{5}{c}{WFC3}&
  \multicolumn{3}{c}{ACS/F814W} &
   \multicolumn{2}{c}{Derived properties} \\
  \cline{2-6}  \cline{7-9} \cline{10-12}  \\
\colhead{}& 
\colhead{$\chi ^2$}& \colhead{host-total flux ratio}& 
\colhead{\Reff}& \colhead{\sersic\ $n$}& 
\colhead{magnitude}&
\colhead{$\chi ^2$}& \colhead{host-total flux ratio}& \colhead{magnitude} &
\colhead{$\log L_R$} &\colhead{$\log M_*$} &\colhead{$\log M_{*, bulge}$} \\
\colhead{}& 
\colhead{(reduced)}& \colhead{}& 
\colhead{($\arcsec$)}& \colhead{}& 
\colhead{(AB system)}& \colhead{(reduced)}& 
\colhead{}& \colhead{(AB system)} &\colhead{($L_{\odot,R}$)} & \colhead{(M$_{\odot}$)} &  \colhead{(M$_{\odot}$)}\\
\colhead{(1)}& 
\colhead{(2)}& \colhead{(3)}& 
\colhead{(4)}& \colhead{(5)}& 
\colhead{(6)}& \colhead{(7)}& 
\colhead{(8)}& \colhead{(9)}&
\colhead{(10)}& \colhead{(11)} &  \colhead{(12)}
}
\setlength{\tabcolsep}{20pt}
\renewcommand{\arraystretch}{1.5}
\startdata 
CID1174 & $2.307$ & $42\%\pm12\%$ & $0\farcs{}31\pm0\farcs{}07$ & $1.9\pm1.3$ & $21.48\substack{+0.37\\-0.28}$ & $2.496$ & $11\%\pm1\%$ & $23.21\substack{+0.11\\-0.10}$ & $11.01\substack{+0.15\\-0.11}$ &$10.63\substack{+0.18\\-0.15}$ & $10.21\substack{+0.40\\-0.53}$ \\[3pt]
CID1281 & $1.322$ & $49\%\pm14\%$ & $0\farcs{}24\pm0\farcs{}09$ & $3.2\pm1.5$ & $22.88\substack{+0.36\\-0.27}$ & $1.378$ & $19\%\pm8\%$ & $24.83\substack{+0.60\\-0.38}$ & $10.38\substack{+0.15\\-0.11}$ &$10.00\substack{+0.18\\-0.15}$ & $9.77\substack{+0.28\\-0.51}$ \\[3pt]
CID206 & $2.054$ & $35\%\pm24\%$ & $0\farcs{}29\pm0\farcs{}15$ & $3.1\pm2.5$ & $21.82\substack{+1.30\\-0.58}$ & $1.903$ & $8\%\pm2\%$ & $23.67\substack{+0.40\\-0.29}$ & $10.83\substack{+0.52\\-0.23}$ &$10.45\substack{+0.53\\-0.25}$ & $10.14\substack{+0.52\\-0.62}$ \\[3pt]
CID216 & $1.514$ & $94\%\pm5\%$ & $0\farcs{}25\pm0\farcs{}06$ & $6.2\pm1.2$ & $21.51\substack{+0.05\\-0.05}$ & $1.425$ & $35\%\pm2\%$ & $23.45\substack{+0.05\\-0.05}$ & $11.01\substack{+0.03\\-0.03}$ &$10.63\substack{+0.10\\-0.10}$ & $10.52\substack{+0.15\\-0.25}$ \\[3pt]
CID237 & $2.349$ & $30\%\pm6\%$ & $0\farcs{}87\pm0\farcs{}17$ & $4.7\pm1.7$ & $21.28\substack{+0.26\\-0.21}$ & $2.354$ & $3\%\pm2\%$ & $23.72\substack{+1.04\\-0.52}$ & $11.13\substack{+0.11\\-0.09}$ &$10.75\substack{+0.14\\-0.13}$ & $10.60\substack{+0.20\\-0.34}$ \\[3pt]
CID255 & $1.625$ & $19\%\pm5\%$ & $0\farcs{}19\pm0\farcs{}06$ & $4.2\pm1.5$ & $21.61\substack{+0.37\\-0.28}$ & 2.858 & $4\%\pm2\%$ & $22.89\substack{+0.60\\-0.39} $& $11.03\substack{+0.15\\-0.11}$ &$10.65\substack{+0.18\\-0.15}$ & $10.48\substack{+0.23\\-0.35}$ \\[3pt]
CID3242 & $2.751$ & $46\%\pm13\%$ & $0\farcs{}20\pm0\farcs{}16$ & $6.1\pm1.9$ & $21.16\substack{+0.35\\-0.26}$ & $2.596$ & $5\%\pm1\%$ & $23.60\substack{+0.34\\-0.26}$ & $11.12\substack{+0.14\\-0.11}$ &$10.75\substack{+0.17\\-0.15}$ & $10.62\substack{+0.19\\-0.29}$ \\[3pt]
CID3570 & $1.665$ & $77\%\pm2\%$ & $0\farcs{}70\pm0\farcs{}01$ & $0.7\pm0.1$ & $21.16\substack{+0.02\\-0.02}$ & $1.332$ & $86\%\pm2\%$ & $22.97\substack{+0.01\\-0.01}$ & $10.98\substack{+0.02\\-0.02}$ &$10.71\substack{+0.10\\-0.10}$ & $9.78\substack{+0.68\\-0.13}$ \\[3pt]
CID452 & $1.684$ & $75\%\pm4\%$ & $0\farcs{}37\pm0\farcs{}02$ & $1.4\pm0.2$ & $21.18\substack{+0.06\\-0.06}$ & $1.452$ & $38\%\pm1\%$ & $22.73\substack{+0.02\\-0.02}$ & $11.13\substack{+0.03\\-0.03}$ &$10.86\substack{+0.10\\-0.10}$ & $10.13\substack{+0.48\\-0.29}$ \\[3pt]
CID454 & $2.203$ & $36\%\pm3\%$ & $0\farcs{}39\pm0\farcs{}02$ & $0.6\pm0.1$ & $21.20\substack{+0.08\\-0.07}$ & $1.291$ & $9\%\pm1\%$ & $23.35\substack{+0.06\\-0.06}$ & $11.07\substack{+0.04\\-0.04}$ &$10.70\substack{+0.10\\-0.10}$ & $9.77\substack{+0.74\\-0.15}$ \\[3pt]
CID50 & $5.576$ & $17\%\pm9\%$ & $0\farcs{}16\pm0\farcs{}11$ & $3.2\pm2.2$ & $20.93\substack{+0.86\\-0.48}$ & $4.940$ & $5\%\pm3\%$ & $22.50\substack{+1.15\\-0.55}$ & $11.07\substack{+0.35\\-0.19}$ &$10.80\substack{+0.36\\-0.21}$ & $10.51\substack{+0.41\\-0.56}$ \\[3pt]
CID543 & $1.902$ & $31\%\pm10\%$ & $0\farcs{}10\pm0\farcs{}00$ & $0.5\pm0.3$ & $21.99\substack{+0.41\\-0.30}$ & $1.435$ & $5\%\pm2\%$ & $23.77\substack{+0.53\\-0.36}$ & $10.70\substack{+0.16\\-0.12}$ &$10.43\substack{+0.19\\-0.15}$ & $9.55\substack{+0.64\\-0.24}$ \\[3pt]
CID597 & $1.565$ & $42\%\pm17\%$ & $0\farcs{}17\pm0\farcs{}06$ & $1.8\pm0.8$ & $21.87\substack{+0.54\\-0.36}$ & $1.254$ & $12\%\pm1\%$ & $23.56\substack{+0.13\\-0.11}$ & $10.73\substack{+0.22\\-0.15}$ &$10.46\substack{+0.24\\-0.18}$ & $9.98\substack{+0.44\\-0.49}$ \\[3pt]
CID607 & $1.692$ & $44\%\pm18\%$ & $0\farcs{}21\pm0\farcs{}09$ & $3.4\pm1.1$ & $21.19\substack{+0.58\\-0.37}$ & $2.590$ & $5\%\pm2\%$ & $23.57\substack{+0.51\\-0.35}$ & $11.02\substack{+0.23\\-0.15}$ &$10.75\substack{+0.25\\-0.18}$ & $10.57\substack{+0.29\\-0.41}$ \\[3pt]
CID70 & $2.041$ & $20\%\pm5\%$ & $0\farcs{}42\pm0\farcs{}10$ & $3.6\pm1.0$ & $21.86\substack{+0.30\\-0.24}$ & $2.361$ & $2\%\pm1\%$ & $24.63\substack{+0.68\\-0.41}$ & $10.93\substack{+0.12\\-0.10}$ &$10.55\substack{+0.16\\-0.14}$ & $10.38\substack{+0.22\\-0.34}$ \\[3pt]
LID1273 & $1.697$ & $53\%\pm9\%$ & $0\farcs{}30\pm0\farcs{}04$ & $1.2\pm0.5$ & $20.94\substack{+0.21\\-0.18}$ & $2.137$ & $6\%\pm1\%$ & $23.29\substack{+0.15\\-0.13}$ & $11.27\substack{+0.09\\-0.07}$ &$10.89\substack{+0.13\\-0.12}$ & $10.13\substack{+0.58\\-0.28}$ \\[3pt]
LID1538 & $2.362$ & $44\%\pm8\%$ & $0\farcs{}18\pm0\farcs{}04$ & $2.8\pm0.5$ & $21.25\substack{+0.22\\-0.18}$ & $2.173$ & $8\%\pm1\%$ & $23.09\substack{+0.19\\-0.16}$ & $11.09\substack{+0.09\\-0.08}$ &$10.71\substack{+0.13\\-0.12}$ & $10.52\substack{+0.22\\-0.37}$ \\[3pt]
LID360 & $3.918$ & $18\%\pm2\%$ & $0\farcs{}63\pm0\farcs{}02$ & $0.8\pm0.4$ & $21.46\substack{+0.14\\-0.12}$ & $4.914$ & $4\%\pm1\%$ & $23.25\substack{+0.17\\-0.15}$ & $11.04\substack{+0.06\\-0.05}$ &$10.66\substack{+0.11\\-0.11}$ & $9.76\substack{+0.65\\-0.16}$ \\[3pt]
XID2138 & $1.597$ & $39\%\pm6\%$ & $0\farcs{}50\pm0\farcs{}03$ & $1.2\pm0.4$ & $21.87\substack{+0.17\\-0.15}$ & $2.731$ & $5\%\pm1\%$ & $23.90\substack{+0.31\\-0.24}$ & $10.85\substack{+0.07\\-0.06}$ &$10.48\substack{+0.12\\-0.12}$ & $9.71\substack{+0.57\\-0.28}$ \\[3pt]
XID2202 & $3.23$ & $33\%\pm8\%$ & $0\farcs{}10\pm0\farcs{}00$ & $4.0\pm1.0$ & $21.16\substack{+0.30\\-0.24}$ & $3.852$ & $8\%\pm2\%$ & $22.59\substack{+0.29\\-0.23}$ & $11.11\substack{+0.12\\-0.10}$ &$10.74\substack{+0.16\\-0.14}$ & $10.58\substack{+0.21\\-0.33}$ \\[3pt]
XID2396 & $3.669$ & $24\%\pm11\%$ & $0\farcs{}58\pm0\farcs{}09$ & $0.8\pm1.4$ & $21.40\substack{+0.65\\-0.40}$ & $5.346$ & $2\%\pm1\%$ & $23.36\substack{+0.24\\-0.20}$ & $11.07\substack{+0.26\\-0.16}$ &$10.69\substack{+0.28\\-0.19}$ & $10.00\substack{+0.58\\-0.40}$ \\[3pt]
CDFS-1 & $1.358$ & $65\%\pm20\%$ & $0\farcs{}14\pm0\farcs{}07$ & $4.8\pm1.1$ & $22.47\substack{+0.40\\-0.29}$ & \nodata & \nodata & \nodata & $10.66\substack{+0.16\\-0.12}$ &$10.29\substack{+0.19\\-0.15}$ & $10.14\substack{+0.22\\-0.31}$ \\[3pt]
CDFS-229 & $4.329$ & $18\%\pm2\%$ & $0\farcs{}51\pm0\farcs{}03$ & $0.5\pm0.2$ & $21.57\substack{+0.14\\-0.13}$ & \nodata & \nodata & \nodata & $10.90\substack{+0.06\\-0.05}$ &$10.63\substack{+0.12\\-0.11}$ & $9.71\substack{+0.70\\-0.16}$ \\[3pt]
CDFS-321 & $3.998$ & $25\%\pm12\%$ & $0\farcs{}38\pm0\farcs{}12$ & $2.3\pm2.0$ & $20.34\substack{+0.70\\-0.42}$ & \nodata & \nodata & \nodata & $11.48\substack{+0.28\\-0.17}$ &$11.10\substack{+0.30\\-0.20}$ & $10.72\substack{+0.42\\-0.58}$ \\[3pt]
CDFS-724 & $1.355$ & $35\%\pm15\%$ & $0\farcs{}12\pm0\farcs{}03$ & $1.6\pm1.1$ & $23.70\substack{+0.58\\-0.38}$ & \nodata & \nodata & \nodata & $10.06\substack{+0.23\\-0.15}$ &$9.79\substack{+0.25\\-0.18}$ & $9.25\substack{+0.48\\-0.48}$ \\[3pt]
ECDFS-358 & $2.012$ & $56\%\pm14\%$ & $0\farcs{}36\pm0\farcs{}04$ & $1.7\pm0.5$ & $21.34\substack{+0.30\\-0.24}$ & \nodata & \nodata & \nodata & $11.11\substack{+0.12\\-0.10}$ &$10.73\substack{+0.16\\-0.14}$ & $10.22\substack{+0.44\\-0.47}$ \\[3pt]
SXDS-X1136 & $1.937$ & $41\%\pm8\%$ & $0\farcs{}10\pm0\farcs{}00$ & $2.0\pm0.5$ & $21.92\substack{+0.23\\-0.19}$ & \nodata & \nodata & \nodata & $10.75\substack{+0.09\\-0.08}$ &$10.49\substack{+0.14\\-0.13}$ & $10.10\substack{+0.36\\-0.48}$ \\[3pt]
SXDS-X50 & $1.423$ & $41\%\pm9\%$ & $0\farcs{}19\pm0\farcs{}04$ & $1.7\pm0.6$ & $21.99\substack{+0.27\\-0.21}$ & \nodata & \nodata & \nodata & $10.80\substack{+0.11\\-0.09}$ &$10.54\substack{+0.15\\-0.13}$ & $10.01\substack{+0.45\\-0.46}$ \\[3pt]
SXDS-X717 & $1.426$ & $61\%\pm9\%$ & $0\farcs{}26\pm0\farcs{}07$ & $5.6\pm1.4$ & $21.76\substack{+0.18\\-0.15}$ & \nodata & \nodata & \nodata & $10.77\substack{+0.07\\-0.06}$ &$10.51\substack{+0.12\\-0.12}$ & $10.38\substack{+0.18\\-0.29}$ \\[3pt]
SXDS-X735 & $2.203$ & $32\%\pm9\%$ & $0\farcs{}22\pm0\farcs{}06$ & $2.0\pm1.0$ & $20.92\substack{+0.33\\-0.25}$ & \nodata & \nodata & \nodata & $11.16\substack{+0.13\\-0.10}$ &$10.78\substack{+0.17\\-0.14}$ & $10.42\substack{+0.36\\-0.56}$ \\[3pt]
SXDS-X763 & $2.376$ & $6\%\pm4\%$ & $0\farcs{}69\pm0\farcs{}53$ & $2.4\pm0.8$ & $24.13\substack{+1.17\\-0.55}$ & \nodata & \nodata & \nodata & $9.95\substack{+0.47\\-0.22}$ &$9.68\substack{+0.48\\-0.24}$ & $9.35\substack{+0.48\\-0.56}$ \\[3pt]
SXDS-X969 & $1.613$ & $29\%\pm11\%$ & $0\farcs{}11\pm0\farcs{}02$ & $2.1\pm1.1$ & $21.59\substack{+0.52\\-0.35}$ & \nodata & \nodata & \nodata & $10.99\substack{+0.21\\-0.14}$ &$10.61\substack{+0.23\\-0.17}$ & $10.21\substack{+0.40\\-0.54}$ \\[3pt]
\enddata
\label{tab:result_sersic}
\tablecomments{
Column~1: Object ID.
Column~2-6: WFC3 inference. 
Column~2: Reduced $\chi ^2$ value by the best PSF in the library.
Column~7-9: ACS inference. 
Column~10: Observed host luminosity in the rest-frame R band.
Column~11: Host total stellar mass.
Column~12: Bulge stellar mass, using \sersic\ index as B/T proxy, see Section~\ref{sec:bh_bulge} for the details.
}
\end{deluxetable*}

\begin{table}
\centering
    \caption{Summary of $\gamma$ inferences}\label{table:gamma_sf}
     \resizebox{8cm}{!}{
     \begin{tabular}{cccc}
     \hline
     Sample & Selection effects &  \mbh-\smass & \mbh-\lhost \\
     &&&\\
     \hline\hline
32 AGNs + intermediate & No &  0.72$\pm$0.20 & 0.64$\pm$0.17 \\
32 AGNs + intermediate& Yes & 0.90$\pm$0.40 & 0.50$\pm$0.50 \\
32 AGNs & No & 1.03$\pm$0.25 & 1.07$\pm$0.23\\
32 AGNs& Yes & 1.50$\pm$0.80 & 1.80$\pm$0.80 \\
     \hline
     \end{tabular}}
    \tablecomments{
    Entire sample includes the 32 AGNs and the intermediate redshift AGNs from the reference in Section~\ref{sec:compare_sample}. Note that the adopted \lhost\ have been transferred to today assuming the passive evolution scenario using Equation~\ref{eq:L_relation}.
The results of selection effects used the uniform (flat) prior of \sint.
}
\end{table}

\begin{table}
\centering
    \caption{Observed $\gamma$ for \mbh-\smass\ without selection effect corrections}\label{table:diff_sample_gam}
     \resizebox{5.5cm}{!}{
     \begin{tabular}{cc}
     \hline
     Sample combination & $\gamma$ \\
     &\\
     \hline\hline
32 AGNs & 1.03$\pm$0.25\\    
30 AGNs (excluded outliers) &  0.89$\pm$0.27\\    
32 AGNs + intermediate &   0.72$\pm$0.20\\
32 AGNs bulge only &  2.09$\pm$0.30\\
     \hline
     \end{tabular}}
\tablecomments{These values are directly inferred for the observation, without corrections due to selection effects.}
\end{table}

\begin{table}
\centering
\caption{Data summary for the comparison \mbh-\smass\ sample}\label{tab:comp_sample}
\resizebox{18cm}{!}{
\begin{minipage}{0.5\textwidth}
\begin{tabular}{l c  p{2cm}c   c  }
\hline\hline
Object ID$^a$ & $z$ & $\log$\smass\ & $\log$\mbh \\
&&$(M_{\odot}$)&$(M_{\odot}$)\\ \\
\hline\hline
\multicolumn{2}{l}{Local sample by \citet{H+R04}}\\
\hline 
M87 & 0.0037$^b$ & 11.78$^c$$\pm$0.18 & 9.48$\pm$0.2 \\
NGC1068 & 0.0035 & 10.36$\pm$0.18 & 7.15$\pm$0.1 \\
NGC3379 & 0.0025 & 10.83$\pm$0.18 & 8.00$\pm$0.1 \\
NGC4374 & 0.0043 & 11.56$\pm$0.18 & 8.63$\pm$0.2 \\
NGC4261 & 0.0073 & 11.56$\pm$0.18 & 8.72$\pm$0.3 \\
NGC6251 & 0.0243 & 11.75$\pm$0.18 & 8.72$\pm$0.3 \\
NGC7052 & 0.0136 & 11.46$\pm$0.18 & 8.52$\pm$0.2 \\
NGC4742 & 0.0036 & 9.79$\pm$0.18 & 7.15$\pm$0.2 \\
NGC821 & 0.0056 & 11.11$\pm$0.18 & 7.57$\pm$0.2 \\
IC1459 & 0.0068 & 11.46$\pm$0.18 & 9.40$\pm$0.4 \\
M31 & 0.0002 & 10.57$\pm$0.18 & 7.65$\pm$0.2 \\
M32 & 0.0002 & 8.90$\pm$0.18 & 6.40$\pm$0.3 \\
NGC1023 & 0.0027 & 10.84$\pm$0.18 & 7.64$\pm$0.5 \\
NGC2778 & 0.0053 & 10.88$\pm$0.18 & 7.15$\pm$0.2 \\
NGC3115 & 0.0023 & 11.08$\pm$0.18 & 9.00$\pm$0.2 \\
NGC3245 & 0.0049 & 10.83$\pm$0.18 & 8.32$\pm$0.3 \\
NGC3377 & 0.0026 & 10.20$\pm$0.18 & 8.00$\pm$0.6 \\
NGC3384 & 0.0027 & 10.30$\pm$0.18 & 7.20$\pm$0.4 \\
NGC3608 & 0.0053 & 10.99$\pm$0.18 & 8.28$\pm$0.3 \\
NGC4291 & 0.0061 & 11.11$\pm$0.18 & 8.49$\pm$0.2 \\
NGC4342 & 0.0036 & 10.08$\pm$0.18 & 8.48$\pm$0.2 \\
NGC4473 & 0.0037 & 10.96$\pm$0.18 & 8.04$\pm$0.6 \\
NGC4564 & 0.0035 & 10.64$\pm$0.18 & 7.75$\pm$0.4 \\
NGC4594 & 0.0023 & 11.43$\pm$0.18 & 9.00$\pm$0.3 \\
NGC4649 & 0.0039 & 11.69$\pm$0.18 & 9.30$\pm$0.0 \\
NGC4697 & 0.0027 & 11.04$\pm$0.18 & 8.23$\pm$0.6 \\
NGC5845 & 0.0060 & 10.57$\pm$0.18 & 8.38$\pm$0.1 \\
NGC7332 & 0.0053 & 10.18$\pm$0.18 & 7.11$\pm$0.2 \\
NGC7457 & 0.0031 & 9.85$\pm$0.18 & 6.54$\pm$0.1 \\
Milky Way & 0.0000 & 10.04$\pm$0.18 & 6.57$\pm$0.2 \\
\hline\hline
\multicolumn{2}{l}{Local sample by \citet{Bennert++2011}}\\
\hline
0121$-$0102 & 0.0540 & 10.12$\pm$0.24 & 7.44$\pm$0.4 \\
0206$-$0017 & 0.0430 & 10.95$\pm$0.23 & 8.01$\pm$0.4 \\
0353$-$0623 & 0.0760 & 10.33$\pm$0.22 & 7.44$\pm$0.4 \\
0802$+$3104 & 0.0410 & 10.38$\pm$0.23 & 7.17$\pm$0.4 \\
0846$+$2522 & 0.0510 & 10.50$\pm$0.23 & 8.52$\pm$0.4 \\
1042$+$0414 & 0.0524 & 10.32$\pm$0.23 & 6.98$\pm$0.4 \\
1043$+$1105 & 0.0475 & 9.83$\pm$0.24 & 7.67$\pm$0.4 \\
1049$+$2451 & 0.0550 & 10.41$\pm$0.23 & 7.84$\pm$0.4 \\
1101$+$1102 & 0.0355 & 10.33$\pm$0.22 & 7.77$\pm$0.4 \\
1116$+$4123 & 0.0210 & 10.20$\pm$0.22 & 7.37$\pm$0.4 \\
1144$+$3653 & 0.0380 & 10.26$\pm$0.22 & 7.36$\pm$0.4 \\
1210$+$3820 & 0.0229 & 9.94$\pm$0.24 & 7.50$\pm$0.4 \\
1250$-$0249 & 0.0470 & 10.14$\pm$0.22 & 7.65$\pm$0.4 \\
1323$+$2701 & 0.0559 & 9.65$\pm$0.22 & 6.98$\pm$0.4 \\
1355$+$3834 & 0.0501 & 10.11$\pm$0.23 & 7.96$\pm$0.4 \\
1405$-$0259 & 0.0541 & 10.04$\pm$0.23 & 7.08$\pm$0.4 \\
1419$+$0754 & 0.0558 & 10.73$\pm$0.21 & 7.51$\pm$0.4 \\
1434$+$4839 & 0.0365 & 10.30$\pm$0.24 & 7.48$\pm$0.4 \\
1535$+$5754 & 0.0304 & 10.24$\pm$0.24 & 7.83$\pm$0.4 \\
1545$+$1709 & 0.0481 & 9.92$\pm$0.22 & 7.28$\pm$0.4 \\
1554$+$3238 & 0.0483 & 10.00$\pm$0.23 & 7.63$\pm$0.4 \\
1557$+$0830 & 0.0465 & 9.82$\pm$0.23 & 7.55$\pm$0.4 \\
1605$+$3305 & 0.0532 & 9.95$\pm$0.23 & 7.78$\pm$0.4 \\
1606$+$3324 & 0.0585 & 10.33$\pm$0.22 & 7.22$\pm$0.4 \\
1611$+$5211 & 0.0409 & 10.33$\pm$0.22 & 7.34$\pm$0.4 \\
\hline\hline
 \\\\ \\\\ \\

\end{tabular}
\end{minipage} \hfill

\begin{minipage}{0.6\textwidth}
\begin{tabular}{l c p{3cm}c c  }
\hline\hline
Object ID & $z$ & $\log$\smass\ & $\log$\mbh \\
&&$(M_{\odot}$)&$(M_{\odot}$)\\ \\
\hline\hline
\multicolumn{3}{l}{Intermediate sample by \citet{Bennert11}}\\
\hline
J033252$-$275119 & 1.227 & 10.58(9.83)$^d$$\pm$0.2 & 8.87$\pm$0.4 \\
J033243$-$274914 & 1.900 & 10.64(10.64)$\pm$0.2 & 9.17$\pm$0.4 \\
J033239$-$274601 & 1.220 & 10.54(10.54)$\pm$0.2 & 8.24$\pm$0.4 \\
J033226$-$274035 & 1.031 & 10.78(9.53)$\pm$0.2 & 7.85$\pm$0.4 \\
J033225$-$274218 & 1.617 & 10.61(10.61)$\pm$0.2 & 8.08$\pm$0.4 \\
J033210$-$274414 & 1.615 & 10.45(10.45)$\pm$0.2 & 8.30$\pm$0.4 \\
J033200$-$274319 & 1.037 & 9.62(9.62)$\pm$0.2 & 7.75$\pm$0.4 \\
J033229$-$274529 & 1.218 & 10.71(10.71)$\pm$0.2 & 8.37$\pm$0.4 \\
J123553$+$621037 & 1.371 & 10.90(9.99)$\pm$0.2 & 8.27$\pm$0.4 \\
J123618$+$621115 & 1.021 & 10.96(9.29)$\pm$0.2 & 8.35$\pm$0.4 \\
J123707$+$622147 & 1.450 & 10.74(10.74)$\pm$0.2 & 8.77$\pm$0.4 \\
\hline\hline
\multicolumn{2}{l}{Intermediate sample by \citet{SS13}}\\
\hline
158 & 0.717 & 10.66(10.66)$\pm$0.2 & 6.94$\pm$0.4 \\
170 & 1.065 & 9.80(9.50)$\pm$0.2 & 6.84$\pm$0.4 \\
271 & 0.960 & 10.49(9.85)$\pm$0.2 & 7.12$\pm$0.4 \\
273 & 0.970 & 10.29(10.29)$\pm$0.2 & 7.99$\pm$0.4 \\
305 & 0.544 & 11.01(11.01)$\pm$0.2 & 8.38$\pm$0.4 \\
333 & 1.044 & 10.33(9.87)$\pm$0.2 & 7.66$\pm$0.4 \\
339 & 0.675 & 10.96(10.96)$\pm$0.2 & 7.60$\pm$0.4 \\
348 & 0.569 & 10.55(10.15)$\pm$0.2 & 7.64$\pm$0.4 \\
379 & 0.737 & 10.66(10.17)$\pm$0.2 & 8.90$\pm$0.4 \\
413 & 0.664 & 10.19(9.80)$\pm$0.2 & 6.81$\pm$0.4 \\
417 & 0.837 & 10.39(9.91)$\pm$0.2 & 8.11$\pm$0.4 \\
465 & 0.740 & 10.76(10.14)$\pm$0.2 & 7.78$\pm$0.4 \\
516 & 0.733 & 10.88(10.71)$\pm$0.2 & 7.66$\pm$0.4 \\
540 & 0.622 & 10.88(10.57)$\pm$0.2 & 7.38$\pm$0.4 \\
597 & 1.034 & 10.91(10.80)$\pm$0.2 & 7.87$\pm$0.4 \\
712 & 0.841 & 11.34(11.19)$\pm$0.2 & 8.41$\pm$0.4 \\
\hline\hline
\multicolumn{2}{l}{Intermediate sample by \citet{Cisternas2011}}\\
\hline
J095817.54$+$021938.5 & 0.73 & 10.30$\pm$0.2 & 7.72$\pm$0.4 \\
J095819.88$+$022903.6 & 0.34 & 11.23$\pm$0.2 & 8.29$\pm$0.4 \\
J095831.65$+$024901.6 & 0.34 & 10.65$\pm$0.2 & 8.08$\pm$0.4 \\
J095840.61$+$020426.6 & 0.34 & 11.02$\pm$0.2 & 8.39$\pm$0.4 \\
J095845.80$+$024634.0 & 0.35 & 10.54$\pm$0.2 & 7.39$\pm$0.4 \\
J095902.76$+$021906.5 & 0.34 & 11.14$\pm$0.2 & 8.66$\pm$0.4 \\
J095909.53$+$021916.5 & 0.38 & 10.68$\pm$0.2 & 7.77$\pm$0.4 \\
J095928.31$+$022106.9 & 0.35 & 10.95$\pm$0.2 & 7.24$\pm$0.4 \\
J100002.21$+$021631.8 & 0.85 & 11.07$\pm$0.2 & 8.29$\pm$0.4 \\
J100012.91$+$023522.8 & 0.70 & 11.17$\pm$0.2 & 8.15$\pm$0.4 \\
J100014.55$+$023852.7 & 0.44 & 10.57$\pm$0.2 & 7.79$\pm$0.4 \\
J100017.54$+$020012.6 & 0.35 & 10.47$\pm$0.2 & 7.59$\pm$0.4 \\
J100025.25$+$015852.2 & 0.37 & 10.57$\pm$0.2 & 8.58$\pm$0.4 \\
J100028.63$+$025112.7 & 0.77 & 10.86$\pm$0.2 & 8.49$\pm$0.4 \\
J100029.69$+$022129.7 & 0.73 & 11.01$\pm$0.2 & 8.03$\pm$0.4 \\
J100033.38$+$015237.2 & 0.83 & 10.81$\pm$0.2 & 8.07$\pm$0.4 \\
J100033.49$+$013811.6 & 0.52 & 10.54$\pm$0.2 & 8.01$\pm$0.4 \\
J100037.29$+$024950.6 & 0.73 & 10.36$\pm$0.2 & 7.41$\pm$0.4 \\
J100043.15$+$020637.2 & 0.36 & 11.28$\pm$0.2 & 8.07$\pm$0.4 \\
J100046.72$+$020404.5 & 0.55 & 11.08$\pm$0.2 & 7.75$\pm$0.4 \\
J100058.71$+$022556.2 & 0.69 & 10.66$\pm$0.2 & 7.91$\pm$0.4 \\
J100118.52$+$015543.0 & 0.53 & 10.84$\pm$0.2 & 8.22$\pm$0.4 \\
J100141.09$+$021300.0 & 0.62 & 10.53$\pm$0.2 & 7.35$\pm$0.4 \\
J100146.49$+$020256.7 & 0.67 & 10.75$\pm$0.2 & 7.73$\pm$0.4 \\
J100202.22$+$024157.8 & 0.79 & 10.53$\pm$0.2 & 8.24$\pm$0.4 \\
J100205.03$+$023731.5 & 0.52 & 11.15$\pm$0.2 & 8.38$\pm$0.4 \\
J100212.11$+$014232.4 & 0.37 & 10.48$\pm$0.2 & 7.70$\pm$0.4 \\
J100218.32$+$021053.1 & 0.55 & 11.20$\pm$0.2 & 8.61$\pm$0.4 \\
J100230.06$+$014810.4 & 0.63 & 10.73$\pm$0.2 & 7.50$\pm$0.4 \\
J100230.65$+$024427.6 & 0.82 & 10.77$\pm$0.2 & 7.82$\pm$0.4 \\
J100232.13$+$023537.3 & 0.66 & 11.03$\pm$0.2 & 8.19$\pm$0.4 \\
J100243.96$+$023428.6 & 0.38 & 11.08$\pm$0.2 & 8.25$\pm$0.4 \\
\hline
\end{tabular}
\end{minipage}
}
\tablecomments{\\$^a$:  The object name is directly adopted from the reference.
\\$^b$:  For the \citet{H+R04} sample, the redshift value is calculated from the Galaxy distance, assuming perfect Hubble flow.
\\$^c$:  \smass\ for the local sample are the bulge masses.
\\$^d$:  If applicable, the values in brackets are the bulge mass.
}
\end{table}

\end{document}